\documentclass{article}
\usepackage[utf8]{inputenc}
\usepackage{fullpage}
\usepackage{amsmath,amsthm,amssymb}
\usepackage{amsfonts}
\usepackage{enumitem}
\usepackage{thmtools,thm-restate}
\usepackage{xcolor}
\usepackage{graphicx}
\usepackage{subcaption}
\usepackage[british]{babel}
\usepackage[autostyle]{csquotes}
\usepackage{mathtools}
\usepackage{relsize}
\usepackage{braket}
\usepackage{xfrac}
\usepackage[bookmarks=false]{hyperref}
\definecolor{darkgreen}{rgb}{0,0.5,0}
\hypersetup{linktocpage=true,colorlinks=true,citecolor=red,linkcolor=darkgreen}

\makeatletter
\newcommand{\leqnos}{\tagsleft@true\let\veqno\@@leqno}
\newcommand{\reqnos}{\tagsleft@false\let\veqno\@@eqno}
\reqnos
\makeatother

\newcommand{\sH}{\mathsf{H}}
\newcommand{\sM}{\mathsf{M}}
\newcommand{\sA}{\mathsf{A}}
\newcommand{\sZ}{\mathsf{Z}}
\newcommand{\sU}{\mathsf{U}}
\newcommand{\sI}{\mathsf{I}}
\newcommand{\polylog}{\mathrm{polylog}}
\newcommand{\eps}{\epsilon}
\newcommand{\tOmega}{\widetilde{\Omega}}

\newcommand{\wgt}{\mathsf{wt}}
\newcommand{\fhat}{\hat{f}}
\newcommand{\BE}{{\mathlarger{\mathbb{E}}}}
\newcommand{\BP}{{\mathlarger{\mathbb{P}}}}

\newcommand{\ind}{\mathbf{1}}
\newcommand{\sign}{\mathsf{sign}}

\newcommand{\BR}{\mathbb{R}}
\newcommand{\BN}{\mathbb{N}}
\newcommand{\ti}{\underline{i}}

\newcommand{\diag}{\mathsf{diag}}

\newcommand{\del}[2]{\frac{\partial #1}{\partial #2}}

\newcommand{\G}{U}

\newcommand{\tG}{V}
\newcommand{\mathscr}[1]{\mathcal{#1}}

\newcommand{\CQ}{\mathscr{Q}}

\newcommand{\CB}{\mathscr{B}}
\newcommand{\CN}{\mathscr{N}}

\newcommand{\Z}{Z}
\newcommand{\tj}{\underline{j}}

\newcommand{\free}{\mathsf{free}}
\newcommand{\fix}{\mathsf{fix}}
\newcommand{\CE}{\mathcal{E}}

\newcommand{\B}[1]{\mathbf{#1}}
\newcommand{\forr}{\mathsf{forr}}

\newcommand{\tO}{\widetilde{O}}
\newcommand{\bits}{\{\pm 1\}}
\newcommand{\cov}{\mathsf{\Sigma}}
\newcommand{\trnc}{\varphi}
\newcommand{\op}{\mathsf{op}}

\newif\ifnotes\notestrue

\ifnotes
 \usepackage{color}
 \definecolor{mygrey}{gray}{0.50}
 \newcommand{\notename}[2]{{\footnotesize{\bf (#1:} {#2}{\bf ) }}}

\else
 
 \newcommand{\notename}[2]{{}}
 
\fi

\newtheorem{theorem}{Theorem}[section]
\newtheorem{claim}[theorem]{Claim}
\newtheorem{proposition}[theorem]{Proposition}
\newtheorem{lemma}[theorem]{Lemma}
\newtheorem{corollary}[theorem]{Corollary}

\numberwithin{equation}{section}

\newcommand{\expref}[2]{{\texorpdfstring{\hyperref[#2]{#1~\ref{#2}}}{#1~\ref{#2}}}} 
\newcommand{\secref}[1]{\expref{Section}{#1}}
\newcommand{\thmref}[1]{\expref{Theorem}{#1}}
\newcommand{\clmref}[1]{\expref{Claim}{#1}}

\newcommand{\lref}[1]{\expref{Lemma}{#1}}
\newcommand{\corref}[1]{\expref{Corollary}{#1}}

\newcommand{\pref}[1]{\expref{Proposition}{#1}}
\newcommand{\figref}[1]{\expref{Figure}{#1}}

\title{$k$-Forrelation Optimally Separates Quantum and Classical Query Complexity}
\author{Nikhil Bansal\thanks{CWI Amsterdam and TU Eindhoven, \texttt{N.Bansal@cwi.nl}. Supported by the NWO VICI grant 639.023.812.} \and 
Makrand Sinha\thanks{CWI Amsterdam, \texttt{makrand@cwi.nl}. Supported by 
the NWO VICI grant 639.023.812.}
}
\date{}

\begin{document}

\maketitle

\begin{abstract}
    Aaronson and Ambainis (SICOMP `18) showed that any partial function on $N$ bits that can be computed with an advantage $\delta$ over a random guess by making $q$ quantum queries, can also be computed classically with an advantage $\delta/2$ by a randomized decision tree making ${O}_q(N^{1-\frac{1}{2q}}\delta^{-2})$ queries. Moreover, they conjectured the $k$-Forrelation problem --- a partial function that can be computed with $q = \lceil k/2 \rceil$ quantum queries --- to be a suitable candidate for exhibiting such an extremal separation. \medskip
    
     We prove their conjecture by showing a tight lower bound of $\widetilde{\Omega}(N^{1-1/k})$ for the randomized query complexity of $k$-Forrelation, where the advantage $\delta = 2^{-O(k)}$. By standard amplification arguments, this gives an explicit partial function that  exhibits an $O_\epsilon(1)$ vs $\Omega(N^{1-\epsilon})$ separation between bounded-error quantum and randomized query complexities, where $\epsilon>0$ can be made arbitrarily small. Our proof also gives the same bound for the closely related but non-explicit $k$-Rorrelation function introduced by Tal (FOCS `20). \medskip
     
     Our techniques rely on classical Gaussian tools, in particular, Gaussian interpolation and Gaussian integration by parts, and in fact, give a more general statement.  We show that to prove lower bounds for $k$-Forrelation against a family of functions, it suffices to bound the $\ell_1$-weight of the Fourier coefficients between levels $k$ and $(k-1)k$. We also prove new interpolation and integration by parts identities that might be of independent interest in the context of rounding high-dimensional Gaussian vectors. 
\end{abstract}

\newpage

\section{Introduction}
The last couple of decades have given us ample evidence to suggest that quantum computers can be exponentially more powerful in solving certain computational tasks than their classical counterparts. The \emph{black-box} or \emph{query} model offers a concrete setting to provably show such exponential speedups. In this model, a quantum algorithm has ``black-box access'' to the input and seeks to compute a function of the input while minimizing the number of queries. Most well-known quantum algorithms, such as Grover's search \cite{G96}, Deutsch-Josza's algorithm \cite{DJ92}, Bernstein-Vazirani's algorithm \cite{BV97}, Simon's Algorithm \cite{Simon97} or Shor's period-finding algorithm \cite{Shor97}, are captured by this black-box access model. There are slightly different models of black-box access to the input and in this work, we consider the most basic access model where each query returns a bit of the input. In this case, the classical counterpart is also commonly known as a randomized decision tree. There are many connections between the settings of quantum and randomized query complexity and for more details, we refer the reader to the survey by Buhrman and de Wolf \cite{BdW02}.

 The above raises a natural question that was first asked by Buhrman, Fortnow, Newman and R\"ohrig \cite{Bplus08}: what is the maximal possible separation between quantum and classical query complexities? Translating the results from slightly different query models to the setting where the queries return a bit of the input, Simon's problem \cite{Simon97} and a work of Childs et al. \cite{Cplus03} exhibited a separation of $O(\log^{2} N)$ quantum queries vs $\tOmega(\sqrt{N})$ randomized queries for partial functions on $N$ bits, while another work of de Beaudrap, Cleve and Watrous \cite{BCW02} implied a $1$ vs $\Omega(N^{1/4})$ separation. However, these works left open the possibility of a $O(1)$ vs $\Omega(N)$ separation, and towards answering this question, Aaronson and Ambainis \cite{AA14} showed that for $q = O(1)$, any $q$-query quantum algorithm can be simulated by a randomized algorithm making $O(N^{1-\frac{1}{2q}})$ queries, thus ruling out the possibility of a $O(1)$ vs $\Omega(N)$ separation. In particular, they proved the following fundamental simulation result.

\begin{theorem}[\cite{AA14}]
\label{thm:aa}
Let $\CQ$ be a quantum algorithm that makes $q$ queries to an input $x \in \{\pm1\}^N$. Then, with high probability, one can estimate $\BP[\CQ \text{ accepts } x]$ up to an additive $\delta$ factor by making $O(4^q N^{1-\frac{1}{2q}}\delta^{-2} )$ classical randomized queries to $x$. Moreover, these queries are also non-adaptive.
\end{theorem}

In the same paper, Aaronson and Ambainis showed that the (standard) Forrelation problem, exhibits a $1$ vs $\tOmega(\sqrt{N})$ separation, improving upon a $1$ vs $\Omega(N^{1/4})$ separation shown earlier by Aaronson \cite{S10} where the standard Forrelation problem was introduced. Given the above theorem and ignoring $\polylog(N)$ factors, this is the maximal separation possible when $q=1$. 

\cite{AA14} asked if \thmref{thm:aa} is also tight for any $q>1$. If true, this would imply an $O(1)$ vs $\Omega(N^{1-\eps})$ separation where $\eps=O(1/q)$ could be made arbitrarily small. Towards this end, they suggested a natural generalization of the standard Forrelation problem, that they called $k$-Forrelation, which we introduce next in a slightly more general setting.  


\paragraph{$(\delta,k)$-Forrelation.} Let $\sH = \sH_N$ denote the $N\times N$ Hadamard matrix where $N=2^n$ for $n \in \BN$ and $\sH$ is normalized to have orthonormal columns, and hence operator norm $1$. Let $k \geq 2$ be an integer and let $\ti = (i_1, \cdots, i_k) \in [N]^k$, and $z := (z_1, \cdots, z_k) \in \{\pm1\}^{kN}$. Define the  function $\forr_k : \{\pm1\}^{kN} \to \BR$ as follows 
\begin{equation}\label{eqn:forr}
    \ \forr_k(z) = \frac{1}{N} \sum_{\ti \in [N]^k} z_1({i_1}) \cdot \sH_{i_1,i_2} \cdot z_2({i_2}) \cdot \sH_{i_2,i_3}\cdots \cdot z_{k-1}(i_{k-1})\cdot \sH_{i_{k-1},i_{k}}\cdot z_k({i_k}).
\end{equation}
Observe that this function can be written as the following quadratic form:
\begin{equation}\label{eqn:forr-qform}
   \forr_k(z) = \frac1N \cdot z_1^\top (\sH\cdot \sZ_2\cdot \sH \cdot \sZ_3\cdots \cdot \sH\cdot \sZ_{k-1} \cdot \sH)z_k,
\end{equation}
where $\sZ_i = \diag(z_i)$ for $i \in \{2,\ldots, k-1\}$ is the diagonal matrix with $z_i \in \bits^{N}$ on the diagonal. From the above quadratic form description, it follows that $\forr_k(z) \in [-1,1]$ always, since $z_1/\sqrt{N}$ and $z_k/\sqrt{N}$ are unit vectors, and the operator norm of the matrix appearing in the quadratic form is at most $1$.

For a parameter $0<\delta<1$, the $(\delta,k)$-\emph{Forrelation function} is then defined in terms of $\forr_k$ as the following partial boolean function:
\begin{equation}\label{eqn:forrpartial}
    \ \forr_{\delta,k}(z) = \begin{cases} 1 ~~~~\text{ if }~~~~~\forr_k(z) ~\ge~ \delta, \text{ and } \\
    0 ~~~~\text{ if }~~~~ |\forr_k(z)| ~\le~ {\delta}/2.
    \end{cases}
\end{equation}
We overload the notation $\forr$ above to denote the real function $\forr_k$, as well as the partial boolean function $\forr_{\delta,k}$, but the reader should not have any ambiguity as to what is meant. Note that the standard Forrelation promise problem of \cite{AA14} is obtained by taking $\delta=3/5$ and $k=2$.

As already observed by \cite{AA14}, there is a simple and efficient quantum circuit that makes $\lceil k/2 \rceil$ queries and computes  $(\delta,k)$-Forrelation in the following manner.
\begin{proposition}[\cite{AA14}]
    \label{prop:aa}
    There exists a quantum circuit $\CQ$ that makes $\lceil k/2 \rceil $ queries and uses $O(k \log N)$ gates, such that for any input $z \in \bits^{kN}$, it holds that $\BP[\CQ \text{ accepts } z] = \tfrac{1}{2}(1+\forr_k(z))$.
\end{proposition}
The above implies a $\delta/{4}$ gap between the acceptance probabilities on the $1$-inputs and $0$-inputs for $(\delta,k)$-Forrelation. Standard tricks can then be used to show that with $\lceil k/2 \rceil$
 quantum queries and a quantum circuit of $O(k \log N)$ size, one can compute $(\delta,k)$-Forrelation with error at most $\frac12 - \delta/16$ on any input.

Combined with \thmref{thm:aa}, this also shows that the $(\delta,k)$-Forrelation function can be computed by making $O(2^{k} N^{1-1/k}\delta^{-2})$ classical randomized queries\footnote{For even $k$ this follows from \thmref{thm:aa} as $\lceil k/2 \rceil = k/2$. The bound also holds for odd $k$ as the proof of \thmref{thm:aa} in fact shows that any bounded \emph{block-multilinear} degree-$d$ polynomial can be approximated up to $\delta$ additive error with $O(2^{d} N^{1-1/d} \delta^{-2})$ randomized queries, and $\forr_k$ is a degree-$k$ block-multilinear polynomial for all $k$. The connection with query complexity arises as the acceptance probability of any $q$-query quantum algorithm can be written as such a polynomial of  degree $2q$.}, even non-adaptively. For even values of $k$, this exactly matches the bound in \thmref{thm:aa} (upto $\polylog(kN)$ factors assuming $k=O(\log\log N)$) and Aaronson and Ambainis \cite{AA14} proposed $(\delta,k)$-Forrelation as a candidate for extremal separations between classical and quantum query complexities. 

On the lower bound side, as mentioned before, Aaronson and Ambainis \cite{AA14} showed that $\Omega(\sqrt{N}/\log N)$ classical queries are required for standard Forrelation.  They also showed a slightly weaker lower bound of $\Omega(\sqrt{N}/\log^{7/2} N)$ for $(\delta, k)$-Forrelation, for $\delta=3/5$ and $k>2$. One can improve this lower bound slightly by observing the following: in the quadratic form description \eqref{eqn:forr-qform} above, if we take $z_2, \cdots , z_{k-1}$ to be the all-one strings, and $k$ is even, then $(\delta,k)$-Forrelation reduces to standard Forrelation as $\sH^{r} = \sH$ if $r$ is an odd natural number. So, the same $\Omega(\sqrt{N}/\log N)$ lower bound holds for $(\delta=3/5,k)$-Forrelation as well, if $k$ is even. Similarly, although not obvious, one can also design an input distribution achieving the same lower bound for odd $k$.

Thus, the current lower bounds for $(\delta,k)$-Forrelation do not exhibit a better than $O(1)$ vs $\tOmega(\sqrt{N})$ separation, still leaving whether \thmref{thm:aa} is tight for $q > 1$ wide open. 

\vspace{-2mm}

\paragraph{Beyond $O(1)$ vs.~$\tOmega(\sqrt{N})$ separation.}  Recently, motivated by this question,
Tal \cite{T20} considered a different variant of the $(\delta,k)$-Forrelation problem, that he refers to as \emph{$k$-Rorrelation}, to show a $\lceil k/2 \rceil$ vs $\tOmega(N^{{2/3} - O(1/k)})$ separation. In particular, Tal shows that if one replaces the Hadamard matrix $\sH$ in \eqref{eqn:forr} and \eqref{eqn:forrpartial} by a random orthogonal matrix $\sU$, then to compute the resulting random  partial function,
 one requires $\widetilde{\Omega}\left(N^{2(k-1)/(3k-1)}\right)$ classical queries  with high probability for parameters $(\delta=2^{-k},k)$. Moreover, any such function can still be computed with $\lceil k/2\rceil$ quantum queries, 
giving the $\lceil k/2 \rceil$ vs $\tOmega(N^{{2/3} - O(1/k)})$ separation.

While this breaks the $\sqrt{N}$ barrier,
the $k$-Rorrelation function is not explicit, and even though it is computable with a small number of quantum queries, the corresponding unitaries may not be efficiently implementable as a quantum circuit. This is in contrast to  $(\delta,k)$-Forrelation, 
where the resulting quantum query algorithms can also be efficiently implemented as a quantum circuit of polylogarithmic size. Tal's proof does not imply a better lower bound  for $(\delta,k)$-Forrelation than the $\widetilde{\Omega}(\sqrt{N})$ bound mentioned before, as it relies on various strong properties of random orthogonal matrices that the Hadamard matrix does not satisfy.

\subsection{Our Results}
In this work, we confirm the conjecture of Aaronson and Ambainis that $(\delta,k)$-Forrelation does exhibit an extremal separation between classical and quantum query complexities by proving the following lower bound. 
\begin{restatable}{theorem}{clower}
\label{thm:lower}
    Let $k \ge 2$ and $\delta=2^{-5k}$. Then, any randomized decision tree that computes $(\delta,k)$-Forrelation with error at most $\frac12-\frac{\eta}{2}$, must make at least the following number of queries, $$\Omega\left(  \frac{1}{k^{28}} \cdot \left(\frac{N}{\log (kN)}\right)^{1-\frac1k} \cdot \frac{\eta^2}{\log(1/\delta)}\right) = \Omega\left(  \frac{1}{k^{29}} \cdot \left(\frac{N}{\log (kN)}\right)^{1-\frac1k} \cdot \eta^2\right).$$  
\end{restatable} 

Note that for an even $k = O(1)$ and an advantage $\eta = \delta/16$, the above lower bound is $\tOmega(N^{1-1/k})$ and it matches the upper bound for $(\delta=\eps^k,k)$-Forrelation implied by \thmref{thm:aa}, up to a $\polylog(kN)$ factor. The bound is also tight for odd $k$, as  mentioned before. 

The previous statement gives a lower bound for randomized algorithms that have a $\Theta(\delta)$ advantage, since we wish to compare it to the advantage of the $\lceil k/2 \rceil$-query quantum algorithm which has a success probability of $1/2 + \Theta(\delta)$. If one wants a success probability of at least $2/3$, by using standard amplification tricks, the quantum query complexity of $(\delta,k)$-Forrelation becomes $O(k \cdot \delta^{-2}) = 2^{O(k)}$. This gives us that there exists an explicit partial boolean function on $M = kN$ bits that can be computed with error at most $1/3$ by quantum circuits of $O(2^{O(k)} \log M)$ size, making  $2^{O(k)}$ queries, but requires $M^{1-1/k}$ randomized queries.

For $k=O(1)$, this gives an $O(1)$ vs $\Omega(N^{1-\eps})$ bounded-error separation and taking $k=\alpha(N)$ where $\alpha$ is an arbitrarily slowly growing function
 of $N$, this yields an $\alpha(N)$ vs $\Omega(N^{1-o(1)})$ bounded-error separation between the quantum vs classical query complexity of an explicit partial function. More precisely, we have the following.

\begin{corollary}[Bounded Error Separation]\label{cor:smallerr}
Let $k \ge 2$ and  $\delta={2^{-5k}}$. Then, there exists a quantum circuit with $O(k\cdot 2^{10k} \cdot \log N)$ gates, making $O(k \cdot 2^{10k})$ queries that computes $(\delta,k)$-Forrelation with error at most $1/3$. On the other hand, any randomized decision tree that computes $(\delta,k)$-Forrelation with error at most $1/3$, needs
$\Omega\left(  \dfrac{1}{k^{29}} \cdot \left(\dfrac{N}{\log (kN)}\right)^{1-\frac1k} \right)$ queries. 
\end{corollary}

\emph{Remark. }{Our proof also works even if one replaces the Hadamard matrix $\sH$ in \eqref{eqn:forr} and \eqref{eqn:forrpartial} by an arbitrary orthogonal matrix $\sU$ where all entries are $\tO(N^{-1/2})$ in magnitude. In particular, the $\tOmega(N^{1-1/k})$ lower bound given above also holds for $k$-Rorrelation as all entries of a random orthogonal matrix are $O((N/\log N)^{-1/2})$ with high probability.}

\medskip
Next, we discuss some applications of our results. 

\vspace{-2mm}
\newcommand{\fn}{\mathsf{f}}
\paragraph{Query Separation for Total Boolean Functions.} Our results also imply an improved separation for total boolean functions. Let $Q(\fn)$ (resp.~$R(\fn)$) denote the minimum number of queries made by a quantum (resp. randomized) algorithm  to compute a (partial or total) boolean function $\fn$ with probability at least $2/3$. 

Then, the results of Aaronson, Ben-David and Kothari \cite{ABK16} imply that an $M^{o(1)}$ vs $M^{1-o(1)}$ separation between the quantum and randomized query complexity of a partial boolean function on $M$ bits implies the existence of a \emph{total} boolean function with cubic separation between the two measures. Combined with our results, this yields the following corollary.

\begin{corollary}
    There exists an explicit total boolean function $\fn$ for which $R(\fn) \ge Q(\fn)^{3-o(1)}$.
\end{corollary}

The recent work of Aaronson, Ben-David, Kothari, Rao and Tal \cite{ABKRT20} conjectures that for any total boolean function $\fn$, it always holds that $R(\fn) = O(Q(\fn)^3)$, so if true, the above separation is optimal up to $o(1)$ factors in the exponent. The current best upper bound is a $4^\text{th}$ power relationship which holds even for deterministic query algorithms: denoting by $D(\fn)$ the deterministic query complexity of $\fn$,   \cite{ABKRT20} prove that $D(\fn) = O(Q(\fn)^4)$. The above is tight for deterministic query algorithms due to an example of Ambainis et al. \cite{Aplus17}.

\paragraph{Separations in Communication Complexity.} 
\newcommand{\indx}{\mathsf{ip}}
\newcommand{\RCC}{\mathsf{RCC}}
\newcommand{\QCC}{\mathsf{QCC}}

Using the query to communication lifting theorem of  Chattopadhyay,  Filmus, Koroth, Meir and Pitassi~\cite{CFK19}, our results also imply analogous improved separations between quantum and classical communication complexity. In particular, let $\indx(x,y)$ be the inner product function where $x,y \in \bits^{2^{15}\log m}$. Then, for any function $f: \bits^{m} \to \bits$, the results of \cite{CFK19} imply that for the composed two-party function
$$ F(x,y) = f \circ \indx^m ~(x,y) := f(\indx(x_1,y_1), \ldots, \indx(x_m,y_m)),$$
the randomized communication complexity of $F$ with error at most $1/3$, denoted by $\RCC( F)$, satisfies $\RCC(F) = \Omega( \log m \cdot R(f))$ where $R(f)$ is the randomized query complexity of $f$.

Using the above with our results and denoting by $\QCC(F)$ the quantum communication complexity of $F$ with error $1/3$,  we have the following corollary. 
\begin{corollary} \begin{enumerate}[label=(\alph*)]
    \item There exists an explicit partial boolean function $F$ on $M$ bits, such that $\QCC(F) = O_\eps(\log M)$ while $\RCC(F) = \Omega(M^{1-\eps})$, where $\eps>0$ can be made arbitrarily small.
    \item  There exists an explicit total boolean function $F$ for which $\RCC(F) \ge \QCC(F)^{3-o(1)}$.
\end{enumerate}
\end{corollary}

The above results give a near optimal separation between quantum vs classical communication for partial functions improving upon the previous best known separation of $O(\log M)$ vs $\tOmega(\sqrt{M})$ for explicit partial functions (see \cite{KR11, G16}), or an $O(\log M)$ vs $\tOmega(M^{2/3-\eps})$ separation implied by the work of \cite{T20} for non-explicit functions. We remark that whether a polynomial relation holds between the quantum and classical communication complexity of a \emph{total} boolean function remains a very interesting open problem.

\subsection{Overview and Techniques}\label{sec:overview}
 Our proof of \thmref{thm:lower} is based on classical Gaussian tools, and builds on the stochastic calculus approach of Raz and Tal \cite{RT19}  for their breakthrough result on oracle separation between BQP and PH (see also the simplification of the results of \cite{RT19} by Wu \cite{Wu20}).
 
In fact, the input  distribution that \cite{RT19} use is a slight variant of the distribution used for standard Forrelation ($k=2$) by \cite{AA14}.
However, as also noted by \cite{T20},
 it is unclear how to use stochastic calculus already for $k=3$, as the hard input distribution for randomized query algorithms has a non-linear structure involving the product of two Gaussians (we elaborate more on this later). 

To get around this, our proof relies on using multilinearity of functions on the discrete cube and the properties of the underlying input distribution in a careful way, together with additional tools such as Gaussian interpolation and Gaussian integration by parts.
In this overview, we first focus on the special case of $k=3$ and  restrict to the simpler setting where the advantage $\delta = 1/\polylog^{k} (N)$. This setting will already suffice to illustrate the main difficulties in extending the previous approaches to prove lower bounds for $(\delta,k)$-Forrelation. 

\vspace{-2mm}

\paragraph{The case of $k=3$.} In this case, for $\ti=(i_1,i_2,i_3) \in [N]^3$ and $z = (z_1,z_2,z_3) \in \BR^{3N}$, we have
\[\forr_3(z) = \frac{1}{N} \sum_{\ti \in [N]^3}  z_1(i_1) \cdot \sH_{i_1,i_2} \cdot z_2(i_2) \cdot \sH_{i_2,i_3} \cdot z_3(i_3).\]
It is not hard to see that the uniform distribution on $\bits^{3N}$ is mostly supported on $0$-inputs for $\forr_3(z)$.  We will give a distribution $p_1(Z)$ on $\bits^{3N}$ --- a variant of the distribution considered in \cite{RT19, T20} --- that is mostly supported on $1$-inputs. 

 Given an arbitrary randomized decision tree making $d$ queries, let $f(z)$ be the acceptance probability of the decision tree on input $z$. To prove a lower bound it suffices to show that for any such $f$, the distinguishing advantage  $\left|\BE_{p_1}[f(Z)] - f(0)\right|$ is small,
 as $f(0)$ is exactly the average acceptance probability under the uniform distribution.\\

\noindent{\bf The distribution $p_1(Z)$.} 
Consider the $2N \times 2N$ covariance matrix $\cov =  \eps \begin{pmatrix} \sI_N & \sH_N\\ \sH_N & \sI_N \end{pmatrix}$ with $\eps = \Theta(1/\log N)$. A random Gaussian vector distributed as $\CN(0, \cov)$ will typically lie inside the cube $[-1/2,1/2]^{2N}$ as the variance of each coordinate is $O(1/\log N)$, and in this overview we assume that this is always the case, to avoid technicalities that can be dealt with truncating and bounding the error separately. Then, $p_1(Z)$ is the following distribution: 
Take two independent $2N$-dimensional Gaussian vectors $G = (U_1,V_1)$ and $B=(U_2,V_2)$ distributed as $\CN(0,  \cov)$ and  obtain a vector $Z \in \bits^{3N}$ by rounding each coordinate independently to $\pm 1$ with bias given by $(U_1, U_2 \odot V_1, V_2) \in [-1/2,1/2]^{3N}$.
Here $\odot$ denotes the Hadamard product\footnote{For $u,v \in \BR^m$, the Hadamard product is the vector $u \odot v \in \BR^m$ defined as $u \odot v = (u(1) \cdot v(1), \cdots, u(m) \cdot v(m))$.} of two vectors.  In other words, for  $i \in [N]$,
\begin{equation}\label{eqn:round}
    \ \BE_{p_1}[Z_1(i) \mid G, B] = U_1(i) \text{ and } \BE_{p_1}[Z_2(i) \mid G, B] = U_2(i) V_1(i) \text{ and } \BE_{p_1}[Z_3(i) \mid G, B] = V_2(i).
\end{equation}
Therefore,  we have
\begin{align}\label{eqn:roundadv}
    \ \BE_{p_1}[\forr_3(Z)] &= \frac{1}{N} \sum_{\ti \in [N]^3} \BE[U_1(i_1) \cdot \sH_{i_1,i_2} \cdot V_1(i_2)G_2(i_2)\cdot \sH_{i_2,i_3} \cdot V_2(i_3)] \notag \\
    \ & =  \frac{\eps^2}{N} \sum_{\ti \in [N]^3} \sH_{i_1,i_2}^2\sH_{i_2,i_3}^2 = \Theta\left(\frac{1}{\log^2 N}\right),
\end{align}
as $\BE[U_1(i)V_1(j)] = \BE[U_2(i)V_2(j)]=\eps \cdot \sH_{i,j}$, and since each entry of $\sH$ is $\pm \frac1{\sqrt{N}}$ and $\eps = \Theta(1/\log N)$.
 
 Extending $f$ from $\bits^{3N}$ to a function from $\BR^{3N}$ to $\BR$, by identifying it with its Fourier expansion, 
 and using the multilinearity of $f$ and the equalities in \eqref{eqn:round}, our task then reduces to showing that
 \begin{equation}\label{eqn:k3}
      \ \left|\BE_{p_1}[f(Z)] - f(0)\right| = \left|\BE[f(U_1, U_2 \odot V_1, V_2)] - f(0)\right| \ll 1/\log^2 N.
 \end{equation}

 \vspace{-2mm}

\paragraph{Previous approaches and their limitations.} This is the starting point of all\footnote{We remark that the original approach of \cite{AA14} does not fit in this framework and it is not clear how to generalize it either for $k>2$.} previous approaches to bounding the above, which essentially proceed in the following two ways.\\

\noindent{\bf (a) Bounding all moments and Fourier weight of all levels.}  
As $f(z) = \sum_{S \subseteq [3N]} \fhat(S) \chi_S(z)$ where $\{\chi_S(z)\}_{S \subseteq [3N]}$ are Fourier characters, 
one can bound 
\[ \left|\BE[f(U_1, U_2 \odot V_1, V_2)] - f(0)\right| \le \sum_{\ell=1}^d \wgt_\ell(f) \cdot \max_{|S|=\ell} \left|\BE[\chi_S(U_1, U_2 \odot V_1, V_2)]\right|,\]
writing $\wgt_\ell(f) = \sum_{|S|=\ell} |\fhat(S)|$, as the $\ell_1$-weight of the Fourier coefficients at level $\ell$.

This approach needs a bound on the Fourier weight $\wgt_\ell(f)$ for all levels $\ell \le d$, as well as a bound on all the moments $\left|\BE[\chi_S(U_1, U_2 \odot V_1, V_2)]\right|$, and consequently suffers from two drawbacks. First, the currently known bounds on $\wgt_\ell$ for decision trees degrade as $\ell$ gets large --- \cite{T20} shows that if $f$ is computable by a randomized decision tree of depth $d$, then $\wgt_\ell(f) \le \tO(d)^{\ell/2}$, which becomes weaker than the trivial bound of $\binom{d}{\ell}$ when $\ell \gg \sqrt{d}$.
For this reason the bound  of \cite{T20} for Rorrelation does not go beyond $\tOmega(N^{2/3-O(1/k)})$. 

Second, the moments can be very large for the Hadamard matrix (e.g.~due to very large submatrices with all $1/\sqrt{N}$ entries). This is not an issue if a random orthogonal matrix is used instead 
(which allows \cite{T20} to go beyond $N^{1/2}$ for Rorrelation). Another limitation is that using a worst case bound for the moment given by each Fourier character does not exploit the non-trivial cancellations that can occur for various terms in the sum.
In fact, it is not even clear how to obtain the $\tOmega(N^{1/2})$ bound for $k=2$ using this approach.\\

\noindent{\bf (b) Stochastic Calculus/Gaussian Interpolation.}
The second approach is 
based on utilizing the special properties of Gaussians and using tools from stochastic calculus \cite{RT19, Wu20}. In this paper, we describe an alternate approach using the classical method of Gaussian interpolation which can also be recovered by stochastic (It\^o) calculus. Gaussian interpolation is a way to continuously interpolate between jointly Gaussian random variables with different covariance structures. By choosing a suitable path to interpolate and controlling the derivatives along this path, one can compute functions of Gaussians with a more complicated covariance structure in terms of an easier one. Talagrand \cite{T11} dubs this the \emph{smart path method} to stress the important of choosing the right path. 

In particular, let $G \in \BR^m$ be a multivariate Gaussian and for an interpolation parameter $t \in (0,1)$, define $\B{G}(t) = \sqrt{t} \cdot G$. Then, the Gaussian interpolation formula (see \secref{sec:interpol}) implies that for any {\em reasonable} function $h: \BR^m \to \BR$ one has 
\begin{equation}
\label{eq:gauss-inter}
    \BE[h({G})] - h(0) = \int_0^1 \frac{d}{dt} 
    \left(\BE[h(\B{G}(t)]\right) dt = \frac12 \sum_{ij} \BE[G_iG_j] ~\int_0^1 \BE\left[\partial_{ij}h(\B{G}(t))\right] dt.
\end{equation}
in terms of the covariance of $G$ and the second derivatives $\partial_{ij}$ of $h$.

Note that if $h$ is a multilinear polynomial, then $\partial_{ij}h(0) = \hat{h}(ij)$ if $i \neq j$ while $\partial_{ii}h$ is identically zero. The right-hand side above involves partial derivatives at arbitrary points $\B{G}(t)$, but these can be reduced to derivatives at $0$ (and hence level-two Fourier coefficients $\hat{h}(ij)$) by a clever random restriction. In particular, the derivative
$\partial_{ij}h(\mu)$ at any $\mu \in [-1/2,1/2]^{m}$ can be interpreted as a Fourier coefficient with respect to a biased product measure (details given later). Thus, this approach only requires a bound on the level-two weight $\wgt_2(f)$, and works very nicely for $k=2$, as in that case our function is a multilinear function of a Gaussian and all the corresponding covariance entries in \eqref{eq:gauss-inter} where $i \neq j$ are $\pm \frac1{\sqrt{N}}$ (as opposed to the covariance entries where $i = j$ which are large). This gives a final bound of $\frac{\wgt_2(f)}{\sqrt{N}}$ for the expression in \eqref{eq:gauss-inter}.

 However for $k=3$, as also noted by \cite{T20}, it is not immediately clear how to use the interpolation approach to bound the expression in \eqref{eqn:k3}, as it involves a product of Gaussians. In particular, the second block of coordinates consists of products of coordinates of Gaussians $U_2$ and $V_1$.

\subsubsection{Our Approach}
Our main insight is that the advantage of $f$ in \eqref{eqn:k3} can essentially be bounded in terms of the Fourier weights of $f$ between levels three and six. For the particular distribution $p_1(Z)$ given by \eqref{eqn:round}, we can in fact bound the advantage of $f$ only in terms of the third and sixth level Fourier weights  (see \eqref{eq:k3form} for the precise statement). More generally for any $k\ge 3$, the advantage of $f$ can be bounded in terms of the Fourier weight of $f$ between  levels $k$ and $(k-1)k$.

 To show this in the simpler setting of the input distribution given by \eqref{eqn:k3}, we use Gaussian interpolation as in \eqref{eq:gauss-inter}. In particular, for $k=3$, given that our vector is of the form $(U_1, U_2 \odot V_1, V_2)$ and $f$ is a multilinear polynomial, we can treat the function $h$ in \eqref{eq:gauss-inter} as a function of the $4N$-dimensional Gaussian vector $(U_1,U_2, V_1, V_2)$. Similarly, for an arbitrary $k$, using a suitable generalization of the distribution $p_1(Z)$, we get a function $h$ of a $2(k-1)N$-dimensional Gaussian vector. The resulting expression in \eqref{eq:gauss-inter} is then a $k-1$ dimensional integral, which leads to partial derivatives of order $2k-2$ instead of $\partial_{ij}$ in \eqref{eq:gauss-inter} above. However, due to the interactions between the variables of $U_i$ and $V_{i-1}$ (an issue which does not arise for $k=2$), the partial deriatives with respect to $U_i$ and $V_{i-1}$ do not necessarily correspond to derivatives of $f$ (with respect to its coordinates), and a key technical idea is to use Gaussian integration by parts to relate them. In particular, the order $2k-2$ derivatives of $h$ can be related to derivatives of $f$ of order between $k$ and $(k-1)k$. 

We remark that a recent work of Girish, Raz and Zhan \cite{GRZ20} used a similar multi-dimensional stochastic walk to prove a lower bound for a different setting: they considered the partial function obtained by taking an XOR of multiple copies of the standard Forrelation problem, and their main focus was to prove a lower bound for quasipolynomially small advantage. The analysis for this setting is closer to the previously mentioned approaches of \cite{RT19,Wu20} for the standard Forrelation problem. In particular, the technical challenges that arise while trying to prove a better than $\tOmega(\sqrt{N})$ lower bound for $k$-Forrelation for $k>2$ do not arise in that case.
\\

\noindent{\bf The case of $k=3$ and polylogarithmic $\delta$.}
We explain the idea for $k=3$ and polylogarithmic $\delta$ first, which is quite a bit simpler, and then sketch the additional ideas needed for higher $k$ and for improving the advantage $\delta$ to $2^{-O(k)}$.
We will crucially leverage the multilinearity of the function $f$ and the specific structure of the random vector $(U_1, U_2 \odot V_1, V_2) \in \BR^{3N}$. In particular, let $S = S_1 \sqcup S_2 \sqcup S_3$ where $S_r$ for $r \in [3]$ is the projection of the subset on the $r^\text{th}$ block of coordinates and $\sqcup$ denotes the disjoint union of the sets. Consider the  monomial $\chi_S(z)$ in the multilinear representation of $f$. Using the multiplicativity of the characters, we have that 
    \[ \chi_S(U_1, U_2 \odot V_1, V_2) = \chi_{S_1}(U_1)\chi_{S_2}(U_2) \cdot \chi_{S_2}(V_1)\chi_{S_3}(V_2) .\]

Our starting point is that as $G=(U_1,V_1)$ and $B=(U_2,V_2)$ are independent, one can interpolate them separately, which leads to a two-dimensional integral in \eqref{eq:gauss-inter}, and the integrand on the right side ranges over the following derivatives
\begin{align} & \BE\left[\del{}{u_1(i_1)\partial v_1(j_{2})}\chi_{S_1}(\B{U}_1(t_1))\chi_{S_2}(\B{V}_1(t_1))\right] \BE\left[\del{}{u_2(i_2)\partial v_2(j_{3})}\chi_{S_2}(\B{U}_2(t_2))\chi_{S_3}(\B{V}_2(t_2))\right] \notag\\[1em] 
              =\, & \BE\,[{\chi_{S_1{\setminus i_1}} (\B{U}_1(t_1))\chi_{S_{2}\setminus j_{2}}(\B{V}_1(t_1))}] \cdot  \BE\,[{\chi_{S_2{\setminus i_2}} (\B{U}_2(t_2))\chi_{S_{3}\setminus j_{3}}(\B{V}_2(t_2))}] \notag\\[1em]
              =\, & \BE\,[{\chi_{S_1{\setminus i_1}} (\B{U}_1(t_1))\chi_{S_{2}\setminus j_{2}}(\B{V}_1(t_1))} \cdot  {\chi_{S_2{\setminus i_2}} (\B{U}_2(t_2))\chi_{S_{3}\setminus j_{3}}(\B{V}_2(t_2))}] , \label{eq:expl:k=3} 
\end{align}
         where $(i_1,i_2) \in S_1 \times S_2$, and $(j_2,j_3) \in S_2 \times S_3$, and $t_1,t_2 \in (0,1)$ are interpolation parameters which we will drop from the notation henceforth.

The main difference now from the $k=2$ case is that because of the presence of products $U_2 \odot V_1$, the above derivatives can not be interpreted in general as derivatives $\del{f}{z_A}(z)$ evaluated at $(\B{U}_1, \B{U}_2 \odot \B{V}_1, \B{V}_2)$.

Let us consider this more closely. Suppose that $i_2 = j_2$. In this case, \eqref{eq:expl:k=3} becomes
\[    \BE\,[{\chi_{S_1{\setminus i_1}} (\B{U}_1)\cdot\chi_{S_{2}\setminus j_{2}}( \B{U}_2 \odot \B{V}_1) \cdot \chi_{S_{3}\setminus j_{3}}(\B{V}_2)}],  \]
which corresponds to a third derivative of $\chi_{S}(z)$ evaluated at $z=(\B{U}_1,\B{U}_2 \odot \B{V}_1,\B{V}_2)$.

However, if $i_2 \neq j_2$, then the term in \eqref{eq:expl:k=3} does not correspond to a derivative of $f(z)$ with respect to $z$. To handle this, we note that $\chi_{S_{2}\setminus j_{2}}(\B{V}_1) \cdot \chi_{S_{2\setminus i_2}}(\B{U}_2)$ can be written as $\chi_{S_{2}\setminus\{i_2, j_{2}\}}( \B U_2 \odot \B V_1) \cdot \B U_2(j_2) \cdot \B V_1(i_1)$, and hence \eqref{eq:expl:k=3} becomes
\begin{equation}\label{eqn:tradeoff}
    \   \BE\,[{\chi_{S_1{\setminus i_1}} (\B U_1)\chi_{S_{2}\setminus\{i_2, j_{2}\}}( \B U_2 \odot \B V_1)  \chi_{S_{3}\setminus j_{3}}(\B V_2)}  \cdot \B U_2(j_2) \cdot \B V_1(i_1) ],
\end{equation}

In particular, the term in the expectation corresponds to the derivative of $\chi_S(\B U_1,\B U_2 \odot \B V_1,\B V_2)$ with respect to $J = \{i_1,i_2,j_2,j_3\}$
times the variables $\B U_2(j_2)$ and $\B V_1(i_2)$.
However, this exactly fits the form required to use the Gaussian integration 
by parts formula (see \secref{sec:interpol}), which says that for correlated real-valued Gaussians $B, G_1, \ldots, G_m$, and any reasonable function $h$ in the variables $x_1, \ldots, x_m$, the following holds:
\begin{align}\label{eqn:intbypartsintro}
     \BE[B \cdot h(G_1, \ldots, G_m)] = \sum_{i=1}^m \BE[BG_i]~ \BE\left[\del{h}{x_i}(G_1, \ldots, G_m)\right].
\end{align}

In particular, in \eqref{eqn:tradeoff}, one can trade off the factors $\B U_2(j_2)$ and $\B V_1(i_2)$ for one additional derivative each, giving us the sixth order derivatives for $\chi_S$.
Both the cases above eventually allow us to 
bound the function in terms of Fourier weight of $f$ at levels three and six. \\

To state the bound we obtain more formally, for $\mu \in [-1/2,1/2]^{3N}$, consider the product measure on $\bits^{3N}$ where the $i$-th bit is $1$ with probability $(1+\mu_i)/2$ and $-1$ with probability $(1-\mu_i)/2$, so that its bias is exactly $\mu_i$. Define the level-$\ell$ Fourier weight with respect to bias $\mu$ as  $\wgt_\ell^\mu(f) = \sum_{|S|=\ell} |\fhat^\mu(S)|$,
where $\fhat^\mu(S)$ is the Fourier coefficient with respect to the biased product measure above (see \secref{sec:fourier} for a formal definition). Then, we show the following key result towards bounding \eqref{eqn:k3}.
\begin{equation}
\label{eq:k3form}
    \left|\BE[f(U_1, U_2 \odot V_1, V_2)] - f(0)\right| \lesssim \sup_{\mu \in [-1/2,1/2]^{3N}} \frac{\eps}{N} \cdot \wgt^\mu_3(f) + \frac{\eps^2}{N^2} \cdot \wgt^\mu_6(f).
\end{equation}

By a random-restriction argument similar to that in previous works, the level-$\ell$ Fourier weight for a decision tree with respect to biased measures is essentially the same as the Fourier weight with respect to the uniform measure (see \corref{cor:wt} later) and hence at most $\tO(d)^{\ell/2}$ by the bounds in \cite{T20}.

Plugging these  bounds in \eqref{eq:k3form} above, yields that for a depth-$d$ randomized decision tree, the advantage is at most
\[\left|\BE[f(U_1, U_2 \odot V_1, V_2)] - f(0)\right| \le  \frac{\eps}{N} \cdot \tO(d)^{3/2} + \left(\frac{\eps}{N} \cdot \tO(d)^{3/2}\right)^2,\]
which is small for $d \ll N^{2/3}$. This gives the optimal bound for $(\delta, k=3)$-Forrelation, where $\delta=\Theta(1/\log^2 N)$. 

\paragraph{Arbitrary $k$ and polylogarithmic $\delta$.} For $k>3$, there is an additional complication that is not apparent in the case of $k=3$. 
In this case, a suitable generalization of the distribution $p_1(Z)$ involves $k-1$ independent $2N$-dimensional Gaussian vectors $(U_\kappa,V_\kappa)$, for $\kappa\in [k-1]$ distributed as $\CN(0, \cov)$.
Moreover, there are $k-2$ blocks of the form $U_{\kappa} \odot V_{\kappa-1}$ for $\kappa \in \{2,\ldots,k-1\}$ (see \secref{sec:outline} for the exact form). Due to this, when we apply Gaussian integration by parts to trade off the (unmatched) factors  $U_\kappa(i)$ and $V_\kappa(j)$ with extra derivatives, this can lead to several more additional factors. 

For example, suppose we apply Gaussian integration by parts to remove the factor $U_2(i)$, then since $U_2(i)$ is correlated with various $V_2(j)$ and each $V_2(j)$ appears together with a $U_3(j)$ in $V_2 \odot U_3$, upon differentiating with respect to variables in $V_2$, this leads to multiple new terms with factors $U_3(j)$.
Apriori, it is not obvious if applying Gaussian integration by parts leads to any progress. However, viewing this dynamics as a branching process and exploiting the multilinearity of the function $f$ and the specific structure of the distribution $p_1(Z)$, we can show using a careful counting argument, that this process eventually terminates without giving too many higher order derivative terms.

 In particular, even though the initial terms after the Gaussian interpolation step involve derivatives of order at most $2k-2$, we show that the final derivatives obtained after applying all the Gaussian integration by parts steps are of order $k,2k,\ldots,(k-1)k$. This allows us to show an overall bound on the advantage of $f$, in terms of the Fourier weight of $f$ at levels $k,2k,3k, \ldots,(k-1)k$ where the relative contribution of the higher level weights gets progressively smaller. In the end, plugging in the bounds on the Fourier weight, we can show that for an arbitrary $k$, the advantage a randomized depth-$d$ decision tree has is at most 
 \begin{equation}\label{eqn:kintro}
     \ \left|\BE_{p_1}[f(Z)] - f(0)\right| ~~\le~~  \sum_{m=1}^{k-1} \left(\frac{\eps}{N}\right)^{m(k-1)/2} \cdot \tO(d)^{mk/2} ~~=~~\sum_{m=1}^{k-1} \left( \left(\frac{\eps}{N}\right)^{1-1/k} \cdot \tO(d) \right)^{mk/2},
 \end{equation}
which is negligible if $d \ll N^{1-1/k}$. This gives the result for general $k$ when $\delta = 1/\polylog^k(N)$. For a detailed proof along the above lines (for the setting of $\delta = 1/\polylog^k(N)$), we refer to the previous version \cite{BSv1} of our paper which might be more accessible for an unfamiliar reader since the analysis is simpler.

In the present version of the paper, we work with a different distribution, where $\delta = 2^{-O(k)}$. This requires additional ideas that make the current analysis more involved and also leads to a bound in terms of the Fourier weight of all the levels between $k$ and $(k-1)k$ (see \thmref{thm:level}), as opposed to only the levels $k,2k, 3k, \ldots, (k-1)k$ that appear in \eqref{eqn:kintro} while analyzing the previous input distribution that had a polylogarithmic advantage.

\paragraph{Improving $\delta$ to $2^{-O(k)}$ with new  Interpolation and Integration by parts Identities.} To improve $\delta$ from $1/\polylog^k(N)$ to $2^{-O(k)}$, we need to revisit the issues that arise from rounding. Recall that eventually we want to generate an input distribution on the discrete hypercube $\bits^{kN}$.
One natural approach to do this is to truncate the high-dimensional Gaussians to $[-1/2,1/2]^{kN}$ so that one can round them to $\bits^{kN}$ as in \eqref{eqn:round}.

The choice of a suitable truncation function is crucial to be able to analyze the resulting quantities. In the proof overview given above, as well as in the previous version of our paper, this was achieved by scaling the Gaussians so that each coordinate has variance $\Theta(1/\log N)$. This way  the Gaussians themselves lie in $[-1/2,1/2]^{kN}$ typically, and then one can just work with the underlying Gaussian distribution directly in the analysis up to a small error that can be bounded separately. Revisiting \eqref{eqn:roundadv}, this results in the advantage being $1/\polylog^k(N)$. 

To improve the advantage to $2^{-O(k)}$, we want the underlying Gaussians to have constant variance, but in this case working with the Gaussians directly causes a large rounding error, so that the previous proof strategy does not give any bounds.

This necessitates working with a different truncation function. A natural choice is the function $\frac12 \cdot \sign: \BR \to \left\lbrace-\frac12, \frac12 \right\rbrace$ \footnote{Note that we truncate to $\left[-\frac12, \frac12\right]$ since the Fourier weights under the biased and unbiased measures are essentially the same if the bias is bounded away from $\pm1$ (see \thmref{thm:fwt}). Such a statement might still be true even if the bias is arbitrarily close to $\pm1$, but this seems more challenging to prove and is not needed for our analysis.}.  This is difficult to analyze directly (although this can perhaps be done using the techniques presented here in conjunction with the work of Eldan and Naor \cite{EN19}), and since for our application the exact constants are not so important, we work with the following truncation function: let us define $\trnc: \BR \to \left[-\frac12,\frac12\right]$  as
 \begin{align}\label{eqn:trncintro}
\ \trnc(s) = \Phi(s) - \frac12 = \int_0^s \gamma(s)ds,
 \end{align}
 where $\gamma$ and $\Phi$ are the density and cumulative distribution functions for the standard Gaussian in $\BR$ (see \secref{sec:prelims}).
 
We show that if $G$ is a multivariate Gaussian in $\BR^n$, then $\trnc(G) : = \trnc(G_1), \ldots, \trnc(G_n)$ morally behaves like a Gaussian for our analysis and satisfies analogous interpolation and integration by parts identities. For instance, we show that the following remarkable analogue of \eqref{eqn:intbypartsintro} holds: if $B, G_1, \ldots, G_m$ are real-valued random variables that are jointly Gaussian, then for any reasonable function $h$, we have 
\begin{align}\label{eqn:intbypartsintrophi}
     \BE[\trnc(B) \cdot h(G_1, \ldots, G_m)] = \sum_{i=1}^m \BE[BG_i]~ \BE\left[\Psi(B)\cdot \del{h}{x_i}(G_1, \ldots, G_m)\right],
\end{align}
where $\Psi$ is a non-negative function that is always bounded by one.  With some additional care, the identity above can be used in lieu of \eqref{eqn:intbypartsintro} to carry out the previous proof strategy even in the case of $\delta=2^{-O(k)}$.  

For more details, and for other related identities, we refer the reader to \secref{sec:id}. These identities might be of independent interest in the context of rounding high-dimensional Gaussian vectors.


\paragraph{Independent Work of Sherstov, Storozhenko and Wu.} In an independent work, Sherstov, Storozhenko and Wu \cite{SSW20} obtained a $\tOmega(N^{1-1/k})$ lower bound on the randomized query complexity of the non-explicit  $k$-Rorrelation partial function with advantage $\delta = 2^{-O(k)}$. The proof of \cite{SSW20} follows the previous approach of \cite{T20} and improves the Fourier bound on level-$\ell$ weight of depth $d$-decision trees from $\sqrt{d^\ell ~O(\log N)^{\ell-1}}$ to $\sqrt{\binom{d}{\ell}~ O(\log N)^{\ell-1}}$ for all levels $\ell \le d$. This was the only bottleneck in the approach of \cite{T20} for $k$-Rorrelation, and thus \cite{SSW20} obtain a $\tOmega(N^{1-1/k})$ lower bound on the randomized query complexity of $k$-Rorrelation. 

Using the new ideas in the current version of our paper (where $\delta$ is improved to $2^{-O(k)}$ from $1/\polylog^k (N)$), our work gives the same $\tOmega(N^{1-1/k})$ lower bound for $k$-Rorrelation. We also obtain the same results for the explicit $k$-Forrelation problem and it is unclear if this can be achieved with the other approach.  %

The techniques of \cite{SSW20} are incomparable to ours as their main focus is on proving optimal bounds on the Fourier weights of decision trees. In contrast, we improve upon a different aspect of the proof --- we show a finer bound on the advantage of any depth-$d$
 decision tree where the requirements are relaxed in two ways. First, we only need low-level Fourier weights, and in this regime the previous bounds of \cite{T20} are already sufficient to give us a tight lower bound on the randomized decision tree complexity of $k$-Forrelation/Rorrelation, and second, the only property of the underlying orthogonal matrix we need is an absolute bound of $\tO(N^{-1/2})$ on the entries, which holds for the Hadamard matrix as well as for a random unitary matrix, as compared to the approach of \cite{T20} and \cite{SSW20} which requires strong bounds on the operator norm of all submatrices --- the latter being the main reason why our approach works for the Hadamard matrix. 
 
 In addition, there have been significant recent breakthroughs in analyzing functions over the discrete cube with continuous methods, such as a stochastic characterization of Goemens-Williamson rounding \cite{EN19}, or the work of Eldan and Gross \cite{EG20} that proved a conjecture of Talagrand in the analysis of boolean functions. The new interpolation and integration by parts identities we prove here give us additional tools that might be useful in further application of continuous techniques in  theoretical computer science and mathematics.

\subsubsection{Organization}
The rest of the paper is organized as follows. We introduce the notation and basic preliminaries in \secref{sec:prelims}. \secref{sec:outline} gives the input distribution, shows that the chosen input distribution has a large support on the $1$ and $0$ inputs of $(\delta,k)$-Forrelation, and also gives a formal outline of the main proof. \secref{sec:id}  introduces new interpolation and integration by parts identities that will be used repeatedly in the proof. \secref{sec:lower} contains the proof of the lower bound on randomized query algorithms.

\section{Preliminaries}\label{sec:prelims}

\noindent{\bf Notation.}
 Throughout this paper, $\log$ denotes the natural logarithm unless the base is explicitly mentioned. We use $[k]$ to denote the set $\{1,2,\dotsc, k\}$. For a singleton set $\{x\}$, we sometimes write $x$ for brevity. The set of natural numbers including zero is denoted by $\BN_0$. Matrices are denoted by capital serif fonts (e.g. $\sA$).
 
 For a random vector (or bit-string) $z$ in $\BR^n$, we will use $z_i$ or $z(i)$ to denote the $i$-th coordinate of $z$, depending on whether we need to use the subscript for another index. If $z \in \BR^{kn}$, then we will write $z=(z_1,\ldots,z_k)$ where $(z_\kappa)_{\kappa \in [k]}$ are vectors in $\BR^n$ to denote the projections on the coordinates $\{(\kappa-1)n,\ldots,\kappa n\}$ --- in this case, we will explicitly mention that $(z_\kappa)_{\kappa \in [k]}$ are vectors so that there is no ambiguity that $z_\kappa$ refers to a coordinate of $z$. The operator and Frobenius norms of a matrix $\sM$ are denoted by $\|\sM\|_{\op}$ and $\|\sM\|_F$.
 
Random variables are denoted by capital letters (e.g.\ $A$) and values they attain are denoted by lower-case letters possibly with subscripts and superscripts (e.g.\ $a,a_1,a'$, etc.). Events in a probability space will be denoted by script letters (e.g.\ $\CB$). We use $\ind_\CB$ or $\ind[\CB]$ to denote the indicator random variable for the event $\CB$. Given a random variable $X$ in a probability space $p$, we write $p(X)$ to denote the distribution of $X$ in the probability space. For random variables $X,Y$, we write $p(X,Y)$ to denote the joint distribution and $p(X)$ to denote the marginal distribution. We write $p(\CB)$ to denote the probability of the event $\CB$. For $\lambda \in [0,1]$, we use $\lambda p(X) + (1-\lambda) p'(X)$ to denote the convex combination of the two distributions, where the random variable $X$ is sampled from $p(X)$  with probability $\lambda$,  and  from $p'(X)$ with probability $1-\lambda$.

For a real valued function $f$, we write $\BE_{p}[f(X)]$ to denote the expectation of the random variable $f(X)$ where $X$ is in the probability space $p$. Similarly, $\BE_p[f(X) \mid Y]$ denotes the conditional expectation of $f(X)$ with respect to $Y$. If the probability space $p$ is clear from the context, we simply write  $\BE[f(X)]$ and $\BE[f(X) \mid Y]$. 
We use $\CN(0,\sigma^2)$ to denote a Gaussian random variable in $\BR$ with mean zero and variance $\sigma^2$. For a positive semi-definite matrix $\cov \in \BR^{m \times m}$, we write $\CN(0,\cov)$ to denote a centered (mean-zero) Gaussian random variable in $\BR^m$ with covariance $\cov$. We call an $m$-dimensional Gaussian standard, if $\cov$ is the identity matrix $\sI_m$.

\subsection{Gaussian Tools}

\paragraph{Gaussian Concentration.}

Let us denote the density and cumulative distribution function for the standard Gaussian $\CN(0,1)$ by
\[
\gamma(s)=\frac{1}{\sqrt{2\pi}}e^{-s^2/2} ~~~\text{ and }~~~ \Phi(s)= \int_{-\infty}^s \gamma(t)dt.\]
The following estimate is standard.
\begin{proposition}[Gaussian Concentration]\label{prop:gconc}
    For any $a>0$, we have $1-\Phi(a) \le \frac12 e^{-a^2/2}.$
\end{proposition}
\begin{proof}
We have that 
    \[ 1-\Phi(a) = \frac{1}{\sqrt{2\pi}}\int\limits_a^\infty e^{-s^2/2} ds = \frac{1}{\sqrt{2\pi}}\int\limits_0^\infty e^{-(a+s)^2/2} ds \le \frac{e^{-a^2/2}}{\sqrt{2\pi}} \int\limits_0^\infty e^{-s^2/2}ds = \frac12 e^{-a^2/2}. \qedhere\]
\end{proof}

Recalling the double factorial notation,  $(2k+1)!! = (2k+1)\cdot(2k-1)\cdot \cdots \cdot 3\cdot 1$ for any non-negative integer $k$, the following series expansion for the normal CDF will be very convenient. 
\begin{proposition}[Series Expansion]\label{prop:ser}
    For every $a \in \BR$, we have that
    $$\Phi(a) = \frac12 + \gamma(a)\sum\limits_{k=0}^\infty \frac{a^{2k+1}}{(2k+1)!!}.$$
\end{proposition}
\begin{proof}
    We write 
    \begin{align}\label{eqn:ser1}
    \ \Phi(a) - \frac12 = \frac1{\sqrt{2\pi}} \int\limits_0^a e^{-s^2/2}ds = \frac{e^{-a^2/2}}{\sqrt{2\pi}} \cdot \underbrace{e^{a^2/2}\int\limits_0^a e^{-s^2/2}ds}_{:=\eta(a)} = \gamma(a) \cdot \eta(a). 
    \end{align}

    Note that the Taylor series of $\eta(a)$ has an infinite radius of convergence, so let $\eta(a) = \sum_{j=0}^\infty \beta_j a^j$ for real coefficients $\beta_j$. We note that $\eta'(a) = 1 + a\eta(a)$ which implies that $\beta_{j+2} = \frac{\beta_j}{j+2}$. Moreover, since $\eta(0)=0$ and $\eta'(0)=1$, for any non-negative integer $k$, we have that $\beta_{2k}=0$ and $\beta_{2k+1}=\frac{1}{(2k+1)!!}$. This implies that  $\eta(a) = \sum_{k=0}^\infty \frac{a^{2k+1}}{(2k+1)!!}$. Plugging it in \eqref{eqn:ser1} completes the proof.
\end{proof}

\paragraph{Gaussian Derivatives.} From the definition of Hermite polynomials, we have that
\begin{equation}\label{eqn:hermite}
\ \gamma^{(n)}(s) = \frac{d}{ds^n}\gamma(s) = {(-1)^n} \cdot h_n(s) \gamma(s),
\end{equation}
where the $h_n(s)$ are the (probabilists') Hermite polynomials. Moreover, it is also well known (see \cite{I61}) that $|h_n(s)| \le \sqrt{ n!} \cdot e^{s^2/4}$. This implies that 
\begin{equation}\label{eqn:hermitebound}
    \ |\gamma^{(n)}(s)| \le \frac{1}{\sqrt{2\pi}} \cdot \sqrt{n!} \le n^{n/2} \text{ for every } s \in \BR. 
\end{equation}

\paragraph{Owen's-$T$ function.}

Owen's-$T$ function \cite{O56} is defined as 
\begin{align}\label{eqn:owen1}
    \ T(h,\sigma) = \frac{1}{2\pi} \left( \arctan \sigma - \int\limits_0^h \int\limits_0^{\sigma x} e^{-(x^2+y^2)/2}dy dx \right).
\end{align}
Note that $T(-h, \sigma) = T(h, \sigma)$ and  $T(h, -\sigma) = - T(h, \sigma)$. Moreover, for $h,\sigma \geq 0$, the value $T(h,a)$ equals the probability $\BP[X \geq h \text{ and }  0 \leq Y \leq \sigma X]$ where $(X,Y)$ is standard Gaussian in $\BR^2$.

An alternate expression (see (3.3) in \cite{O56}) for $T(h,\sigma)$, involving only a single integral is 
\begin{align}\label{eqn:owen2}
    \ T(h,\sigma) = \frac{1}{2\pi} \int_0^\sigma \frac{e^{-h^2(1+x^2)/2}} { 1+x^2 } dx.
\end{align}
To see that the expressions in \eqref{eqn:owen2} and \eqref{eqn:owen1} are equal, one can differentiate the right hand side of \eqref{eqn:owen2} with respect to $h$ and integrate it back after a substitution.

\vspace{-2mm}

\paragraph{Gaussian Interpolation and the Smart Path Method.}\label{sec:interpol}
We refer to Talagrand's book \cite{T11} for a nice exposition, and in particular, \S 1.3 and Appendix A.4 there, for proofs of the lemmas given below.

Let $f: \BR^n \to \BR$ be an infinitely differentiable function. We say that $f$ is of moderate growth if all partial derivatives of $f$ satisfy the following
\begin{equation}\label{eqn:growth}
 \lim_{\|x\| \to \infty} \left|\partial_{\ti}f(x)\right|\: e^{-a\|x\|^2} = 0 \text{ for every } \ti = (i_1, \cdots, i_n) \in \BN_0^{n} \text{ and } a \in \BR_{>0},
\end{equation}
where $\partial_{\ti}$ denotes the partial derivative $\displaystyle \del{}{x_1^{i_1}}\cdots\del{}{x_n^{i_n}}$ and $\|\cdot\|$ is the Euclidean norm. One can check that multivariate polynomials are always of moderate growth, and also, the truncation function $\trnc$ given in $\eqref{eqn:trncintro}$ is of moderate growth, since all its derivatives are bounded as shown by \eqref{eqn:hermitebound}. Moreover, if $f, g: \BR^n \to \BR$ are of moderate growth, then so is $f(x)g(x)$. Lastly, if $f$ is a multivariate polynomial and $q: \BR \to \BR$ satisfies  \eqref{eqn:growth}, then $f(q(x_1), \cdots, q(x_n))$ also satisfies the moderate growth condition of \eqref{eqn:growth}.\\

Consider $f: \BR^n \to \BR$ satisfying the moderate growth condition and consider two centered jointly Gaussian random vectors ${G}$ and ${B}$ in $\BR^n$. Let us define $\B{G}(t) = (\B{G}_i(t))_{i \le n}$ where 
\begin{equation}
    \B{G}_i(t) = \sqrt{t} ~G_i + \sqrt{1-t} ~B_i,
\end{equation}
so that ${G}=\B{G}(1)$ and ${B}=\B{G}(0)$ and consider the function
\begin{equation}
    \zeta(t) =  \BE[f({\B{G}}(t))].
\end{equation}
For clarity, we will use boldface font to refer to the interpolating Gaussian.

\begin{lemma}[Gaussian Interpolation]\label{lem:interpol}
For $0<t<1$ we have
\[\zeta'(t) = \frac12 \sum_{ij} \left(\BE[G_iG_j] - \BE[B_iB_j]\right)~ \BE\left[\del{f}{x_i\partial x_j}(\B{G}(t))\right].\]
\end{lemma}
Choosing the covariance of ${B}$ to be the all zero matrix, we have that $\B{G}(t)=\sqrt{t}~{G}$, and the following useful identity follows from the previous lemma by the fundamental theorem of calculus
\begin{align*}
    \BE[f({G})] - f(0) = \int_0^1 \zeta'(t)dt = \frac12 \sum_{ij} \BE[G_iG_j] ~\int_0^1 \BE\left[\del{f}{x_i\partial x_j}(\B{G}(t))\right] dt.
\end{align*}
 We remark that one can derive the same formula using It\^o calculus.

Another important tool that we will use is the multivariate Gaussian integration by parts formula.
\begin{lemma}[Gaussian Integration by Parts]\label{lem:intbyparts}
If $B, G_1, \ldots, G_n$ are real-valued random variables that are jointly Gaussian and $f:\BR^n \to \BR$ is of moderate growth, then
\[ \BE[B \cdot f(G_1, \ldots, G_n)] = \sum_{i=1}^n \BE[BG_i]~ \BE\left[\del{f}{x_i}(G_1, \ldots, G_n)\right].\]
\end{lemma}

Note that this formula replaces the expectation of the product of a Gaussian random variable with the function $f$, with a weighted sum of expectation of the derivatives of $f$.

The Gaussian integration by parts formula can be used to prove \lref{lem:interpol} and it turns out that  it also uniquely characterizes the multivariate Gaussian distribution.

\subsection{Fourier Analysis on the Discrete Cube}
\label{sec:fourier}
We give some facts from Fourier analysis on the discrete cube that we will need, and for more details we refer to the book \cite{OD14}.
Every boolean function $f: \{\pm1\}^m \to \BR$ can be written uniquely as a sum of monomials $\chi_S(x) = \prod_{i \in S} x_i$,
\begin{align}\label{eqn:expansion}
    f(x) = \sum_{S \subseteq [m]} \fhat(S) \chi_S(x),
\end{align}
where $\fhat(S) = \BE_{p}[f(X)\chi_S(X)]$ is the Fourier coefficient with respect to the uniform measure $p$ on $\{\pm 1\}^m$. The monomials $\chi_S(x) = \prod_{i \in S} x_i$ form an orthonormal basis for real-valued functions on $\{\pm1\}^m$, called the \emph{Fourier basis}.

Any function on  $\{\pm 1\}^m$ can be extended to $\BR^m$ by identifying it with the multilinear polynomial given by \eqref{eqn:expansion}, which is also called the \emph{harmonic extension} of $f$ and is unique. We will denote the harmonic extension of $f$ also by $f$ and in general, we have the following identity by interpolating the values of $f$ on the vertices of the discrete hypercube.
\begin{align}\label{eqn:extension}
    f(x) = \sum_{y \in \{\pm 1\}^m} w_x(y) f(y), ~~~\text{ where }~~~ w_x(y) = \prod_{i=1}^m \frac{1+x_iy_i}{2} \text{ for any } x\in \BR^m.
\end{align}
The above implies that for a boolean function $f:\{\pm 1\}^m\rightarrow [-1,1]$, the harmonic extension of $f$ also satisfies $\max_{x \in [1,1]^m}|f(x)| \le 1$.

The discrete derivative of a function on the hypercube $\{\pm 1\}^m$ is given by  
\[ \partial_i f(x) = \frac{1}{2} (f(x^{i\to 1}) - f(x^{i\to -1})),\]
where $x^{i\to b}$ is the same as $x$ except that the $i$-th coordinate is set to $b$. It is easily checked that the harmonic extension of $\partial_i f(x)$ is the real partial derivative $\del{}{x_i}$ of the harmonic extension of $f$ and we will identify it as such.  
Furthermore, for a boolean function $f:\{\pm 1\}^m\rightarrow [-1,1]$, 
the discrete derivative at any point $x \in \bits^m$ also satisfies $|\partial_i f(x)| \leq 1$ and hence \eqref{eqn:extension} implies that $\max_{x \in [1,1]^m}|\partial_A f(x)| \le 1$ for any $A \subseteq [m]$ identifying $\partial_A f$ as the harmonic extension of the real partial derivative of $f$. Moreover, from \eqref{eqn:expansion}, it also follows that
\begin{equation}\label{eqn:derivative}
    \partial_A f(x) = \sum_{S: S \supseteq A}\fhat(S) \chi_{S \setminus A}(x)
\end{equation}
for any subset $A \subseteq [m]$. The above also implies that $\partial_A f(0) = \fhat(A)$.

The level-$\ell$ Fourier weight of $f$ is defined as $\wgt_\ell(f) = \sum_{|S|=\ell} |\fhat(S)|$.

 For a function $f(x_1,\ldots,x_m)$, a restriction $\rho \in \{-1,1,\star\}^m$ gives a partial assignment to the variables $(x_i)_{i \le m}$. We denote the set of coordinates of $\rho$ whose value is $\star$ as $\free(\rho)$ while the set of coordinates that are fixed to $\pm 1$ is denoted by $\fix(\rho)$. We use $f_\rho$ to denote the function obtained from $f$ by setting the variables in $\fix(\rho)$ to the values given by $\rho$.

\vspace{-2mm}
\paragraph{Fourier basis for biased measures.}

For a proofs of the results below, see Chapter 8 in \cite{OD14}.
Given any $\mu \in (-1,1)^m$, let $p_\mu(X)$ be the biased product distribution over $\{\pm1\}^m$ such that each coordinate of $X \in \bits^m$ is sampled independently so that $X_i=1$ with probability $(1+\mu_i)/2$ and $X_i=-1$ with probability $(1-\mu_i)/2$. So the expectation and the variance of $X_i$ are
\[ \BE_{p_\mu}[X_i] = \mu_i, ~~\text{ and }~~ \BE_{p_\mu}[(X_i- \mu_i)^2]= 1-\mu_i^2. \]
Then, the Fourier basis with respect to the biased product measure $p_\mu$ is given by the following functions indexed by subsets $S \subseteq [n]$:
\[\phi^\mu_S(x) = \prod_{i \in S} \phi^\mu_i(x), ~~~~\text{ where } ~~~~\phi^\mu_i(x) = \frac{x_i-\mu_i}{\sigma_i},\]
with $\sigma_i = (1-\mu_i^2)^{1/2}$ being the standard deviation of the biased random bit $X_i$.
Note that 
\[ \BE_{p_\mu}[\phi^\mu_S(X)^2] ~~=~~ \prod_{i\in S}~ \BE_{p_\mu}[\phi^\mu_i(X)^2] ~~=~~  \prod_{i \in S}~ \frac1{\sigma_i^2} \cdot {\BE_{p_\mu}[(X_i - \mu_i)^2]}~~=~~ 1,\]
and that $\BE_{p_\mu}[\phi^\mu_S(X)\phi^\mu_T(X)] = 0$ if $S \neq T$. So the functions $\phi_S^\mu(x)$ form an orthonormal basis for real-valued functions on $\bits^m$ with respect to the inner product obtained by taking expectation under $p_\mu$.
The Fourier expansion with respect to the biased product measure $p_\mu$ is given by 
\begin{align}\label{eqn:biased-expansion}
    f(x) = \sum_{S \subseteq [n]} \fhat^\mu(S) \phi^\mu_S(x),
\end{align}
where $\fhat^\mu(S) = \BE_{p_\mu(x)}[f(x)\phi^\mu_S(x)]$ are the Fourier coefficients with respect to $p_\mu$.

The discrete derivative with respect to $\phi_i^\mu$ is defined as 
\begin{equation}\label{eqn:bias-derivative}
    \ \partial_i^\mu f(x) :=  \frac{f(x^{i\to 1}) - f(x^{i\to -1})}{\phi^\mu_i(1) - \phi^\mu_i(-1)} = \sigma_i \cdot \frac{f(x^{i\to 1}) - f(x^{i\to -1})}{2} = \sigma_i\cdot \partial_i f(x),
\end{equation}
where $\partial_i f(x)$ is the discrete derivative with respect to the standard Fourier basis (with respect to the uniform measure over $\bits^m$).

Since $\partial_i f$ can be viewed as the real partial derivative of the harmonic extension of $f$, using the chain rule for taking derivatives, $\displaystyle  \del{f}{\phi^{\mu}_i} = \sigma_i \cdot \partial_i f$, so one can identify $\partial_i^\mu f$ as the real partial derivative $\displaystyle  \del{f}{\phi^{\mu}_i}$ for the harmonic extension of $f$. Moreover, from \eqref{eqn:biased-expansion}, it also follows that $\partial^\mu_S f(\mu) = \fhat^\mu(S)$ for any subset $S \subseteq [n]$, so $\mu$ acts as the origin with respect to the biased measure.

The level-$\ell$ Fourier weight of $f$ with respect to bias $\mu$ is defined as $\wgt_\ell^\mu(f) = \sum_{|S|=\ell} |\fhat^\mu(S)|.$

\section{Input Distribution and the Proof Outline}
\label{sec:outline}
We now give a formal outline of the proof. We first give an input distribution for which $(\delta,k)$-Forrelation is easy to compute using quantum queries, but hard for classical queries. Our distribution is a variant of that used in \cite{T20} with a different truncation function. 

To define the distribution we first introduce some notation. Recall the truncation function $\trnc: \BR \to \left[-\frac12,\frac12\right]$ defined as
 \begin{align}\label{eqn:trnc}
\ \trnc(s) = \Phi(s) - \frac12 = \int_0^s \gamma(s)ds.
 \end{align}
For notational convenience, we will write $\trnc(s_1,\ldots,s_m)$ to denote $(\trnc(s_1),\ldots,\trnc(s_m))$. Let us also introduce the following \emph{block shifted Hadamard product} of two vectors: given vectors $x := (x_1,\cdots,x_{k-1}) \in \BR^{(k-1)N}$ and $y := (y_1,\cdots,y_{k-1}) \in \BR^{(k-1)N}$, we define $x \diamond y$ to be the following vector in $\BR^{kN}$,
\begin{equation}\label{eqn:had}
    \ x \diamond y = (x_1, \cdots, x_{k-1}, \ind) \odot (\ind, y_1, \cdots, y_{k-1}) = (x_1, y_1\odot x_2, y_2\odot x_3, \ldots, y_{k-2}\odot x_{k-1},y_{k-1}),
\end{equation}
where $\ind$ is the all ones vector in $\BR^N$ and $\odot$ is the Hadamard product of two vectors. The above product will allow a natural generalization of the input distribution described in \secref{sec:overview} to the case of arbitrary $k$. To see some examples, for $k=2$ and vectors $x,y \in \BR^n$, we have that $x \diamond y = (x,y)$; while for $k=3$, we have that $x \diamond y = (x_1, y_1 \odot x_2, y_2) = (x_1, x_2 \odot y_1, y_2)$ reminiscent of the expression appearing in \eqref{eqn:k3}.

We can now describe the input distribution. Recall that $\delta=2^{-5k}$ and let $\cov = \begin{pmatrix} \sI_N & \sH_N\\ \sH_N & \sI_N \end{pmatrix}$. Then, our input distribution $p(\Z)=\frac12 p_0(\Z) + \frac12 p_1(\Z)$ where $p_0(Z)$ and $p_1(Z)$ are defined in \figref{fig:input}.

\begin{figure}[!h]
\begin{tabular}{|l|}
\hline
\begin{minipage}{\textwidth}
\vspace{1ex} 
\begin{description}
    \item[Distribution $p_0(\Z)$:] $\Z$ is uniform over $\bits^{kN}$.
    \item[Distribution $p_1(\Z)$:]  Let $(\G_\kappa,\tG_\kappa)_{\kappa \in [k-1]}$ be independent random variables in $\BR^{2N}$ that are distributed as $\CN(0, \cov)$. Write $U = (U_\kappa)_{\kappa \in [k-1]}$ and $V = (V_\kappa)_{\kappa \in [k-1]}$ and define $W=\trnc(U) \diamond \trnc(V) $ where $W \in [-1/2,1/2]^{kN}$. Let $Z = (Z_1, \ldots, Z_k) \in \bits^{kN}$ be obtained by rounding each coordinate of the vector $W$ independently to $\pm1$ by interpreting them as means, i.e., for each coordinate $i \in [kN]$, we have $\BE[Z(i) \mid U,V] = W(i).$
\end{description}
\vspace{0.1ex}
\end{minipage}\\
\hline
\end{tabular}
\caption{Input Distributions $p_0(Z)$ and $p_1(Z)$}
\label{fig:input}
\end{figure}

 We now show that  $p_b(Z)$ for $b \in \{0,1\}$ has a large support on $b$-inputs for $(\delta,k)$-Forrelation. 
\begin{restatable}{theorem}{inputdist}
\label{thm:input}
For the input distribution defined in \figref{fig:input}, 
\[ p_0(\forr_{\delta,k} \text{ outputs } 0) \ge 1-\frac{4}{\delta^2 N} ~\text{ and }~ p_1(\forr_{\delta,k} \text{ outputs } 1) \ge 6\delta.\]
\end{restatable}

\begin{proof}
We first consider 
$p_0$.  Since $p_0(z)$ is uniform on $\bits^{kN}$ and $\forr_k(z)$ is a multilinear and homogeneous polynomial, clearly $\BE_{p_0(z)}[\forr_k(z)]=0$.
Next, we claim that $\BE_{p_0}[\forr_k(Z)^2]\le 1/N$.
To see this, we use the quadratic form description \eqref{eqn:forr-qform}. Fix any values  $z_2,\ldots,z_{k-1}$, and let $\sA = \sH \cdot \diag(z_2)  \cdots \cdot \sH \cdot \diag(z_{k-1})\cdot \sH$ be the matrix appearing in the quadratic form which satisfies $\|\sA\|_{\op} \le 1$.   Then, we have
    \begin{align*}
          \ \BE_{p_0}[\forr_k(Z)^2] &= \frac{1}{N^2}\BE_{p_0}[(Z^\top_1\sA Z_k)^2] =\frac{1}{N^2} \sum_{ij,rs} \BE_{p_0}[{\sA_{ij}\sA_{rs}\cdot  Z_1(i)Z_k(j)Z_1(r)Z_k(s)}] \\
          \ & = \frac{1}{N^2} \sum_{ij} \sA_{ij}^2 = \frac{\|\sA\|_F^2}{N^2} \le \frac{N \|\sA\|^2_{\op}}{N^2} \le \frac1N. \
    \end{align*}
By Chebshev's inequality, it follows that 
$p_0(\forr_{\delta,k}(Z) \text{ outputs } 1) \leq p_0(|\forr_k(Z)| \ge \delta/2) \le (4/\delta^2N).$\\

    We now consider $p_1$. As $\forr_k(z)$ is a multilinear polynomial, from the description of $p_1(Z)$, we have that $\BE_{p_1}[\forr_k(z) \mid U,V] = \forr_k(\trnc(U) \diamond\trnc(V)) $. Defining $X = \trnc(U)$ and $Y=\trnc(V)$, \lref{lem:corr} proved in \secref{sec:id}, implies that $\BE[X_\kappa(i)\cdot \sH_{i,j} \cdot Y_\kappa(j)] \ge \frac{1}{32} \cdot \sH_{ij}^2$ for any $i,j \in [N]$ and $\kappa \in [k-1]$ since $\BE[U_\kappa(i) V_\kappa(j)] = \sH_{ij}$. Therefore,
    \begin{align*}
          \ &\BE_{p_1}[\forr_k(Z)] = \BE_{p_1}[\forr_k(X \diamond Y)]\\
          \ & = \frac{1}{N} \sum_{\ti} \BE[X_1(i_1) \cdot \sH_{i_1,i_2} \cdot Y_1(i_2)X_2(i_2) \cdot \sH_{i_2,i_3} \cdots \sH_{i_{k-2},i_{k-1}} \cdot Y_{k-2}(i_{k-1})X_{k-1}(i_k)\cdot \sH_{i_{k-1},i_k} \cdot Y_{k-1}(i_k)] \\
          \ &=\frac{1}{N} \sum_{\ti} \BE[X_1(i_1) \cdot \sH_{i_1,i_2} \cdot Y_1(i_2)] \cdot \BE[X_2(i_2) \cdot \sH_{i_2,i_3} \cdot Y_2(i_3)] \cdots \BE[X_{k-1}(i_k)\cdot \sH_{i_{k-1},i_k} \cdot Y_{k-1}(i_k)] \\
          \ & \ge \frac{1}{N} \sum_{\ti} \left(\frac{1}{32}\right)^{k-1} \cdot \sH_{i_1,i_2}^2 \cdots \sH_{i_{k-1},i_k}^2 = \frac{1}{N} \sum_{\ti} \left(\frac{1}{32}\right)^{k-1}  \cdot \frac{1}{N^{k-1}}  =  \left(\frac{1}{32}\right)^{k-1},
    \end{align*}
    where the second equality used that $(X_\kappa,Y_\kappa)$ are independent for different values of $\kappa$, the inequality follows from the implication of \lref{lem:corr} discussed above, and the fourth equality follows since each entry of $\sH$ is $\pm\frac{1}{\sqrt{N}}$ and the sum is over $N^k$ indices. It thus follows that $\BE_{p_1}[\forr_k(Z)] \ge  \left(\frac{1}{32}\right)^{k-1} = 32\delta$.

Let $\alpha = p_1(\forr_k(Z) \ge \delta)$.
    Recalling \eqref{eqn:forr-qform}, we have that $|\forr_k(z)| \leq 1$ for $z \in [-1,1]^{kN}$. So, the above gives that $\alpha + (1-\alpha)\delta \geq 32 \delta$ and hence in particular that $\alpha \geq 6 \delta$, as $\delta \le 1/2^{10}$.
\end{proof}

To prove a lower bound for classical query algorithms (decision trees), we show that the advantage of any bounded real-valued function on $\bits^{kN}$ can be computed in terms of the low-level Fourier weight of the function $f$ with respect to biased measures, as mentioned in \secref{sec:overview}. In particular, for $\mu \in [-1/2,1/2]^{kN}$, consider the product measure $p_\mu$ induced on $Z \in \bits^{kN}$ by sampling each bit independently so that $\BE_{p_\mu}[Z_i] = \mu_i$. Then, we prove the following which is the main contribution of this work.

\begin{restatable}{theorem}{level}
\label{thm:level}
Let $f: \bits^{kN} \to [0,1]$. Then,
    \[ \left|\BE_{p_1}[f(Z)] -  \BE_{p_0}[f(Z)]\right| \le \sup_{\mu \in [-\frac12,\frac12]^{kN}} \sum_{\ell=k}^{k(k-1)} \left(\frac{1}{\sqrt{N}}\right)^{\ell\left(1-\frac1k\right)} \cdot {(8k)^{14\ell}} \cdot  \wgt_{\ell}^\mu(f).\]
\end{restatable}
 
Note that in the previous work of \cite{RT19} for the standard Forrelation problem $(k=2)$, one only gets an upper bound in terms of the level-2 weight of the function $f$, but here we have an upper bound in terms of level $\ell$ weights where $\ell$ is between $k$ and $k(k-1)$. We stress that the weight of the higher levels ($\ell > k$) can be much larger than the level-$k$ weight, but the extra $1/\sqrt{N}$ factors in the above theorem takes care of it.

To bound the level-$\ell$ Fourier weight with respect to biased measures, we use the following bound proven in \cite{T20} for Fourier weights under the uniform measure.

\begin{theorem}[\cite{T20}]
    \label{thm:tal}
    Let $f: \bits^{m} \to [0,1]$ be the acceptance probability function of a randomized depth-$d$ decision tree. Then, for any $\ell \le d$, the following holds for a universal constant $c$,
    \[ \wgt_\ell(f) \le \big((cd)^\ell \log^{\ell-1} m\big)^{1/2},\]
     where the Fourier weight $\wgt_\ell(f)$ is with respect to the uniform measure on $\bits^{m}$. 
\end{theorem}

 We prove the following general statement showing that if a function and all its restrictions have a small Fourier weight on level-$\ell$ with respect to the uniform measure, then the Fourier weight with respect to an arbitrary bias $\mu \in \left[-\frac12,\frac12\right]^{m}$ is also small. 

\begin{restatable}{theorem}{fwt}
  \label{thm:fwt}
  Let $f: \bits^{m} \to \BR$ and $\ell \in [m]$. Let $w$ be such that for any restriction $\rho \in \{-1,1,\star\}^m$, we have $\wgt_\ell(f_\rho) \le w$ where the Fourier weight is with respect to the uniform measure. Then, for any $\mu \in \left[-\frac12,\frac12\right]^{m}$, we have $\wgt^\mu_\ell(f) \le 4^\ell w$. 
\end{restatable}

Since depth-$d$ decision trees are closed under restrictions, combining \thmref{thm:fwt} with \thmref{thm:tal} gives us that the level-$\ell$ weight of depth-$d$ decision trees with an arbitrary bias $\mu \in \left[-\frac12,\frac12\right]^m$ is also bounded by $\big((cd)^\ell \log^{\ell-1} (m)\big)^{1/2}$. 

\begin{restatable}{corollary}{corwt}
  \label{cor:wt}
  Let $f: \bits^{m} \to [0,1]$ be the acceptance probability function of a randomized depth-$d$ decision tree. Then, for any $\mu \in \left[-\frac12,\frac12\right]^{m}$ and $\ell \le d$, we have $\wgt^\mu_\ell(f) \le \big((cd)^\ell \log^{\ell-1} m\big)^{1/2}$ for a universal constant $c$. 
\end{restatable}

Combined with \thmref{thm:level}, the above implies that if the depth $d$ of the decision tree satisfies $d \ll N^{1-1/k}$, then the advantage of $f$ would be much smaller than $\delta$.

\section{Interpolation and Integration by Parts Identities}\label{sec:id}

As mentioned before, we will use the truncation function $\trnc: \BR \to [-\frac12,\frac12]$ defined in \eqref{eqn:trnc} to truncate vectors in $\BR^m$ to $[-\frac12, \frac12]^m$. This necessitates generalizing the Gaussian integration by parts and Gaussian interpolation identities to handle expressions of the form $\BE[f(\trnc(G_1), \cdots, \trnc(G_m))]$ or $\BE[\trnc(U) \cdot f(G_1, \cdots, G_m)]$ where $U, G_1, \cdots, G_m$ are jointly Gaussian. In this section we prove some such identities that will be repeatedly used throughout the paper. 

Below if $s = s_1, \ldots, s_m$, then for brevity, we write $\trnc(s)$ to denote $\trnc(s_1), \ldots, \trnc(s_m)$.  The first lemma gives us an interpolation formula for functions of the form $f(\trnc(s))$. 
\begin{lemma}[Interpolation]\label{lem:interpolphi}
    Let $\cov = \begin{pmatrix} \sI_n & \sM \\ \sM^\top & \sI_n\end{pmatrix}$ where $\sM$ is an $n \times n$ orthogonal matrix. Let $(U,V) \in \BR^{2n}$ be distributed as $\CN(0, \cov)$ and for $t \in (0,1)$, define the interpolation
        $$ \zeta(t) =  \BE[f(\trnc(\B{U}, \B{V}))] \text{ where } (\B{U}, \B{V}):=(\B{U}(t), \B{V}(t)) = \sqrt{t} \cdot (U,V).$$ Then, for any multilinear polynomial $f(x,y) = f(x_1, \ldots, x_n, y_1, \ldots, y_n)$, the following holds 
    \[ \zeta'(t) = \frac{1}{1+t} \cdot \sum_{i,j=1}^n \sM_{ij}  \cdot  \BE\left[ \del{f}{x_i\partial y_j}(\trnc(\B{U} ,\B{V}))\gamma(\B{U}_i)\gamma(\B{V}_j)\right].\]
\end{lemma}
\begin{proof}
    Consider the function $q(u, v) = f(\trnc(u,v))$ in the variables $u=u_1, \ldots, u_n$ and $v = v_1, \ldots, v_n$. Then, since $\trnc'(s)= \gamma(s)$ and $\gamma'(s) = -s \gamma(s)$ and $f$ is multilinear, we have that for any $i, j \in [n]$, 
    \begin{align*} 
        \ \frac{\partial^2q}{\partial u_i^2}(u,v) &= - u_i \gamma(u_i) \del{f}{x_i}(\trnc(u,v)) \text { and }  \frac{\partial^2q}{\partial v_j^2}(u,v) = - v_j \gamma(v_j) \del{f}{y_j}(\trnc(u,v)) \\
        \ \frac{\partial q}{\partial u_i \partial v_j}(u,v) &= \del{f}{x_iy_j}(\trnc(u,v)) \gamma(u_i) \gamma(v_j). 
    \end{align*}

    Since the only non-zero entries of $\cov$ are those on the diagonal (which are always one) or the entries of $\sM$, applying the Gaussian interpolation formula (\lref{lem:interpol}) yields that
    \begin{align}\label{eqn:intbyparts}
        \ \zeta'(t) &= \sum_{i,j=1}^n \sM_{ij}  \cdot  \BE\left[ \del{f}{x_i\partial y_j}(\trnc(\B{U} ,\B{V}))\gamma(\B{U}_i)\gamma(\B{V}_j)\right] \notag \\
        \ &- \frac12 \sum_{i=1}^n \BE\left[\B{U}_i \gamma(\B{U}_i) \cdot  \del{f}{x_i}(\trnc(\B{U},\B{V}))\right] - \frac12 \sum_{j=1}^n \BE\left[\B{V}_j \gamma(\B{V}_j) \cdot  \del{f}{y_j}(\trnc(\B{U}, \B{V}))\right].
    \end{align}
    Note that the factor $\frac12$ in the first term disappears after adding the contributions of  pairs $(i,j)$ and $(j,i)$ which are the same. To take care of the terms involving first order derivatives, we prove the following claim.

\begin{claim}\label{lem:gamma}
    Let $B, G_1, G_2, \cdots, G_n$ be real-valued random variables that are jointly Gaussian and let $h(x_1, \ldots, x_n)$ be a moderately growing function. Then, letting $G=(G_1, \ldots, G_n)$ and $ \BE[B^2] = \sigma^2 $, we have  
    \[ \BE[B \gamma(B) \cdot h(\trnc(G))] = \frac{1}{1+\sigma^2} \cdot \sum_{i=1}^n \BE[BG_i] \cdot  \BE\left[ \del{h}{x_i}(\trnc(G)) \cdot  \gamma(B) \gamma(G_i)\right]. \]
\end{claim}
\begin{proof}
    For variables $g = g_1, \ldots, g_n$, let us write $q(g) = h(\trnc(g))$. Recalling that $\trnc'(s) = \gamma(s)$ and $\gamma'(s) = -s\gamma(s)$, we have $\frac{\partial q}{\partial g_i}(g)= \del{h}{x_i}(g) \cdot \gamma(g_i)$. Thus, applying Gaussian integration by parts (\lref{lem:intbyparts}), we get that
     \[ \BE[B \gamma(B) \cdot h(\trnc(G))] = \sum_{i=1}^n \BE[BG_i] \cdot  \BE\left[ \gamma(B) \cdot \del{h}{x_i}(\trnc(G)) \cdot \gamma(G_i)\right] - \sigma^2 \cdot \BE[B \gamma(B) \cdot h(\trnc(G))].\]
     Rearranging the above gives the claim.
\end{proof}

    Noting that $\BE[\B{U}_i^2]=t=\BE[\B{V}_j^2]$, and $\BE[\B{U}_i \B{V}_j]=\sM_{ij} t$  for all $i,j \in [n]$, and that $\del{f}{x_i}(\trnc(u,v))$ is not a function of the variable $u_i$ as $f$ is multilinear, we can apply \clmref{lem:gamma} to obtain 
    \begin{align*}
        \ \sum_{i=1}^n \BE\left[\B{U}_i \gamma(\B{U}_i) \cdot  \del{f}{x_i}(\trnc(\B{U},\B{V}))\right] &= \frac1{1+t} \sum_{i,j=1}^n \BE[\B{U}_i\B{V}_j] \cdot  \BE\left[ \del{f}{x_i\partial y_j}(\trnc(\B{U} ,\B{V}))\gamma(\B{U}_i)\gamma(\B{V}_j)\right] \\
        \ &= \frac{t}{1+t} \sum_{i,j=1}^n \sM_{ij} \cdot  \BE\left[ \del{f}{x_i\partial y_j}(\trnc(\B{U} ,\B{V}))\gamma(\B{U}_i)\gamma(\B{V}_j)\right],
    \end{align*}
    and similarly for the last term in \eqref{eqn:intbyparts}. Again the terms corresponding to $\BE[\B{U}_i\B{U}_j]$ where $i \neq j$ do not appear since the correlation is zero for these pairs.
    
    Plugging the above in \eqref{eqn:intbyparts}, we get the statement of the lemma.
\end{proof}

The next lemma is an analogue of Gaussian integration by parts for computing expressions of the form $\BE[\trnc(U)f(G_1, \ldots, G_m)]$.

\begin{lemma}[Integration by Parts]\label{lem:intbypartsphi}
    Let $h: \BR^m \to \BR$ be a moderately growing function in the variables $x_1, \ldots, x_m$ and let $B, G_1, \ldots, G_m$ be real-valued random variables that are jointly Gaussian with $\BE[B^2] = \sigma^2$ with $\sigma \in (0,1]$. Then, writing $G = (G_1, \ldots, G_m)$, we have
    \[ \BE[\trnc(B) \cdot h(G)] = \sum_{i=1}^m \BE[B G_i] \cdot \BE\left[\Psi_\sigma(B) \cdot \del{h}{x_i}(G)\right], \]
    where $\Psi_\sigma: \BR \to \left[0, \frac{1}{\sqrt{2\pi}}\right]$ is the non-negative function defined as 
    \begin{equation}\label{eqn:psi}
    \ \Psi_\sigma(s) = \frac{1}{\sqrt{2\pi}} \int\limits_0^1 \frac{e^{-s^2 y^2/2}}{1+\sigma^2 y^2} dy.
    \end{equation}
    Moreover, for every integer $n\ge 1$, the $n^{\text{th}}$ derivative $|\Psi^{(n)}_\sigma(s)| \le n^{n/2}$ for every $s \in \BR$.
\end{lemma}
\begin{proof}
    Recalling the series representation given by \pref{prop:ser}, we see that the function $\trnc(s)/s$ is well-defined, so we can write the left hand side above as $\BE[B \cdot \left(\frac{\trnc(B)}{B}\right) \cdot h(G)]$ and apply Gaussian integration by Parts (\lref{lem:intbyparts}). This however produces a term corresponding to $\BE[G^2]$ and the idea is to repeat this process iteratively to get rid of such terms. For this let us define the following related series for any non-negative integer $j$,
    $$\trnc_j(s) :=  (-1)^j  \cdot \gamma(s) \sum_{k=0}^\infty \frac{(2j-1)!!}{(2k+2j+1)!!} \cdot s^{2k+1},$$
    where $(-1)!!=1$ by convention. Note that $\trnc(s)=\trnc_0(s)$ and $|\trnc_{j+1}(s)| \le |\trnc(s)| \le 1/2$  for every $j$. Moreover, recalling \eqref{eqn:hermite}, the $n^{\text{th}}$-derivatives of $\trnc_j(s)$ and $\trnc_j(s)/s$ can be upper bounded by $|q_{n,j}(s) \cdot (1+ \gamma(s))|$ for some polynomial $q_{n,j}(s)$, so they are moderately growing for every non-negative integer $j$. We claim that 
    \begin{claim}\label{clm:ind} 
        For every non-negative integer $j$, we have $\frac{d}{ds}\left(\frac{\trnc_j(s)}s\right) = \trnc_{j+1}(s)$. Moreover,
         \begin{align*}
             \BE\left[\trnc_j(B) \cdot h(G)\right] = \sum_{i=1}^n &  \BE[UG_i] \cdot  \BE\left[\left(\frac{\trnc_j(B)}{B}\right) \cdot \del{h}{x_i}(G_1,\ldots, G_m) \right] \\
        \ & +  \sigma^2 \cdot  \BE\left[\trnc_{j+1}(B)  \cdot h(G) \right].
             \end{align*}
        
    \end{claim}
    \begin{proof}[Proof of \clmref{clm:ind}]
    Since $\gamma'(s) = -s \gamma(s)$, we have that
        \begin{align*}
            \frac{(-1)^j}{(2j-1)!!} \cdot \frac{d}{ds}\left(\frac{\trnc_j(s)}{s}\right) &= \gamma'(s)\left(\sum_{k=0}^\infty \frac{1}{(2k+2j+1)!!} \cdot s^{2k}\right) + \gamma(s)\left(\sum_{k=1}^\infty \frac{2k}{(2k+2j+1)!!} \cdot s^{2k-1}\right)\\
            \ &= \gamma(s)\left(- \sum_{k=0}^\infty \frac{1}{(2k+2j+1)!!} \cdot s^{2k+1}\right) + \gamma(s)\left(\sum_{k=0}^\infty \frac{2k+2}{(2k+2j+3)!!} \cdot s^{2k+1}\right)\\
            \ &= -\gamma(s)\left(\sum_{k=0}^\infty \frac{2j+1}{(2k+2j+3)!!} \cdot s^{2k+1}\right) = \frac{(-1)^j}{(2j-1)!!}  \cdot {\trnc_{j+1}(s)}, 
        \end{align*}
        and thus the first statement in the claim follows. For the moreover part, applying Gaussian integration by parts (\lref{lem:intbyparts}) yields
    \begin{align*}
        \BE\left[B \cdot \left(\frac{\trnc_j(B)}{B}\right)  \cdot h(G)\right] = \sum_{i=1}^n &  \BE[BG_i] \cdot  \BE\left[\left(\frac{\trnc_j(B)}{B}\right) \cdot \del{h}{x_i}(G) \right] \\
        \ & +  \sigma^2 \cdot  \BE\left[h(G) \cdot \frac{d}{ds}\left(\frac{\trnc_j(s)}{s}\right)_{|s=B}\right].
    \end{align*}
    Using the first part of the claim, the statement follows.
    \end{proof}

    Starting with $j=0$ and applying the above claim recursively for each $j$ yields that 
         \begin{align*}
             \BE\left[\trnc(B) \cdot h(G)\right] &= \sum_{j=0}^\infty \sigma^{2j} 
             \left(\sum_{i=1}^n  \BE[BG_i] \cdot  \BE\left[\frac{\trnc_j(B)}{B} \cdot \del{h}{x_i}(G) \right] \right)\\
             \ & = \sum_{i=1}^n  \BE[BG_i] \cdot  \BE\left[\frac{1}{B}\left(\sum_{j=0}^\infty \sigma^{2j} \trnc_j(B)\right) \cdot \del{h}{x_i}(G) \right],
             \end{align*}
        where the interchanging of the summation and the expectation is justified by the dominated convergence theorem since the functions $\trnc_j$ are bounded and alternating in sign. 

    To complete the proof, we show that the series occurring in the right hand side above equals the function $\Psi_\sigma$ in the statement of the lemma. 
    \begin{claim}\label{clm:psi}
        For every $s \in \BR$, we have that
        $\displaystyle \Psi_\sigma(s) = \frac{1}{s} \left(\sum_{j=0}^\infty \sigma^{2j} {\trnc_j(s)}\right)$.
    \end{claim}
    \begin{proof}[Proof of \clmref{clm:psi}]
        Let $q(s)=\frac{1}{s} \left(\sum_{j=0}^\infty \sigma^{2j} {\trnc_j(s)}\right)$. Note that for $\sigma \in (0,1]$, we have 
        \[ q(0) = \frac{1}{\sqrt{2\pi}} \sum_{j=0}^\infty (-1)^j  \frac{\sigma^{2j}}{2j+1} = \frac1{\sqrt{2\pi} \cdot \sigma}\left(\sum_{j=0}^\infty (-1)^j  \frac{\sigma^{2j+1}}{2j+1}\right) = \frac{1}{\sqrt{2\pi}} \frac{\arctan(\sigma)}{\sigma}. \]
        Using \clmref{clm:ind}, we have 
        \begin{align*}
            \ q'(s) &= \sum_{j=0}^\infty \sigma^{2j} \frac{d}{ds}\left(\frac{\trnc_j(s)}s\right) = \sum_{j=0}^\infty \sigma^{2j} \trnc_{j+1}(s) = \frac1{\sigma^2}\left(s q(s) - \trnc(s)\right).
        \end{align*}
       
        Thus, $sq(s) - \sigma^2q'(s) = \trnc(s)$ with the boundary condition $q(0)=\frac{1}{\sqrt{2\pi}} \cdot \frac{\arctan(\sigma)}{\sigma}$. Solving this differential equation yields that
        \begin{align*}
            \ q(s) & =  e^{s^2/2\sigma^2}  \left(\frac{1}{\sqrt{2\pi}} \cdot \frac{\arctan(\sigma)}{\sigma} -  \frac{1}{\sigma^2} \int\limits_0^s \trnc(s) e^{-r^2/2\sigma^2} dr \right) \\
            \ & = \frac{e^{s^2/2\sigma^2}}{\sqrt{2\pi}\cdot \sigma}  \left(\arctan(\sigma) -  \int\limits_0^s \int\limits_0^r e^{-v^2/2} \cdot  e^{-r^2/2\sigma^2} dv \frac{dr}{\sigma} \right).
        \end{align*}
        Making the substitution $w=r/\sigma$, we have $dw=dr/\sigma$ and the upper limit of the outer integral changes to $s/\sigma$. We can now express the above in terms of Owen's-$T$ function as follows
        \begin{align*}
            \ q(s) &= \frac{e^{s^2/2\sigma^2}}{\sqrt{2\pi}\cdot \sigma}  \left(\arctan(\sigma) -  \int\limits_0^{s/\sigma} \int\limits_0^{\sigma w} e^{-v^2/2} e^{-w^2/2} dv dw \right) \stackrel{\eqref{eqn:owen1}}{=} \sqrt{2\pi} \cdot e^{s^2/2\sigma^2} \cdot \frac{1}{\sigma} \cdot T(s/\sigma, \sigma)
        \end{align*}

        Using the other form of Owen's-$T$ function, we have 
        \begin{align*}
            \ q(s) &\stackrel{\eqref{eqn:owen2}}{=} \frac{e^{s^2/2\sigma^2}}{\sqrt{2\pi}} \cdot \frac1\sigma \int\limits_0^\sigma \frac{e^{-s^2(1+x^2)/2\sigma^2}}{1+x^2} dx = \frac{1}{\sqrt{2\pi}} \int\limits_0^\sigma \frac{e^{-s^2x^2/2\sigma^2}}{1+x^2} \frac{dx}{\sigma} = \frac{1}{\sqrt{2\pi}} \int\limits_0^1 \frac{e^{-s^2y^2/2}}{1+\sigma^2y^2} dy = \Psi_\sigma(s). \qedhere 
        \end{align*}
    \end{proof}
    
        From the expression for $\Psi_\sigma$ given by \eqref{eqn:psi}, it is easily seen that $\Psi_\sigma(s) \in \left[0,\frac{1}{\sqrt{2\pi}}\right]$ for every $s \in \BR$ and $\sigma \in (0,1]$ as the integrand is always non-negative and bounded by one. To see the bounds on the derivatives, we see that 
        \[ \frac{d}{ds^n}{\Psi_\sigma(s)} =  \int\limits_0^1 \frac{y^n \gamma^{(n)}(sy)}{1+\sigma^2y^2}dy  .\]
        From the above it follows that $|\Psi_\sigma^{(n)}(s)| \le \max_{r \in \BR} |\gamma^{(n)}(r)| \le n^{n/2}$ using \eqref{eqn:hermitebound}. This completes the proof of the lemma. 
    
 \end{proof}

 The next lemma shows that the truncation function $\trnc$ preserves correlations up to a constant factor.
 \begin{lemma}\label{lem:corr}
     Let $\rho \in (-1,1)$ and $B,G$ be real valued random variables such that $(B,G)$ is distributed as $\CN\left(0, \begin{pmatrix} 1 & \rho\\ \rho & 1\end{pmatrix}\right)$. Then, we have $\rho \cdot \BE[\trnc(B) \trnc(G)] \ge \frac{\rho^2}{32}.$
    \end{lemma}
    
    One can explicitly compute that $\rho \cdot \BE[\trnc(B) \trnc(G)] = \frac{\rho}{2\pi} \arctan\left(\frac\rho{\sqrt{4-\rho^2}}\right)$, from which the above lemma is obvious, but here we take a principled approach and use the integration by parts identity given by \lref{lem:intbypartsphi} which also results in a simpler proof.
    
    \begin{proof}[Proof of \lref{lem:corr}]
        We may assume that $\rho \ge 0$ since otherwise we can replace $(B,G)$ with $(-B,G)$ and the statement of the lemma remains the same. Applying \lref{lem:intbypartsphi}, we have that 
        \[ \rho\cdot \BE[\trnc(B) \trnc(G)] = \rho^2 \cdot \BE[\Psi(B)\gamma(G)],\]
       where 
        $$\Psi(s) := \Psi_1(s) = \frac1{\sqrt{2\pi}} \int\limits_0^1 \frac{e^{-s^2y^2/2}}{1+y^2}dy.$$
        Note that both $\Psi$ and $\gamma$ are non-negative functions and that $\Psi(s) \ge \frac12 \gamma(s)$ for every $s \in \BR$. Therefore,
       \begin{align*}
           \BE[\Psi(B)\gamma(G)] \ge \frac{1}{4\pi} \cdot \BE\left[e^{-(B^2+G^2)/2}\right] = \frac{1}{\pi} \cdot \BE\left[e^{-(B^2+G^2)/2} \cdot \ind_{B\ge 0} \cdot \ind_{G \ge 0}\right].
       \end{align*}
        We lower bound the above by further restricting to the contribution of the event $ B, G \in [0,c]$ where $c$ is a constant that will be optimized later. Since both $B$ and $G$ are standard Gaussians marginally, the probability of the above event by the union bound is at least $1-2(1-\Phi(c))\ge 1-e^{-c^2/2}$ using \pref{prop:gconc}. This gives us that
        \[   \BE[\Psi(B)\gamma(G)] \ge \frac1\pi e^{-c^2} (1-e^{-c^2/2}). \]
        Since the maximum value of $e^{-c^2}(1-e^{-c^2/2})$ is $\frac4{27}$ for $c = \sqrt{\log\left(\frac94\right)}$, we get that the right hand side in the last inequality above is at least $1/32$ which gives us the lemma. 
    \end{proof}

\section{Lower Bound for Decision Trees}\label{sec:lower}

We first prove \thmref{thm:level} that bounds the advantage of the randomized decision tree in terms of biased Fourier weights. Following that, we show how to bound the Fourier weight of a function under a biased measure (\thmref{thm:fwt}) by using a random restriction argument. We assemble all the pieces together to prove \thmref{thm:lower} and \corref{cor:smallerr} following that.

\subsection{Advantage in terms of Fourier weight: Proof of \thmref{thm:level}}

By multiliearity, we have that 
\begin{equation}\label{eqn:adv}
    \ \BE_{p_1}[f(Z)] - f(0) =  \BE_{p_1}[f(\trnc(U) \diamond \trnc(V))] - f(0). 
\end{equation}

To evaluate the first term on the right hand side, we will use interpolation identity given by \lref{lem:interpolphi}. Recall that $(U_\kappa,V_\kappa)_{\kappa \in [k-1]}$ are independent multivariate Gaussians. We will interpolate them separately. In particular, for each $\kappa \in [k-1]$ and $t_\kappa \in (0,1)$, we define
    \[ (\B{U}_\kappa(t_\kappa), \B{V}_\kappa(t_\kappa)) = \sqrt{t_\kappa} \cdot (U_\kappa, V_\kappa).\] 
We will refer to the interpolation parameter $t=(t_1,\cdots,t_{k-1})$ as time and we will drop the time index and just write $\B{U}$ and so on, if there is no ambiguity. We remind the reader of our convention that bold fonts will always refer to the interpolated random variables. 

To use interpolation, we consider the function $\zeta: (0,1)^{k-1} \to \BR$ defined as
\begin{equation*}
    \zeta(t) =  \BE[f(\trnc(\B{U}(t))\diamond \trnc(\B{V}(t)))].
\end{equation*}
For any fixed values of $t_1,\cdots,t_{k-2}$, by the fundamental theorem of calculus we have that 
\begin{align*}
    \ \BE[f(\trnc(\B{U}(t_1,\cdots,t_{k-2},1)) \diamond \trnc(\B{V}(t_1,\cdots,t_{k-2},1)))] &- \BE[f(\trnc(\B{U}(t_1,\cdots,t_{k-2},0)) \diamond \trnc(\B{V}(t_1,\cdots,t_{k-2},0)))] \\
    \ & = \int_0^1 \frac{\partial {\zeta}}{ \partial{t_{k-1}}}(t) d{t_{k-1}}. 
\end{align*}
Repeating the above and fixing each index of the time parameter one by one, we obtain 
\begin{align}
\label{eqn:integral}
    \BE[f(\trnc(U) \diamond \trnc(V))] - f(0) &= \BE[f(\trnc(\B{U}(\ind)) \diamond \trnc(\B{V}(\ind))] -  \BE[f(\trnc(\B{U}(0)) \diamond \trnc(\B{V}(0))] \notag\\
\ &=  \idotsint\limits_{[0,1]^{k-1}} \frac{\partial {\zeta}}{\partial{t_1} \cdots \partial{t_{k-1}}}(t) d{t_{k-1}} \cdots d{t_1},
\end{align}
where $\ind$ is the all ones vector in $\BR^{k-1}$.

To bound the value of the above partial derivative (taken with respect to the time parameters) at any point, we will use \lref{lem:interpolphi}. Since $f(z)$ is a multilinear polynomial, it suffices to compute the derivative of a character and towards this end, we show the following key lemma in terms of derivatives $\partial_J f = \del{f}{z_J}$ where the order of the derivative $|J|$ is always between $k$ and $k(k-1)$.

\begin{lemma}\label{lemma:monomial}
    Let $t \in (0,1)^{k-1}$ and $S \subseteq [kN]$. Defining $\zeta_S(t) = \BE[\chi_S(\trnc(\B{U}(t)) \diamond \trnc(\B{V}(t)))]$, the following holds 
    \[ \frac{\partial {\zeta_S}}{\partial{t_1} \cdots \partial{t_{k-1}}}(t) = \sum_{\ell=k}^{k(k-1)}  \left(\frac{1}{\sqrt{N}}\right)^{\ell\left(1-\frac1k\right)} \cdot \sum_{\substack{J \subseteq S\\|J|=\ell}} \BE\left[\chi_{S\setminus J}(\trnc(\B{U}(t)) \diamond \trnc(\B{V}(t)))~ \cdot ~\theta_J(t, \B{U}(t), \B{V}(t))\right], \]
    where $\theta_J : (0,1)^{k-1} \times \BR^{(k-1)N} \times \BR^{(k-1)N} \to \BR$ is a function that only depends on $J$ (and not on $S$) and satisfies $\max_{t,u,v} |\theta_J(t,u,v)| \le (4k)^{14|J|}$. 
\end{lemma}
We first finish the proof of \thmref{thm:level} and then prove the above lemma. Given \lref{lemma:monomial}, since $\zeta(t) = \sum_{S \subseteq [kN]} \fhat(S) \zeta_S(t)$, by linearity of expectation  and exchanging the order of summation, it follows that for a given time $t$,
\begin{align}\label{eqn:interpol}
    \frac{\partial {\zeta}}{\partial{t_1} \cdots \partial{t_{k-1}}}(t) &= \sum_{\ell=k}^{k(k-1)}   \left(\frac{1}{\sqrt{N}}\right)^{\ell\left(1-\frac1k\right)} \cdot \BE\left[\sum_{\substack{J \subseteq [kN]\\|J|=\ell}} \sum_{S: S\supseteq J} \fhat(S)\chi_{S\setminus J}(\trnc(\B{U}) \diamond \trnc(\B{V})) \cdot \theta_J(t, \B{U}, \B{V}) \right] \notag\\
    \ &=  \sum_{\ell=k}^{k(k-1)}   \left(\frac{1}{\sqrt{N}}\right)^{\ell\left(1-\frac1k\right)} \cdot \BE\left[\sum_{\substack{J \subseteq [kN]\\|J|=\ell}} \theta_J(t, \B{U}, \B{V}) \cdot \partial_J f(\trnc(\B{U}) \diamond \trnc(\B{V}))\right],
\end{align}
where the second equality uses \eqref{eqn:derivative}.

Next, we express the derivatives in \eqref{eqn:interpol} as biased Fourier coefficients. For any fixed values $u, v \in \BR^{(k-1)N}$, define $\mu := \mu(u,v) \in \left[-\frac12,\frac12\right]^{kN}$ as $\mu = \trnc(u) \diamond \trnc(v)$ and recalling the identity \eqref{eqn:bias-derivative}, we see that $\partial_Jf(z) = \sigma_J^{-1}\fhat^\mu(J)$ where $\sigma_J = \prod_{i \in J} \sigma_i$ with $\sigma_i = \sqrt{1-\mu_i^2} \ge 1/2$. Furthermore, as $\max_{t,u,v} |\theta_J(t, u, v)| \le (4k)^{14|J|}$, equation \eqref{eqn:interpol} gives us that the following holds for any $t \in (0,1)^{k-1}$,
\begin{align*}
    \  \left|\frac{\partial {\zeta}}{\partial{t_1} \cdots \partial{t_{k-1}}}(t)\right|  &\le \sup_{u,v \in \BR^{(k-1)N}} \sum_{\ell=k}^{k(k-1)}  \left(\frac{1}{\sqrt{N}}\right)^{\ell\left(1-\frac1k\right)} \sum_{\substack{J \subseteq [kN]\\|J|=\ell}} |\theta_J(t, u, v)| \cdot \sigma_J^{-1} \cdot |\fhat^\mu(J)|.
    \\ & \le \sup_{\mu \in \left[-\frac12,\frac12\right]^{kN}} \sum_{\ell=k}^{k(k-1)} \left(\frac{1}{\sqrt{N}}\right)^{\ell\left(1-\frac1k\right)} \cdot (8k)^{14\ell}\cdot\wgt^{\mu}_{\ell}(f).
\end{align*}

Finally, using \eqref{eqn:integral} and \eqref{eqn:adv}, the above implies that 
 \[ \left|\BE_{p_1}[f(Z)] - f(0)\right|  \le \sup_{\mu \in \left[-\frac12,\frac12\right]^{kN}} \sum_{\ell=k}^{k(k-1)} \left(\frac{1}{\sqrt{N}}\right)^{\ell\left(1-\frac1k\right)} \cdot (8k)^{14\ell}\cdot\wgt^{\mu}_{\ell}(f), \]
 completing the proof of \thmref{thm:level} given \lref{lemma:monomial}, which we prove next.

\subsection{Proof of \lref{lemma:monomial}}
For ease of exposition we first give the proof for the simpler case of $k=3$.
The application of integration by parts is much easier here, as it does not recursively lead to other terms. 
For larger values of $k$, we need more technical care and additional ideas in the form of a careful counting argument.

\subsubsection*{Proof for the $k=3$ Case}
In this case, we shall prove that
\begin{align}\label{eqn:stmt3} 
    \ \frac{\partial {\zeta_S}}{\partial{t_1} \partial{t_{2}}}(t) =  \sum_{\ell=3}^6\sum_{\substack{J \subseteq S\\|J|=\ell}}  \left(\frac{1}{\sqrt{N}}\right)^{2\ell/3} \cdot\BE[\chi_{S \setminus J} (\trnc(\B{U}) \diamond \trnc(\B{V})) \cdot \theta_J(t, \B{U}, \B{V})], 
\end{align}
  where $\theta_J(t,u,v)$ is a function that only depends on $J$ and not on $S$ and satiesfies $\max_{t,u,v}|\theta_J(t,u,v)| \le 12^{14|J|}$.

  Let $S = S_1 \sqcup S_2 \sqcup S_3$ where $S_1 \subseteq [N], S_2 \subseteq \{N+1,\ldots,2N\}$ and $S_3 \subseteq \{2N+1,\ldots,3N\}$. Let us also define $X_\kappa = \trnc(U_\kappa)$ and $Y_\kappa = \trnc(V_\kappa)$ for $\kappa \in [2]$ and analogously we define $\B{X}_\kappa$ and $\B{Y}_\kappa$ in terms of the interpolated Gaussians $\B{U}_\kappa$ and $\B{V}_\kappa$. We first observe that because of the multiplicativity of the characters $\chi_S$ and the definition of block-shifted Hadamard product, we have that for any $x,y \in \BR^{2N}$, 
  \begin{equation}\label{eqn:product}
      \chi_S(x \diamond y) =  \chi_{S_1}(x_1)\chi_{S_2}(y_1) \cdot \chi_{S_2}(y_2)\chi_{S_3}(y_2). 
  \end{equation}

 We will treat $\chi_{S_1}(x_1)\chi_{S_2}(y_1)$ and $\chi_{S_2}(x_2)\chi_{S_3}(y_2)$ as function in the variables $x_1 = (x_1(i))_{i \in S_1}, y_1 = (y_1(j))_{j \in S_2}$ and $x_2 = (x_2(i))_{i \in S_2}, y_2= (y_2(j))_{j \in S_3}$ respectively and write $\del{}{x_1(i)}, \del{}{x_1(j)}$ to denote the corresponding partial derivatives. To prevent any confusion, we clarify that $\partial_i = \del{}{z_i}$ will always denote the derivative with respect to $z$. 
 
  Now, since $(U_1,V_1)$ and $(U_2,V_2)$ are independent Gaussians and they are being interpolated separately, we can apply the interpolation formula given by \lref{lem:interpolphi} separately to compute $\BE[\chi_{S_1}(X_1)\chi_{S_2}(Y_1)]$ and $\BE[\chi_{S_2}(X_2)\chi_{S_3}(Y_2)]$ appearing in \eqref{eqn:product}. Therefore, applying \lref{lem:interpolphi} and using linearity of expectation, we have
          \begin{align}\label{eqn:start3}
              \ \frac{\partial {\zeta_S}}{\partial{t_1} \partial{t_{2}}}(t) \notag &=  \sum_{\ti,\tj}  \frac{\sH_{i_1 ,j_{2}}}{1+t_1} \cdot \frac{\sH_{i_2,j_{3}}}{1+t_2} \cdot \BE\left[\del{}{x_1(i_1)\partial y_1(j_{2})}\chi_{S_1}(\B{X}_1)\chi_{S_2}(\B{Y}_1) \gamma(\B{U}_1(i_1))\gamma(\B{V}(j_2))\right] \\
              \ &\hphantom{horizontal space separation}\cdot \BE\left[\del{}{x_2(i_2)\partial y_2(j_{3})}\chi_{S_2}(\B{X}_2)\chi_{S_3}(\B{Y}_2) \gamma(\B{U}_2(i_2))\gamma(\B{V}_2(j_3))\right] \notag\\
              \ &=  \sum_{\ti,\tj}  \frac{\sH_{i_1 ,j_{2}}}{1+t_1} \cdot \frac{\sH_{i_2,j_{3}}}{1+t_2} \cdot \BE[{\chi_{S_1 {\setminus i_1}} (\B{X_1})\chi_{S_{2}\setminus j_{2}}(\B{Y_1})} \gamma(\B{U}_1(i_1))\gamma(\B{V}(j_2))] \notag \\
              \ &\hphantom{horizontal space separation} \cdot  \BE[{\chi_{S_2{\setminus i_2}} (\B{X_2})\chi_{S_{3}\setminus j_{3}}(\B{Y_2})}  \gamma(\B{U}_2(i_2))\gamma(\B{V}_2(j_3))],  
        \end{align}
         writing $\ti = (i_1,i_2) \in S_1 \times S_2$ and $\tj = (j_2,j_3) \in S_2 \times S_3$. Note that the indices are shifted for $\tj$ to clarify that they lie in the corresponding set $S_r$ and we will keep using this indexing convention.

         We can classify the terms in \eqref{eqn:start3} into two types: terms where $i_2=j_2$ and where $i_2 \neq j_2$. These behave very differently, and we bound their contributions separately.
        
            \vspace{2mm}
        
\noindent {\bf (a) Terms where $i_2=j_2$:} ~ In this case, defining $i_3=j_3$, extending the tuple $\ti = (i_1,i_2,i_3)$, the corresponding terms in \eqref{eqn:start3} are given by
         \begin{align*} 
             \ &\frac{\sH_{i_1 ,j_{2}}}{1+t_1} \cdot \frac{\sH_{i_2,j_{3}}}{1+t_2} \cdot \BE[{\chi_{S_1{\setminus i_1}} (\B{X_1})\chi_{S_{2}\setminus i_{2}}(\B{Y_1} \odot \B{X_2})\chi_{S_{3}\setminus i_{3}}(\B{Y_2})} ~\cdot ~\gamma(\B{U_1}(i_1))  \gamma(\B{V_1}(i_2))  \gamma(\B{U_2}(i_2))  \gamma(\B{V_2}(i_3)) ] \notag \\
               \ &= \sH_{i_1,i_{2}} \cdot  \sH_{i_2,i_{3}} \cdot \BE\left[\chi_{S \setminus \{i_1, i_2, i_3\}}(\B{X} \diamond \B{Y})~\cdot~ \frac{\gamma(\B{U_1}(i_1))  \gamma(\B{V_1}(i_2))  \gamma(\B{U_2}(i_2))  \gamma(\B{V_2}(i_3))}{(1+t_1)(1+t_2)}\right] \\
               \ & = \left(\frac{1}{\sqrt{N}}\right)^2 \cdot \BE[\chi_{S \setminus \{i_1, i_2, i_3\}}(\trnc(\B{U}) \diamond \trnc(\B{V})) ~\cdot~ \theta_{\ti}(t, \B{U}, \B{V})], 
         \end{align*}
          where we used that $\B{X} = \trnc(\B{U})$ and $\B{Y}= \trnc(\B{V})$ and the function $\theta$ is defined as $$\theta_{\ti}(t,u,v) = \sign(\sH_{i_1,i_2} \cdot \sH_{i_2,i_3}) \cdot \dfrac{\gamma(u_1(i_1))  \gamma(v_1(i_2))  \gamma(u_2(i_2))  \gamma({v_2}(i_3))}{(1+t_1)(1+t_2)}.$$ Viewing the tuple $\ti$ as a set $J \subseteq S$ of size $3$, this gives us that the sum of all the terms in \eqref{eqn:start3} where $i_2=j_2$ is exactly
         \begin{align} \label{eq:term3}
             \sum_{\substack{J \subseteq S\\|J|=3}} \left(\frac{1}{\sqrt{N}}\right)^2 \cdot \BE[\chi_{S \setminus J} (\trnc(\B{U}) \diamond \trnc(\B{V})) \cdot \theta_J(t, \B{U}, \B{V})],
         \end{align}
         where $\max_{t,u,v} |\theta_J(t,u,v)| \le 1$.
       
       \vspace{2mm}
       
       \noindent {\bf (b) Terms where $i_2 \neq j_2$:}~ To bound these terms, we use the integration by parts identity of \lref{lem:intbypartsphi} to reduce them to derivatives of orders $4,5$ and $6$. Consider a fixed term where $i_2 \neq j_2$. Then, the corresponding expectation term in \eqref{eqn:start3} is
         \begin{equation}\label{eqn:termsintro}
              \ \BE[{\chi_{S_1{\setminus i_1}} (\B{X_1})\chi_{S_{2}\setminus j_{2}}(\B{Y_1})} \cdot \gamma(\B{U}_1(i_1))\gamma(\B{V}_1(j_2)) ] \cdot  \BE[{\chi_{S_2{\setminus i_2}} (\B{X_2})\chi_{S_{3}\setminus j_{3}}(\B{Y_2})} \cdot \gamma(\B{U}_2(i_2))\gamma(\B{V}_2(j_3))]. 
         \end{equation}
As $i_2 \neq j_2$, the term $\chi_{S_{2}\setminus j_{2}}(\B{Y_1})$ still depends on the random variable $\B{Y}_1(i_2)$ (while $\chi_{S_{2}\setminus j_{2}}(\B{X_2})$ does not).
Since eventually we need a function of $\B{X}_2 \odot \B{Y}_1$, we pull out $\B{Y}_1(i_2)$ and write,
         \begin{align}\label{eqn:first}
             \ & \BE[{\chi_{S_{1\setminus i_1}}  (\B{X_1})\chi_{S_{2}\setminus j_{2}}(\B{Y_1})} \cdot \gamma(\B{U}_1(i_1))\gamma(\B{V}_1(j_2))] \notag \\[1em]
             \ &\hphantom{horizontal space space} =~ \BE\left[\B{Y}_1(i_2) \cdot \chi_{{S_1} \setminus i_1} (\B{X_1})\chi_{{S_2}\setminus \{i_2, j_{2}\}}(\B{Y_1}) \cdot \gamma(\B{U}_1(i_1))\gamma(\B{V}_1(j_2))\right] 
            \end{align}

         Recalling that $\B{X_1} = \trnc(\B{U_1})$ and $\B{Y_1}=\trnc(\B{V_1})$, we can now apply \lref{lem:intbypartsphi} with 
         $$h := h(u,v) = \chi_{{S_1} \setminus i_1} (\trnc{(u_1)})\chi_{{S_2}\setminus \{i_2, j_{2}\}}(\trnc{(v_1)}) \cdot \gamma(u_1(i_1))\gamma(v_1(j_2)),$$
         and $B=\B{Y_1}(i_2)=\trnc(\B{V_1}(i_2))$. Note the very crucial fact that $h$ does not depend on the variable $v_1(i_2)$ as the term $ \chi_{{S_2}\setminus \{i_2, j_{2}\}}(\trnc{(v_1)})$ does not contain it and neither does $\gamma(v_1(j_2))$ as $j_2 \neq i_2$. Thus, to apply \lref{lem:intbypartsphi}, we only need to care about the terms corresponding to $\BE[\B{V_1}(i_2)\B{U_1}(q_1)] = t_1 \sH_{q_1,i_2}$ as the other terms will disappear --- those corresponding to $\BE[\B{V_1}(i_2)\B{V_1}(q_2)]$, where $i_2 \neq q_2$, disappear because $ \BE[\B{V_1}(i_2)\B{V_1}(q_2)]=0$, and the term corresponding to  $\BE[\B{V_1}(i_2)^2]$ disappears as $\del{h}{v_1(i_2)}=0$.
        
          Since, $\BE[\B{V}(i_2)\B{U}_1(q_1)] = t_1 \sH_{q_1,i_2}$, and
          \begin{align*}
           \del{h}{u_1(q_1)} &=  \chi_{{S_1} \setminus \{i_1,q_1\}} (\trnc(u_1))\chi_{{S_2}\setminus \{i_2, j_{2}\}}(\trnc({v_1})) \cdot \gamma(u_1(i_1))\gamma(v_1(j_2)), \text{ if } q_1 \neq i_1, \text{ and}, \\
           \ \del{h}{u_1(i_1)} &=  \chi_{{S_1} \setminus i_1} (\trnc(u_1))\chi_{{S_2}\setminus \{i_2, j_{2}\}}(\trnc({v_1})) \cdot \gamma'(u_1(i_1))\gamma(v_1(j_2)), \text{ otherwise},
          \end{align*}

             \lref{lem:intbypartsphi} gives us that  \begin{align}\label{eqn:firstp}
                \ \eqref{eqn:first} &=\sum_{\substack{q_1 \in S_1,\\ q_1  \neq i_1}} t_1\sH_{q_1,i_2} \cdot \BE\left[\chi_{{S_1} \setminus \{i_1,q_1\}} (\B{X_1})\chi_{{S_2}\setminus \{i_2, j_{2}\}}(\B{Y_1}) ~\cdot~ \gamma(\B{U}_1(i_1)) \gamma(\B{U}_1(q_1)) \Psi_{t_1}(\B{V}_1(i_2)) \gamma(\B{V}_1(j_2)) \right] \notag\\
             \ & \hphantom{horizontal}+ t_1 \sH_{i_1,i_2} \cdot  \BE\left[\chi_{{S_1} \setminus {i_1}} (\B{X_1})\chi_{{S_2}\setminus \{i_2, j_{2}\}}(\B{Y_1}) ~\cdot~ \gamma'(\B{U}_1(i_1)) \Psi_{t_1}(\B{V}_1(i_2)) \gamma(\B{V}_1(j_2)) \right]  
         \end{align}
         
         Analogously, for the second expectation in  \eqref{eqn:termsintro}, the term $\chi_{S_{2}\setminus i_{2}}(\B{X_1})$ still depends on the random variable $\B{X}_2(j_2)= \trnc(\B{U}_2(j_2))$. Therefore, applying  integration by parts (\lref{lem:intbypartsphi}) as above, one gets
         \begin{align}\label{eqn:second}
             & \BE[{\chi_{S_{2\setminus i_2}} (\B{X_2})\chi_{S_{3}\setminus j_{3}}(\B{Y_2})}\cdot \gamma(\B{U}_2(i_2))\gamma(\B{V}_2(j_3)) ] \notag\\[1em]
             \ &=\sum_{\substack{q_3 \in S_3,\\ q_3 \neq j_3}} t_2\sH_{j_2,q_3} \cdot \BE\left[\chi_{{S_2} \setminus \{j_2,i_2\}} (\B{X_2})\chi_{{S_2}\setminus \{j_3, q_{3}\}}(\B{Y_2}) ~\cdot~ \gamma(\B{U}_2(i_2)) \Psi_{t_2}(\B{U}_2(j_2)) \gamma(\B{Y}_2(j_3)) \gamma(\B{V}_2(q_3)) \right] \notag\\
             \ & \hphantom{horizontal}+ t_2 \sH_{j_2,q_3} \cdot  \BE\left[\chi_{{S_2} \setminus \{j_2, i_2\}} (\B{X_2})\chi_{{S_2}\setminus j_{3}}(\B{Y_2}) ~\cdot~ \gamma(\B{U}_2(i_2)) \Psi_{t_2}(\B{U}_2(j_2)) \gamma'(\B{V}_2(j_3)) \right]  
\end{align}
         \newcommand{\tr}{\underline{\alpha}}
         \newcommand{\tm}{\underline{\beta}}
         Combining \eqref{eqn:firstp} and \eqref{eqn:second}, we get the sum of all the terms in \eqref{eqn:start3} where $i_2\neq j_2$. In particular, defining new tuples $\tr = (i_1,j_2,q_3) \in S_1 \times S_2 \times S_3$ and $\tm=(q_1,i_2,j_3) \in S_1 \times S_2 \times S_3$, where we allow $q_1=i_1$ and $q_3=i_3$, we get that the sum of the terms in \eqref{eqn:start3} where $i_2 \neq j_2$ equals
         \begin{align}\label{eq:term6}
             \ &\sum_{{\tr}, {\tm}} \left(\frac{1}{\sqrt{N}}\right)^4 s_{{\tr},{\tm}}\cdot \BE[\chi_{S_1 \setminus \{i_1, q_1\}}(\B{X}_1) \cdot \chi_{S_2 \setminus \{j_2, i_2\}} (\B{Y}_1) \cdot \chi_{S_2 \setminus \{j_2,  i_2\}}(\B{X}_2) \cdot \chi_{S_3 \setminus \{j_3, q_3\}} (\B{Y}_2) ~\cdot~ \nu_{\tr, \tm}(t, \B{U}, \B{V})] \notag \\
               \ &=\sum_{{\tr}, {\tm}} \left(\frac{1}{\sqrt{N}}\right)^4  s_{\tr, \tm} \cdot \BE[\chi_{S_1 \setminus \{i_1, q_1\}}(\B{X}_1) \cdot \chi_{S_2 \setminus \{j_2, i_2\}} (\B{Y}_1 \odot \B{X}_2) \cdot \chi_{S_3 \setminus \{j_3, q_3\}} (\B{Y}_2) ~\cdot~ \nu_{\tr, \tm}(t, \B{U}, \B{V})] \notag\\
               \ &= \sum_{{\tr}, {\tm}} \left(\frac{1}{\sqrt{N}}\right)^4 \BE[\chi_{S \setminus \{i_1,q_1,i_2,j_2,j_3,q_3\}} (\B{X} \diamond \B{Y}) ~\cdot~ \theta_{\tr, \tm}(t, \B{U}, \B{V})],
         \end{align}
where the sum ranges over all possible tuples $\tr, \tm$ satisfying $i_2 \neq j_2$ and $s_{\tr,\tm} := \sign(\sH_{i_1,j_2} \sH_{j_2,q_3} \sH_{q_1,i_2} \sH_{i_2,j_3})$, the function $\nu_{\tr,\tm}(t,u,v)$ is some function that is always bounded by one (since $\gamma, \gamma' \text{ and } \Psi_\sigma$ are all bounded by one in magnitude and $t \in (0,1)^2$),  and $\theta_{\tr, \tm}(t,u,v) := s_{\tr, \tm}  \cdot \nu_{\tr,\tm}(t,u,v)$.

Note that there are at most $8$ possible tuples $\tr, \tm$ that give rise to the set $J = \{i_1,q_1,i_2,j_2,j_3,q_3\}$. It follows that the sum in \eqref{eq:term6} is exactly 
       \begin{align}
           \sum_{\ell=4}^6 \sum_{\substack{J \subseteq S\\|J|=\ell}}  \left(\frac{1}{\sqrt{N}}\right)^4 \cdot \BE[\chi_{S \setminus J} ( \trnc(\B{U}) \diamond \trnc(\B{V})) ~\cdot~\theta_J(t, \B{U}, \B{V})],
       \end{align}
       where $\theta_J(t, u, v)$ only depends on $J$ and $\max_{t,u,v}|\theta_J(t,u,v)| \le 8$. The level four and five weights appear since we allow the possibility that $i_1=q_1$ or $j_3=q_3$.

  Then, plugging in the bounds from \eqref{eq:term3} and \eqref{eq:term6} for the two cases in \eqref{eqn:start3}, we get \eqref{eqn:stmt3}.

\subsubsection*{Proof for Arbitrary $k$}
        Let $S = S_1 \sqcup S_2 \sqcup \cdots \sqcup S_k$ where $S_j \subseteq \{(j-1)N+1,\ldots,j N\}$ for $r \in [k]$.   Let us also define $X_\kappa = \trnc(U_\kappa)$ and $Y_\kappa = \trnc(V_\kappa)$ for $\kappa \in [k-1]$ and analogously we define $\B{X}_\kappa$ and $\B{Y}_\kappa$ in terms of the interpolated Gaussians $\B{U}_\kappa$ and $\B{V}_\kappa$. As before, we first observe that because of the multiplicativity of the characters $\chi_S$ and the definition of block-shifted Hadamard product, we have that for any $x,y \in \BR^{(k-1)N}$, 
        \begin{align}\label{eqn:productk}
        \ \chi_S(x \diamond y) = \prod_{\kappa \in [k-1]} \chi_{S_\kappa}(x_\kappa) \chi_{S_{\kappa+1}}(y_\kappa), 
        \end{align}
        so it decomposes as a product of the $k-1$ functions from $\BR^{2N}$ to $\BR$ where the $\kappa^{\text{th}}$ function is evaluated at $(x_\kappa,y_\kappa)$. Let us treat them as functions in the variables $x_\kappa = (x_\kappa(i))_{i \in S_\kappa}$ and $y_\kappa = (y_\kappa(j))_{j \in S_{\kappa+1}}$ and write the derivatives as $\del{}{x_\kappa(i)}, \del{}{y_\kappa(j)}$.

         Now, since $(U_\kappa,V_\kappa)$ are independent Gaussians for different values of $\kappa$ and they are being interpolated separately, we can apply the interpolation formula of \lref{lem:interpolphi} separately to the functions of $(U_\kappa,V_\kappa)$ appearing in \eqref{eqn:productk}. Since these functions are multilinear in the variables $(x_\kappa,y_\kappa)$, applying \lref{lem:interpolphi} and using linearity of expectation, we have
        \begin{align}\label{eqn:start}
            \frac{\partial {\varphi}}{\partial{t_1} \cdots \partial{t_{k-1}}}(t)  &=  \sum_{\ti,\tj} \prod_{\kappa \in [k-1]} \frac{\sH_{i_\kappa,j_{\kappa+1}}}{1+t_\kappa} \cdot \BE\left[\del{}{x_\kappa(i_\kappa)\partial y_\kappa(j_{\kappa+1})} \left(\chi_{S_\kappa}(\B{X}_\kappa)\chi_{S_{\kappa+1}}(\B{Y}_\kappa)\right)\cdot \gamma(\B{U}_\kappa(i_\kappa)) \gamma(\B{V}_\kappa(j_{\kappa+1}))\right], \notag\\
              \ & =\sum_{\ti,\tj} \prod_{\kappa \in [k-1]} \frac{\sH_{i_\kappa,j_{\kappa+1}}}{1+t_\kappa} \cdot  \BE[{\chi_{S_\kappa{\setminus i_\kappa}} (\B{X_\kappa})\chi_{S_{\kappa+1}\setminus j_{\kappa+1}}(\B{Y_\kappa})} \cdot \gamma(\B{U}_\kappa(i_\kappa)) \gamma(\B{V}_\kappa(j_{\kappa+1}))],  
        \end{align}
         where $\ti = (i_1, \cdots, i_{k-1}) \in S_1 \times \cdots \times S_{k-1}$ and  $\tj = (j_2, \cdots, j_k) \in S_2 \times \cdots \times S_k$ are tuples. As before, the indices are shifted for $\tj$ to clarify that they lie in the corresponding set $S_r$.\\
         
         Unlike the case of $k=3$, there are many types of terms in the above summation. Some of them correspond to derivatives $\partial_J = \del{}{z_J}$ of $\chi_S(z)$ where $\partial_J$ is of order $k$, and others which do not correspond to any such derivative.
         
\paragraph{Terms that correspond to derivatives with respect to $z$.}         Observe that when tuples $\ti$ and $\tj$ satisfy $i_\ell=j_\ell$ for $2\le \ell \le k-2$, then defining $i_k:=j_k$ (this extends the $(k-1)$-tuple $\ti$ to a $k$-tuple), the corresponding term in \eqref{eqn:start} is 
         \begin{align}\label{eqn:oneconf}
             \ &\prod\nolimits_{\kappa \in [k-1]}  \left(\frac{\sH_{i_\kappa,i_{\kappa+1}}}{1+t_\kappa}\right) \cdot  \BE[{\chi_{S_\kappa{\setminus i_\kappa}} (\B{X_\kappa})\chi_{S_{\kappa+1}\setminus i_{\kappa+1}}(\B{Y_\kappa})} \cdot \gamma(\B{U}_\kappa(i_\kappa)) \gamma(\B{V}_\kappa(i_{\kappa+1}))] \notag \\
               \ &=\left(\prod\nolimits_{\kappa} \frac{\sH_{i_\kappa,i_{\kappa+1}}}{1+t_\kappa}\right) \cdot \BE[\chi_{S_1 \setminus i_1}(\B{X}_1) \chi_{S_{2} \setminus i_{2}}(\B{Y}_1) \cdot \chi_{S_2 \setminus i_2}(\B{X}_2) \chi_{S_{3} \setminus i_{3}}(\B{X}_2) \cdots \cdot \chi_{S_{k-1} \setminus i_{k-1}}(\B{X}_{k-1}) \chi_{S_{k} \setminus i_{k}}(\B{Y}_{k-1}) \notag \\
               \ &\hphantom{horizontal spacing text } \cdot \gamma(\B{U}_1(i_1)) \gamma(\B{V}_1(i_{2})) \cdot  \gamma(\B{U}_2(i_2)) \gamma(\B{V}_2(i_{3})) \cdots  \gamma(\B{U}_{k-1}(i_{k-1})) \gamma(\B{V}_1(i_{k}))] \notag \\
             \ &= \left(\frac{1}{\sqrt{N}}\right)^{k-1} \cdot \BE[\chi_{S_1 \setminus i_1}(\B{X}_1) \chi_{S_{2} \setminus i_{2}}(\B{Y}_1 \odot \B{X}_2) \cdot  \chi_{S_{3} \setminus i_{3}}(\B{Y}_2 \odot \B{X}_3) \cdots \cdot \chi_{S_{k} \setminus i_{k}}(\B{Y}_{k-1}) \cdot \theta_{\{i_1, \ldots, i_k\}}(t, \B{U},\B{V})] \notag\\ 
               \ &= \left(\frac{1}{\sqrt{N}}\right)^{k-1} \cdot \BE[\chi_{S \setminus \{i_1, \ldots, i_k\}}(\trnc(\B{U}) \diamond \trnc(\B{V})) \cdot \theta_{\{i_1, \ldots, i_k\}}(t, \B{U},\B{V})],
         \end{align}
         where the last equality holds since $\B{X}=\trnc(\B{U})$ and $\B{Y}=\trnc(\B{V})$ so that the product of the various $\chi_{S_\kappa \setminus i_\kappa}$ terms equals $\chi_{S \setminus \{i_1, \ldots, i_k\}}(\trnc(\B{U}) \diamond \trnc(\B{V}))$ by the definition of the $\diamond$ product and the function $\theta_J(t,u,v)$ for $J=\{i_1,\cdots,i_k\}$ is defined as
\[ \theta_J(t,U,V) = \prod_{\kappa\in [k-1]} \left(\frac{\sign(\sH_{i_\kappa,i_{\kappa+1}})}{1+t_\kappa}\right) \cdot \gamma(\B{U}_\kappa(i_\kappa)) \gamma(\B{V}_\kappa(i_{\kappa+1})),  \]
and is always bounded by one in magnitude.

         This corresponds to taking the partial derivative $\partial_J \chi_S(\trnc(\B{U}) \diamond \trnc(\B{V})) = \chi_{S\setminus J}(\trnc(\B{U}) \diamond \trnc(\B{V}))$. However, the other terms in \eqref{eqn:start} can not be written as such a partial derivative. We will give a process that reduces such terms to a higher order derivative by repeated application of the integration by parts identity of \lref{lem:intbypartsphi}. 

\paragraph{Setup to apply integration by parts.} To describe the process, we will need some additional notation. Let us consider the terms appearing in \eqref{eqn:start} and drop the $\prod_{\kappa \in [k-1]} \frac{\sH_{i_\kappa,j_{\kappa+1}}}{1+t_\kappa}$ scaling factor  for notational convenience. By the independence of $(\B{U}_r,\B{V}_r)$ for different $r$, any term
\begin{align*}
    \ \prod_{r \in [k-1]} \BE[ \chi_{S_r\setminus i_r}&   (\B{X}_r) \cdot \chi_{S_{r+1} \setminus j_{r+1}}(\B{Y}_r) \cdot  \gamma(\B{U}_r(i_r)) \gamma(\B{V}_r(i_{r+1})) ] \\
    \ &= \BE\left[ \prod_{r \in [k-1]}  \chi_{S_r\setminus i_r}   (\B{X}_r) \cdot \chi_{S_{r+1} \setminus j_{r+1}}(\B{Y}_r) \cdot  \gamma(\B{U}_r(i_r)) \gamma(\B{V}_r(i_{r+1}))  \right].
\end{align*}

Recalling \eqref{eqn:oneconf}, one can see that if indices do not match, then one can not write the above in terms of a derivative of $\chi_S(z)$ evaluated at $\B{X} \diamond \B{Y}$. 
The idea will be to keep track of the the set of indices that do not match, and then apply integration by parts (\lref{lem:intbypartsphi}) until these sets become empty. We will need to track carefully how the indices evolve and the various terms that are generated. We will do this by viewing this as a branching process where one branch consists of one application of integration by parts.

To keep track of the indices, let us define the sets $A_1,\ldots,A_{k}$ and $B_1,\ldots,B_k$, where $A_r,B_r \subseteq S_r$ as follows: 
\begin{equation} \label{eqn:startingconf}
    A_r = \{i_r\}, B_{r+1} = \{j_{r+1}\} \text{ for } r \in [k-1] \text{ and } A_k = \emptyset, B_1 = \emptyset.
\end{equation}
 Let us also define $\B{V}_0=\B{U}_{k}=\ind$ where $\ind$ is the all ones vector in $\BR^N$. Furthermore, to each element of the sets $A_r$'s and $B_r$'s, we also associate a non-negative function: for $r \in [k-1]$ and $i_r \in A_r$ define $\eta_{i_r} := \gamma$ and $j_{r+1} \in B_{r+1}$, define $\xi_{j_{r+1}} := \gamma$.

Then the above can be written as 
\begingroup\leqnos
\begin{align}\label{eq:wsab}
    \ &~~~~ \BE \bigg[\prod_{r \in [k-1]}  \chi_{S_r\setminus A_r}   (\B{X}_r) \cdot \chi_{S_{r+1} \setminus B_{r+1}}(\B{Y}_r) \cdot   \prod_{i_r \in A_r} \gamma(\B{U}_r(i_r))  \cdot  \prod_{j_{r+1} \in B_{r+1}} \gamma(\B{V}_r(j_{r+1}))  \bigg] \\
\ &= \BE\Bigg[  \chi_{S_1\setminus A_1}  (\B{X}_1 \odot \B{Y}_0)  \bigg(\prod_{r=2}^{k-1}  \chi_{S_r\setminus A_r}   (\B{X}_r) \cdot \chi_{S_{r} \setminus B_{r}}(\B{Y}_{r-1}) \bigg)  
    \cdot \chi_{S_k\setminus B_k}   (\B{X}_k \odot \B{Y}_{k-1}) \notag\\ 
    \ &\hphantom{horizontal spacing horizontal spacing spacing} \cdot \prod_{r \in [k-1]} \prod_{i_r \in A_r} \eta_{i_r}(\B{U}_r(i_r))  \cdot  \prod_{j_{r+1} \in B_{r+1}} \xi_{j_{r+1}}(\B{V}_r(j_{r+1})) \Bigg]. \notag
\end{align}    
\endgroup

Let us write the middle term in the above expectation in a different way as follows,
\[\Big(\prod_{r=2}^{k-1}  \chi_{S_r\setminus A_r}   (\B{X}_r) \cdot \chi_{S_{r} \setminus B_{r}}(\B{Y}_{r-1}) \Big)  =  \prod_{r=2}^{k-1} \chi_{S_r\setminus (A_r \cup B_r)}   (\B{X}_r \odot \B{Y}_{r-1}) \cdot \chi_{B_r\setminus A_r} (\B{X}_r) \cdot \chi_{A_r \setminus B_r} (\B{Y}_{r-1}).  \]

\begin{sloppypar}Let us denote the list of functions $((\eta_{i_r})_{i_r \in A_r}, (\xi_{j_{r+1}})_{j_{r+1} \in B_r})_{r \in [k-1]}$ by $L$. Then, writing $A= (A_1,\ldots,A_k)$ and $B=(B_1,\ldots,B_k)$, we call $(A,B, L)$ a \emph{configuration}. We will always require a configuration to have $A_k=\emptyset$ and $B_1=\emptyset$ and we call such configurations \emph{valid}. 
\end{sloppypar}

For notational convenience, let us write $A\cup B$ to denote $(A_1\cup B_1) \cup \ldots \cup ( A_k \cup B_k)$ for a configuration $(A,B, L)$. Then, as $A_k=\emptyset, B_1 = \emptyset$, the expression in \eqref{eq:wsab} can be written as
\begin{align*}
    \ \Gamma(S,A,B, L) &:= \BE\bigg[ \chi_{S \setminus (A \cup B)} ( \B{X} \diamond \B{Y})   \cdot   \prod_{r=2}^{k-1} \Big(\chi_{B_r\setminus A_r} (\B{X}_r) \cdot \chi_{A_r \setminus B_r} (\B{Y}_{r-1})  \Big) \notag\\
    \ & \hphantom{horizontal spacing horizontal spacing} \cdot \prod_{r \in [k-1]} \prod_{i_r \in A_r} \eta_{i_r}(\B{U}_r(i_r))  \cdot  \prod_{j_{r+1} \in B_{r+1}} \xi_{j_{r+1}}(\B{V}_r(j_{r+1})) \bigg].
\end{align*}

\begin{figure*}[t!]
    \centering
    \begin{subfigure}[t]{0.4\textwidth}
        \centering
        \includegraphics[height=1.8in]{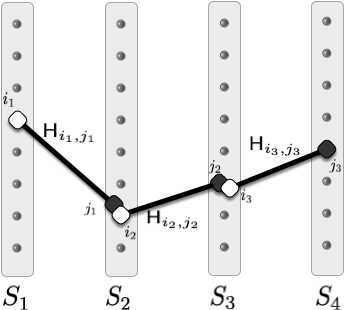}
    \end{subfigure}%
    \qquad\qquad
    \begin{subfigure}[t]{0.4\textwidth}
        \centering
        \includegraphics[height=1.8in]{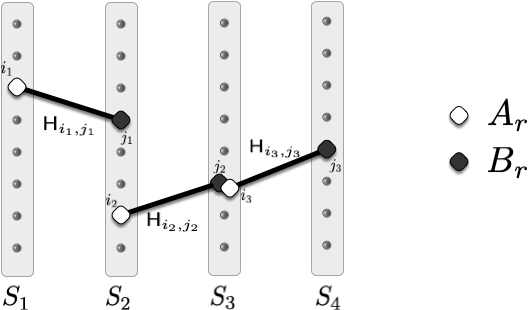}
    \end{subfigure}
    \caption{\footnotesize Examples of configurations that corresponds to terms in \eqref{eqn:start}. The bullets represents the indices in the sets $S_r$'s, while the white and black pebbles represent the indices in the sets $A_r$'s and $B_r$'s respectively. The labels on the edges denote the corresponding Hadamard factors in \eqref{eqn:start}. To each distinct pebble is an associated function $\gamma$ that is not shown in the figure. The left configuration is inactive and corresponds to a derivative of order $k$ as in \eqref{eqn:oneconf}, while the right configuration is active.}\label{fig:conf}
\end{figure*}

Note that for $r \in \{2, \ldots, k-1\}$, the sets $B_r \setminus A_r$ (resp. $A_r \setminus B_r$) keep track of the excess $\B{X}_r$ (resp. $\B{Y}_{r-1}$) variables that can not be absorbed in $\chi_{S \setminus (A \cup B)} ( \B{X} \diamond \B{Y})$. 

For terminology, let us call a configuration \emph{active} if there is some $r \in \{2,\ldots,k-1\}$ such that either $B_r \setminus A_r \neq \emptyset$ or  $A_r \setminus B_r \neq \emptyset$. Any configuration that is not active, is referred to as being \emph{inactive}. Note that the $\Gamma$ value of any inactive configuration has the form $\BE[\chi_{S \setminus J} (\B{X}\diamond \B{Y}) \cdot \theta(t, \B{U}, \B{V})]$ for some  $J \subseteq S$ and some function $\theta$ that is product of $\eta_{i_r}, \xi_{j_{r+1}}$'s. In particular, inactive configurations correspond to derivatives $\partial_{J} \chi_S(z)$ evaluated at $\B{X}\diamond \B{Y}$, for instance, the case of \eqref{eqn:oneconf} corresponds to the inactive configuration with $J = \{i_1,i_2,\ldots,i_k\}$. See \figref{fig:conf} for an illustration.

\vspace{-2mm}

\paragraph{A branching process from integration by parts.} Given an initial active configuration $(A,B, L)$, to compute $\Gamma(S,A,B)$, we will apply integration by parts (\lref{lem:intbypartsphi}). Doing so will lead to several other terms of the same type with different configurations $(A',B', L')$, and we recursively continue this way until all resulting configurations are inactive.
This can be viewed as a branching process, where starting from the configuration $(A,B, L)$, we get a tree, where the leaves correspond to inactive configurations, and $\Gamma(S,A,B, L)$ is a weighted sum of the $\Gamma$ values of the leaf configurations.

Below, we first describe the branching process and how the $\Gamma$ values of the child nodes produced by one step of the process are related to the $\Gamma$ value of the parent. After that, we will describe the properties of the leaf configurations generated by the process and relate it to the left hand side of \eqref{eqn:start}. 

\vspace{-2mm}

\paragraph{(a) The branching process.} Fix the set $S$, and consider an active configuration $(A,B)$. Suppose that $B_q \setminus A_q \neq \emptyset$ for some $q \in \{2,\ldots,k-1\}$. Consider some arbitrary $i_q \in B_q \setminus A_q$. Then, we have the following key lemma.
\begin{lemma}
\label{lem:br1}
For any $q \in \{2,\ldots,k-1\}$ and for any $i_q \in B_q \setminus A_q$, we have that
\[ \Gamma(S,A,B, L) = \sum_{j_{q+1} \in S_{q+1}} \sH_{i_q,j_{q+1}} t_q \cdot  \Gamma(S,A\cup \{i_q\},B \cup\{j_{q+1}\}, L') \]
    where $A \cup \{i_q\}$ is obtained from $A$ by setting $A_q = A_q \cup \{i_q\}$,  and  $B \cup\{j_{q+1}\}$ from $B$ by setting $B_{q+1} = B_{q+1} \cup \{j_{q+1}\}$ and $L'$ is obtained by updating $L$ as follows,
\begin{alignat}{3}
    \ &\eta_{i_q} \leftarrow \Psi_{t_q} \text{ and }\: &&\xi_{j_{q+1}} \leftarrow \xi'_{j_{q+1}} &&\text{ if }j_{q+1} \in B_{q+1}, \text{ and } \label{eqn:typeone1}\\
        \ &\eta_{i_q} \leftarrow \Psi_{t_q} \text{ and }\: &&\xi_{j_{q+1}} \leftarrow \gamma &&\text{ if } j_{q+1} \in S_{q+1} \setminus B_{q+1}. \label{eqn:typetwo1}
    \end{alignat}
\end{lemma}

Note that when $j_{q+1} \in B_{q+1}$, only one new element $\eta_{i_q}$ is added to the list and $\xi_{j_{q+1}}$ is updated to its derivative, otherwise both elements are added to the list.\\

\begin{figure}[ht]
        \centering
        \includegraphics[height=3.8 in]{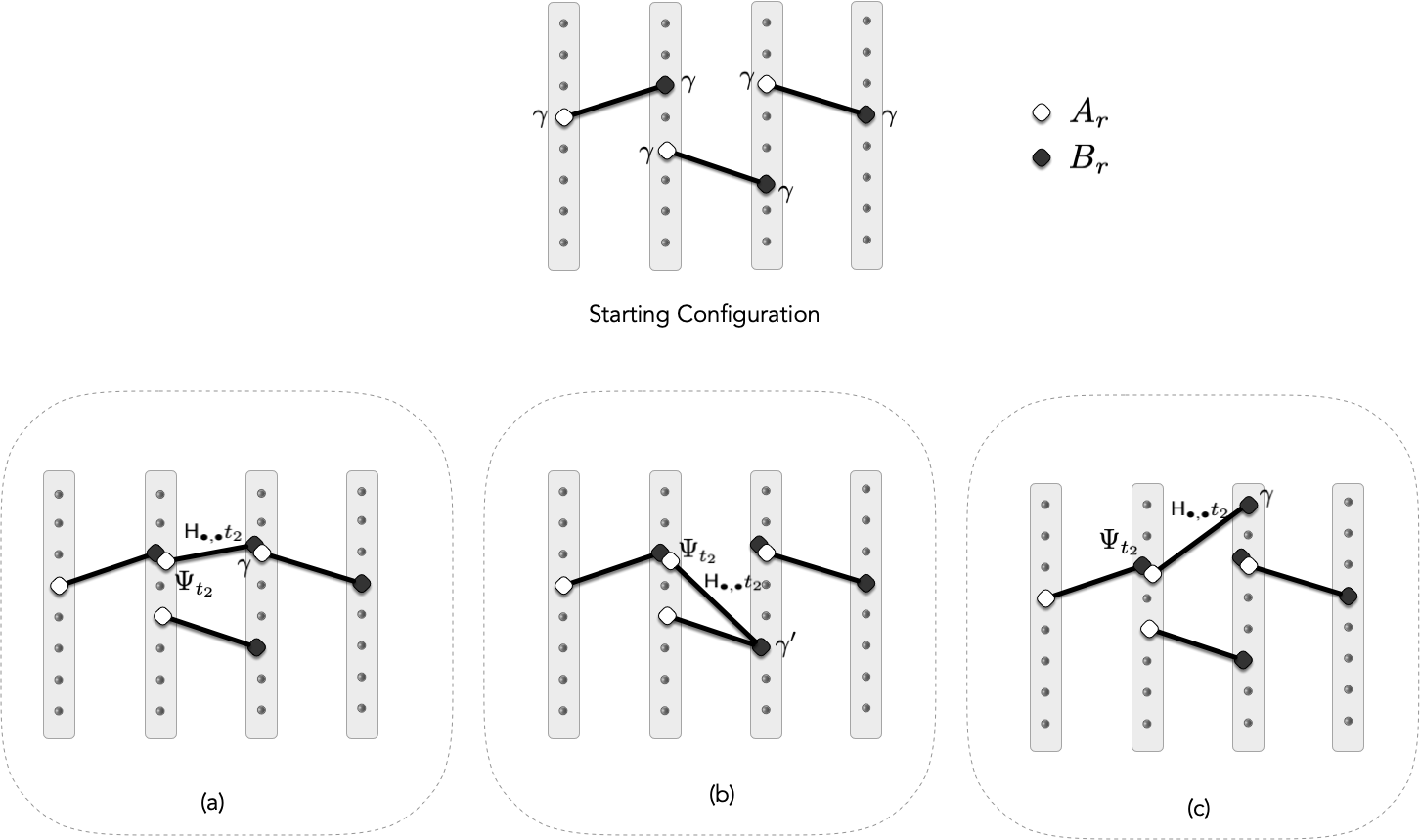}
    \caption{\footnotesize An application of how the configuration can evolve by an application of \lref{lem:br1}. Only the functions that are updated in the list $L$ are shown in the evolved configurations (a),(b) and (c). The function $\Psi_{t_q}$ corresponds to the $i_q$ index while $\gamma$ and $\gamma'$ corresponds to the $j_{q+1}$ index. The $\sH_{\bullet, \bullet} t_{\bullet}$ denotes the factor that is gained by the application of \lref{lem:br1}. The configurations (a) and (b) are type \textrm{I} transitions while (c) is a type \textrm{II} transition. Note that in a type \textrm{I} transition the union of the set of pebbles does not change, while in a type \textrm{II} transition the union of the set of pebbles gains a new element.}\label{fig:dynamics} 
\end{figure}

\vspace{-2mm}

Before proceeding with the proof, we remark that it may be useful to interpret the above lemma in the following way: any configuration $(A',B', L') = (A \cup \{i_q\}, B \cup \{j_{q+1}\}, L')$ that appears on the right hand side above, absorbs excess variables $\B{X}_q(i_q)$ and $\B{Y}_q(j_{q+1})$ into the $\chi_{S \setminus (A'\cup B')}(\B{X} \diamond \B{Y})$ term, but might add an excess variable $\B{X}_{q+1}(j_{q+1})$ in the $\chi_{B_{q+1} \setminus A_{q+1}}(\B{X}_{q+1})$ term depending on whether $j_{q+1} \in A_{q+1}$ or not. If an excess variable is not added (when $j_{q+1} \in A_{q+1}$), we call it a type \textrm{I} transition, otherwise (when $j_{q+1} \notin A_{q+1}$) we call it a type \textrm{II} transition.  See \figref{fig:dynamics} for an illustration.

\begin{proof}[Proof of \lref{lem:br1}]
Recall that $\B{X} = \trnc(\B{U})$ and $\B{Y}=\trnc(\B{V})$. Writing 
$\chi_{B_q\setminus A_q} (\B{X}_q) =     \B{X}_q(i_q) \cdot  \chi_{B_q\setminus (A_q \cup \{i_q\})} (\B{X}_q)$  applying
integration by parts (\lref{lem:intbypartsphi}) gives that 
\begingroup\leqnos
\begin{align}\label{eq:wsab2}
    \Gamma &(S,A,B, L) = \BE\bigg[ \chi_{S \setminus (A \cup B)} ( \B{X} \diamond \B{Y})   \cdot   \prod_{r=2}^{k-1} \Big(\chi_{B_r\setminus A_r} (\B{X}_r) \cdot \chi_{A_r \setminus B_r} (\B{Y}_{r-1})  \Big) \\
    & \hphantom{horizontal spacing horizontal spacing} \cdot \prod_{r \in [k-1]} \prod_{i_r \in A_r} \eta_{i_r}(\B{U}_r(i_r))  \cdot  \prod_{j_{r+1} \in B_{r+1}} \xi_{j_{r+1}}(\B{V}_r(j_{r+1}))\bigg] \notag\\
  ~~~~~~~~~~= & \sum_{\mathclap{j_{q+1} \in S_{q+1}}}~~~ \BE [\B{U}_q(i_q) \B{V}_q(j_{q+1})] \notag\\
    \ & \hphantom{hori} \cdot  \BE\bigg[ \Psi_{t_q}(\B{U}(i_q)) \cdot \frac{\partial}{\partial v_q(j_{q+1})} \bigg(  \chi_{S \setminus (A' \cup B)} ( \B{X} \diamond \B{Y})   \cdot   \prod_{r=2}^{k-1} \Big(\chi_{B_r\setminus A'_r} (\B{X}_r) \cdot \chi_{A'_r \setminus B_r} (\B{Y}_{r-1})  \Big)    \notag\\
    \ & \hphantom{horizontal spacing horizontal space} \cdot \prod_{r \in [k-1]} \prod_{i_r \in A_r} \eta_{i_r}(\B{U}_r(i_r))  \cdot  \prod_{j_{r+1} \in B_{r+1}} \xi_{j_{r+1}}(\B{V}_r(j_{r+1})) \bigg)\bigg] \notag
\end{align}
\endgroup
where we denote $A' = A \cup \{i_q\}$.
In the summation above, we only need to consider $j_{q+1} \in S_{q+1}$ as $\B{U}_q(i_q)$ has non-zero correlation only with coordinates of $\B{V}_{q}$ and with $\B{U}_q(i_q)$.
However, the $\BE[\B{U}_q(i_q)^2]$ term does not contribute to the above sum as the partial derivative with respect to $u_q(i_q)$ is identically zero as $u_q(i_q)$ does not appear in 
    $$\chi_{S \setminus (A' \cup B)} ( \trnc(u) \diamond \trnc(v)) =  \prod_{r=1}^k \chi_{S_r \setminus (A'_r \cup B_r)} ( \trnc(u_r) \odot \trnc(v_{r-1}))$$ or in
    $\chi_{B_q\setminus A'_q} (\trnc(u_q))$ as $i_q \in A'_q$ or in $\prod_{i_q \in A_q} \eta_{i_q}(u_q(i_q))$ as $i_q \notin A_q$.\\ 

Simplifying further, we have $\BE [\B{U}_q(i_q) \B{V}_q(j_{q+1})] = \sH_{i_q,j_{q+1}} t_q$. Next, the expectation containing the derivative simplifies as follows.

\begin{claim}\label{clm:br1}
 For $j_{q+1} \in S_{q+1}$, let $L'$ be obtained by updating $L$ as in the claim. Then, we have that
    \begin{align*}
        \  & \BE \bigg[  \Psi_{t_q}(\B{U}(i_q)) \cdot \frac{\partial}{\partial v_q(j_{q+1})} \bigg(  \chi_{S \setminus (A' \cup B)} ( \B{X} \diamond \B{Y})   \cdot   \prod_{r=2}^{k-1} \Big(\chi_{B_r\setminus A'_r} (\B{X}_r) \cdot \chi_{A'_r \setminus B_r} (\B{Y}_{r-1})  \Big)  \\
        \ &\hphantom{horizontal spacing hori} \cdot \prod_{r \in [k-1]} \prod_{i_r \in A_r} \eta_{i_r}(\B{U}_r(i_r))  \cdot  \prod_{j_{r+1} \in B_{r+1}} \xi_{j_{r+1}}(\B{V}_r(j_{r+1})) \bigg) \bigg] =  \Gamma(S,A',B \cup \{j_{q+1}\}, L').
    \end{align*}
 \end{claim}
 
The statement of \lref{lem:br1} follows from the claim above and \eqref{eq:wsab2}. We now prove \clmref{clm:br1}.
 \begin{proof}[Proof of \clmref{clm:br1}]
     We have three cases, depending on whether $v_q(j_{q+1})$ appears 
     \[ \text{(a) in } \chi_{A'_{q+1} \setminus B_{q+1}} (\trnc(v_{q})),~\text{ (b) in }\chi_{S \setminus (A' \cup B)} (\trnc(u) \diamond \trnc(v)),~ \text{ or (c) in }\xi_{j_{q+1}}(v_q(j_{q+1})).\]
     Note that it can only appear once in either of the three cases. We look at each of the cases separately.

     \begin{enumerate}[label=(\alph*)]
         \item 
             \begin{sloppypar}As $v_{q}(j_{q+1})$ appears in  $\chi_{A'_{q+1} \setminus B_{q+1}} (\trnc(v_{q}))$, upon taking the derivative this term becomes  $\chi_{A'_{q+1} \setminus (B_{q+1} \cup \{j_{q+1}\})} (\trnc(v_{q})) \cdot \gamma(v_q(j_{q+1}))$. 
                 Upon setting $B' = B \cup \{j_{q+1}\}$, note that the terms $\chi_{S \setminus (A' \cup B')} (\trnc(u) \diamond \trnc(v))$ and $\chi_{B'_{q+1}\setminus A'_{q+1}} (\trnc(u_q)) $ remain unchanged as $v_q(j_{q+1})$ does not appear there. Last, observe that the leftover factors $\Psi_{t_q}(\B{U}(i_q)) \cdot \gamma(\B{V}_q(j_{q+1}))$ are captured by the update of $L$ to $L'$. It follows that the left hand side of the claim equals $\Gamma(S,A',B',L') = \Gamma(S, A', B \cup \{j_{q+1}\}, L')$. 
             \end{sloppypar}

\item Now we consider the more interesting case where  $v_{q}(j_{q+1})$ appears in the term  
    \[\chi_{S \setminus (A' \cup B)} (\trnc(u) \diamond \trnc(v)) =   \chi_{S_{q+1} \setminus (A'_{q+1} \cup B_{q+1})} ( \trnc(u_{q+1}) \odot \trnc(v_q)) \cdot \prod_{\substack{r \in [k]\\r \neq q+1}} \chi_{S_r \setminus (A'_r \cup B_r)} ( \trnc(u_r) \odot \trnc(v_{r-1})).\]

             Consider the term $\chi_{S_{q+1} \setminus (A'_{q+1} \cup B_{q+1})} (\trnc(u_{q+1}) \odot \trnc(v_q))$ appearing in the above expression.  Upon taking the derivative $\del{}{v_q(j_{q+1})}$ this becomes 
             \[   \trnc(u_{q+1}(j_{q+1})) \cdot \chi_{ S_{q+1} \setminus (A'_{q+1} \cup B_{q+1} \cup \{j_{q+1}\})} ( \trnc(u_{q+1}) \odot \trnc(v_q)) \cdot \gamma(v_q(j_{q+1})),\]
which has the extra factor $\trnc(u_{q+1}(j_{q+1}))$ apart from $\gamma(v_q(j_{q+1}))$. 
             However, setting $B'= B \cup \{j_{q+1}\}$, this $\trnc(u_{q+1}(j_{q+1}))$ factor is absorbed in $\chi_{B'_{q+1}\setminus A'_{q+1}} (\trnc(u_{q+1}))$ as $j_{q+1} \notin A'_{q+1}$ by our assumption. Note that this does not affect the $\chi_{A'_{q+1}\setminus B'_{q+1}} (\trnc(v_q))$ term. Lastly, the leftover factors $\Psi_{t_q}(\B{U}(i_q)) \cdot \gamma(\B{V}_q(j_{q+1}))$ are captured by the update of $L$ to $L'$ as in the previous case. It follows that the claim holds in this case as well.
         \item Last, consider the case when $v_q(j_{q+1})$ appears in $\xi_{j_{q+1}}(v_q(j_{q+1}))$ which happens when $j_{q+1} \in B_{q+1}$. In this case, taking the derivative, this term changes to $\xi'_{j_{q+1}}(v_q(j_{q+1}))$ and all the other terms are unaffected. In the end we are left with an additional $\Psi_{t_q}(\B{U}(i_q)) \cdot \xi'_{j_{q+1}}(\B{V}_q(j_{q+1}))$ factor which is captured by the update to $L'$. The claim follows since $B_{q+1} \cup \{j_{q+1}\} = B_{q+1}$ since $j_{q+1} \in B_{q+1}$. \qedhere
\end{enumerate}
 \end{proof}
This completes the proof of  \lref{lem:br1}. \end{proof}

A completely analogous lemma holds if $i_q \in A_q \setminus B_q$, and  integration by parts is applied with respect to the variable $\B{Y}_{q-1}(i_q)$. 
In particular we have the following.
\begin{lemma}
\label{lem:br2}
For any $q \in \{2,\ldots,k-1\}$ and for any $i_q \in A_q \setminus B_q$, we have that 
\[ \Gamma(S,A,B, L) = \sum_{j_{q-1} \in S_{q-1}} \sH_{i_q,j_{q-1}} t_{q-1} \cdot  \Gamma(S,A\cup \{j_{q-1}\},B \cup\{i_q\}, L') \]
where $A \cup \{j_{q-1}\}$ is obtained from $A$ by setting $A_{q-1} = A_{q-1} \cup \{j_{q-1}\}$,  and  $B \cup\{i_q\}$ from $B$ by setting $B_{q} = B_{q} \cup \{i_q\}$ and $L'$ is obtained by updating $L$ as follows,
\begin{alignat}{3}
    \ &\xi_{i_q} \leftarrow \Psi_{t_{q-1}} \text{ and }\: &&\eta_{j_{q-1}} \leftarrow \eta'_{j_{q-1}} &&\text{ if }j_{q-1} \in A_{q-1}, \text{ and } \label{eqn:typeone2}\\
        \ &\xi_{i_q} \leftarrow \Psi_{t_{q-1}} \text{ and }\: &&\eta_{j_{q-1}} \leftarrow \gamma &&\text{ if } j_{q-1} \in S_{q-1} \setminus A_{q-1}. \label{eqn:typetwo2}
    \end{alignat}
\end{lemma}

Similar to \lref{lem:br1}, it may be useful to interpret the above lemma as follows: any configuration $(A',B', L') = (A \cup \{j_{q-1}\}, B \cup \{i_{q}\}, L')$ that appears on the right hand side above, absorbs excess variables $\B{X}_{q-1}(j_{q-1})$ and $\B{Y}_{q-1}(i_{q})$ into the $\chi_{S \setminus (A'\cup B')}(\B{X} \diamond \B{Y})$ term, but might add an excess variable $\B{Y}_{q-1}(j_{q-1})$ in the $\chi_{A_{q-1} \setminus B_{q-1}}(\B{Y}_{q-1})$ term depending on whether $j_{q-1} \in B_{q-1}$ or not. If an excess variable is not added (when $j_{q-1} \in B_{q-1})$, we call it a type \textrm{I} transition, otherwise (when $j_{q-1} \notin B_{q-1})$ we call it a type \textrm{II} transition.

Also, we remark that in both \lref{lem:br1} and \lref{lem:br2} above, the resulting configurations $A',B'$ still satisfy $A'_k = B'_1 =  \emptyset$ and hence are valid. In particular, as $q \in \{2,\ldots,k-1\}$,  neither $A_k$ or $B_1$ are ever updated in either of the lemmas.

Finally, note that \lref{lem:br1} and \lref{lem:br2} take different actions --- \lref{lem:br1} chooses an $i_q \in B_q \setminus A_q$ and uses an application of integration by parts using the variable $\B{X}_q(i_q)$; on the other hand, \lref{lem:br2} chooses an $i_q \in A_q \setminus B_q$ and applies integration by parts using the variable $\B{Y}_{q-1}(i_q)$. However, both \lref{lem:br1} and \lref{lem:br2} allow us to express the value $\Gamma(S,A,B, L)$ for a configuration $(A,B,L)$ as a weighted sum of $\Gamma$ values of other configurations. Hence, given a starting configuration $(A,B, L)$ and applying \lref{lem:br1} and \lref{lem:br2} alternately gives the claimed branching process. Note that a branch of the process terminates at a leaf configuration which is inactive. We remark that the same configuration may appear multiple times as different nodes of the branching tree, but we will treat each node of the branching tree as a separate configuration.

\vspace{-2mm}

\paragraph{(b) Properties of the leaf configurations.} We next show some properties of the configurations that arise in the branching process where each initial configuration is given by \eqref{eqn:startingconf}. Note that an initial configuration $(A,B,L)$ corresponds uniquely to a tuple $\ti,\tj$ appearing in \eqref{eqn:start}. We define the weight of an initial configuration $(A,B, L)$ that corresponds to the tuple $\ti,\tj$ as 
 \[ \wgt(S,A,B, L) := \prod_{\kappa \in [k-1]} \frac{\sH_{i_\kappa,j_{\kappa+1}}}{1+t_\kappa}.\]

Note that left hand side of \eqref{eqn:start} equals the weighted sum of $\Gamma$ values of all the initial configurations where the weights are given by the $\wgt$ values. For each initial configuration, we will start a separate branching process in parallel, and we will always maintain the invariant that the left hand side of \eqref{eqn:start} always equals the weighted sum of the $\Gamma$ values of all the configurations generated at any intermediate step. To do this we define the weight of each node in one such branching tree in the following way: if a configuration $(A \cup \{i_q\},B \cup \{j_{q+1}, L'\})$ is a child of $(A,B, L)$ in the branching tree, then the weight of $(A \cup \{i_q\},B \cup \{j_{q+1}, L'\})$ is defined as
\[ \wgt(S, A \cup \{i_q\},B \cup \{j_{q+1}, L'\}) =  \sH_{i_q,j_{q+1}}t_q \cdot \wgt(S,A,B,L), \]
where $\sH_{i_q,j_{q+1}}t_q$ is the factor appearing in front of $\Gamma(S, A \cup \{i_q\},B \cup \{j_{q+1}, L'\})$ in \lref{lem:br1} or \lref{lem:br2}. Note that the weight of a node in the branching tree depends on the path from the initial configuration to that node in the tree. 

The following proposition is an immediate consequence of the definition of weight and of \lref{lem:br1} and \lref{lem:br2}.
\newcommand{\CL}{\mathscr{L}}
\begin{proposition}\label{prop:total}
    Let $\CL(S)$ denote the collection of all the leaf configurations  (viewed as a multiset) generated by the parallel branching processes starting from all possible initial configurations. Then the left hand side of \eqref{eqn:start} equals
    \[ \sum_{(A,B, L) \in \CL(S)} \wgt(S,A,B, L) \cdot \Gamma(S,A,B, L).\]
\end{proposition}

Recall that for any final inactive leaf configuration $(A,B, L)$, we have that $\Gamma(S,A,B, L) = \BE[\chi_{S \setminus J} ( \B{X} \diamond \B{Y}) \cdot \theta(t, \B{U}, \B{V})]$ where $J = A \cup B$ and $\theta$ consists of the product of functions in $L$ evaluated at the appropriate coordinates. This will exactly give us the derivative $\partial_J \chi_S(z)$ evaluated at $\B{X} \diamond \B{Y}$. Next, towards bounding \eqref{eqn:start}, we compute the contribution of each leaf configuration $(A,B, L)$ in terms of these derivatives. Note that the same set  $J$ may correspond to multiple leaf configurations $(A,B, L)$ appearing with potentially different weights and different accompanying functions $\theta$.

\vspace{-2mm}

\paragraph{Contribution of a Leaf Configuration.}
Given a fixed $S$, let $(A^{(0)},B^{(0)}, L^{(0})$ denote some initial configuration, and consider some path in the branching tree starting from $(A^{(0)},B^{(0)}, L^{(0)})$ and ending in $(A^{(T)},B^{(T)}, L^{(T)})$. Consider a step on this path where the configuration changes from $(A^{(\tau)},B^{(\tau)}, L^{(\tau)})$ to $(A^{(\tau+1)},B^{(\tau+1)}, L^{(\tau+1)})$ and also recall that in either application of \lref{lem:br1} or \lref{lem:br2} at each step, there are two types of transitions --- either of type \textrm{I} or type \textrm{II}. We note that if  $(A^{(\tau+1)},B^{(\tau+1)}, L^{(\tau+1)})$ is derived from a type \textrm{I} transition, then $|A^{(\tau+1)} \cup B^{(\tau+1)}| = |A^{(\tau)} \cup B^{(\tau)}|$ while if it is derived from a type \textrm{II} transition, then  $|A^{(\tau+1)} \cup B^{(\tau+1)}| = |A^{(\tau)} \cup B^{(\tau)}|+1$ (recall \figref{fig:dynamics}).

The following lemma shows that each leaf configuration corresponds to a derivative of order between $k$ and $k(k-1)$ and also gives a bound on the contribution of each leaf configuration towards \eqref{eqn:start}.

\begin{lemma}\label{lem:potential}
    Consider the branching tree started at the initial configuration $(A^{(0)},B^{(0)}, L^{(0)})$. For any inactive leaf configuration $(A^{(T)},B^{(T)}, L^{(T)})$ in this branching tree, defining $J = J(A^{(T)},B^{(T)}, L^{(T)}) = |A^{(T)} \cup B^{(T)}|$, we have that $k \le |J| \le k(k-1)$. Moreover, in this case $|J| \left(1-\frac1k\right) -(k-1) \le T \le  3|J|$ and the list of functions $L^{(T)}$ only contains functions from the set $\{ \gamma^{(d)}, \Psi_{t_r}^{(d)} \mid d \in \{0\} \cup [k], r \in [k-1]\}$ where $h^{(d)}$ denotes the $d^\text{th}$ derivative of $h$. 

    Finally, there exists a function $\theta(t,u,v)$ satisfying $\max_{t,u,v} |\theta(t,u,v)| \le (4k)^{2k}$ such that
    \begin{equation}\label{eqn:finalconf}
    \ \wgt(S,A^{(T)},B^{(T)}, L^{(T)}) \cdot \Gamma(S,A^{(T)},B^{(T)}, L^{(T)}) = \left(\frac{1}{\sqrt{N}}\right)^{|J|\left(1-\frac1k\right)} \cdot  \BE[\chi_{S\setminus J}(\B{X}(t) \diamond \B{Y}(t)) \cdot \theta(t, \B{U}, \B{V})],
    \end{equation}
    where $\theta$ depends only on the path from $(A^{(0)},B^{(0)}, L^{(0)})$ to $(A^{(T)},B^{(T)}, L^{(T)})$ in the branching tree.
\end{lemma}
\begin{proof}[Proof of \lref{lem:potential}]
\newcommand{\ellm}{\beta}
    In the initial configuration $(A^{(0)},B^{(0)}, L^{(0)})$, recall that $|A^{(0)}_r|=|B^{(0)}_{r+1}| = 1$ for $r \in [k-1]$ and $|A^{(0)}_k|=|B^{(0)}_1|=0$.
Let $\ellm$ denote the number of blocks $r \in \{2,\ldots,k-1\}$ for which $|A^{(0)}_r\setminus B^{(0)}_r|=0$ (or equivalently $|A^{(0)}_r \cup B^{(0)}_r| =1$.

For a configuration $(A,B, L)$, consider the following potential
which nicely captures many properties of the dynamics of the configurations generated by the branching process.
\[ \CE(A,B, L)  = |A \cup B|  + \sum_{r=2}^{k-1}  (k-r) |B_r \setminus A_r|  + \sum_{r=2}^{k-1} (r-1) |A_r \setminus B_r|.\]
    \begin{claim}\label{clm:a}
    For an initial configuration, $\CE(A^{(0)},B^{(0)},L^{(0)}) = k(k-1-\ellm)$. 
\end{claim}
\begin{proof}
We compute each of the terms in the potential.
First
\[ |A^{(0)} \cup B^{(0)}| = 1 + 2(k-2 -\ellm) + \ellm + 1 = 2k-2 -\ellm\]
as $|A^{(0)}_1 \cup B^{(0)}_1|=|A^{(0)}_k\cup B^{(0)}_k|=1$, and
 for $r \in \{2,\ldots,k-1\}$, we have $|A^{(0)}_r \cup B^{(0)}_r|=1$ for $\ellm$ indices and $2$ for the remaining $k-2-\ellm$ indices.

We now consider the terms  $(k-r) |B_r \setminus A_r| + (r-1) |A_r \setminus B_r|$ for $r \in \{2,\ldots,k-1\}$. This contributes exactly  $(k-r) + (r-1) = k-1 $ whenever $B^{(0)}_r \neq A^{(0)}_r$, which happens for $k-2-\ellm$ indices, and contributes $0$ for all other indices. This gives
    \[ \CE(A^{(0)},B^{(0)}, L^{(0)})= 2k-2 -\ellm  + (k-2-\ellm) (k-1)  =  k(k-1 -\ellm). \qedhere \]
\end{proof}
Next, we show how $\CE$ evolves along any edge of the branching tree.
    \begin{claim}\label{clm:b}
    Consider any transition  $(A^{(\tau)},B^{(\tau)}, L^{(\tau)})$ to $(A^{(\tau+1)},B^{(\tau+1)}, L^{(\tau+1)})$.
    If this is type \textrm{I} transition then the potential decreases by at least one and at most $k$, otherwise for a type \textrm{II} transition the potential remains unchanged. \end{claim}
\begin{proof}
    We first consider the type \textrm{I} transition, and consider the setting of \lref{lem:br1}, where $i_q \in B_q \setminus A_q$. Then, $|A^{(\tau)} \cup B^{(\tau)}|$ does not change and $|B_q \setminus A_q|$ decreases by exactly $1$ (as $i_q$ is added to $A_q$) and $|A_{q+1} \setminus B_{q+1}|$ either decreases by exactly $1$  (if $j_{q+1} \in  A_{q+1} \setminus B_{q+1}$) or remains the same (if $j_{q+1} \in B_{q+1}$). Thus, the only change in the potential comes from the contribution of the corresponding terms.

    Thus, the potential change $\CE(A^{(\tau+1)},B^{(\tau+1)})  - \CE(A^{(\tau)},B^{(\tau)})$ is either $-(k-q)  - (q+1-1) = -k$ when both terms decrease, or $-(k-q)$ otherwise. Since $q \in \{2, \cdots, k-1\}$, it follows that the potential decreases by at least $1$ and at most $k$ for a type \textrm{I} transition. An exactly analogous argument works for type \textrm{I} transition corresponding to the setting of \lref{lem:br2}.

For type \textrm{II} transition, again consider the setting of  \lref{lem:br1}.
Then $|A^{(\tau+1)} \cup B^{(\tau+1)}| = |A^{(\tau)} \cup B^{(\tau)}|+1$. As $j_{q+1}$ is added to $B^{(\tau)}_{q+1}$ (and $j_{q+1}$ is not in $A^{(\tau)}_{q+1}$), 
the quantity $|B^{(\tau)}_{q+1} \setminus A^{(\tau)}_{q+1}|$ increases by $1$,
and as $i_q$ is added to $A^{(\tau)}_q$ (and $i_q$ is in $B^{(\tau)}_q\setminus A^{(\tau)}_q$ before it is added to $A^{(\tau)}_q$) the quantity $|B^{(\tau)}_{q+1} \setminus A^{(\tau)}_{q+1}|$ decreases by $1$. As the coefficient of these terms in the potential is $k-(q+1)$ and $k-q$ respectively, overall we have that
\[ \CE(A^{(\tau+1)},B^{(\tau+1)})  - \CE(A^{(\tau)},B^{(\tau)}) = 1 + (k-(q+1)) - (k-q)  = 0 \]
The setting of \lref{lem:br2} is exactly analogous.
\end{proof}

    The next claim concerns type \textrm{I} transitions and the structure of the list $L$.

    \begin{claim}\label{clm:c} There can be at most $2k-2$ type \textrm{I} transitions from any initial configuration to any final leaf configuration. Moreover, for any function $\lambda \in L^{(T)}$ either $\lambda=\gamma^{(d_\lambda)}$ or $\lambda=\Psi_{t_q}^{(d_\lambda)}$ for some non-negative integer $d_\lambda$ and $q \in \{2, \ldots, k-1\}$. Moreover, the total order of all the derivatives in $L$ satisfies $\sum_{\lambda \in L^{(T)}} d_\lambda \le 2k-2$. 
\end{claim}
    \begin{proof} For any $q \in \{2, \ldots, k-1\}$, let us call an index $i_q \in B_q\setminus A_q$ an active-$B$ index and any index $i_q \in A_q\setminus B_q$ an active-$A$ index. Note that in any initial configuration there are at most $k-1$ active $A$ and $B$ indices. We claim that the number of active indices remain the same on type \textrm{II} transitions and decrease by $1$ on type \textrm{I} transitions. To see this, consider the setting of \lref{lem:br1} where an active-$B$ index $i_q \in B_q \setminus A_q$ is chosen. 
        \begin{itemize}
            \item If it is a type \textrm{II} transition, then the active index $i_q \in B_q \setminus A_q$ is removed and another active-$B$ index $j_{q+1}$ is added to $B_{q+1}\setminus A_{q+1}$, so the number of active indices does not change.
            \item If it is a type \textrm{I} transition, then the active index $i_q \in B_q \setminus A_q$ is removed, but no new active indices are added, so the number of active indices decrease by $1$.
        \end{itemize}
        The setting of \lref{lem:br2} is analogous. Since there are no active indices in a leaf configuration, it follows that there can be at most $2k-2$ type \textrm{I} transitions proving the first part of the claim. 

        To see the second part of the claim, we note that in any type \textrm{II} transition, two new elements $\gamma$ and $\Psi_{t_q}$ are added to the list (the update is according to \eqref{eqn:typetwo1} or \eqref{eqn:typetwo2}) and in a type \textrm{I} transition, one of the elements of the list is updated to its derivative and a new factor $\Psi_{t_q}$ is added (the update is according to \eqref{eqn:typeone1} or \eqref{eqn:typeone2}). It follows that the list only contains derivatives of $\gamma$ or $\Psi_{t_q}$'s and the total order of the derivatives is the number of type \textrm{I} transitions which is at most $2k-2$ from the first part of the claim.
    \end{proof}

We can now prove the lemma.
First, note that for an inactive configuration
as $|A_r \setminus B_r| = |B_r \setminus A_r |=0$ for $r \in \{2,\ldots,k-1\}$, and hence, the value of the potential is exactly $|A\cup B|$.

    To see the first statement, consider some path from $(A^{(0)},B^{(0)},L^{(0)})$ to $(A^{(T)},B^{(T)}, L^{(T)})$. As $|A^{(0)},B^{(0)}| = k$ and any transition can only add new elements to $J$, so the quantity $|J|$ can only increase and $|J| = |A^{(T)} \cup B^{(T)}| \ge k$.
    
    To see that $|J|\le k(k-1)$, we use \clmref{clm:a} and \clmref{clm:b} as follows:
    \[ |J|=|A^{(T)} \cup B^{(T)}| = \CE(A^{(T)},B^{(T)}, L^{(T)}) \le  \CE(A^{(0)},B^{(0)}, L^{(0)}) = k(k-1-\ellm)  \le k(k-1).\]
 
    Next we prove the bounds on depth of the branching process $T$. For this let $\nu_1$ be the number of type \textrm{I} transitions along the path, and observe that $|A\cup B|$ rises by $1$ exactly for $T-\nu_1$ steps (at type \textrm{II} transitions). 
As  $|A^{(0)}\cup B^{(0)}| = 2k-2-\ellm$ initially, we have that 
\begin{equation}
\label{eq:t-nu}
    T-\nu_1 = |J|  - (2k-2-\ellm).
\end{equation} 
This gives the upper bound that $T \le \nu_1 + |J| \le 3|J|$, as  $\nu_1 \le 2k-2$ by \clmref{clm:b}, which is at most $2|J|$ as  $|J| \geq k$.

To obtain the lower bound on $T$, we shall show that  
\[T+(k-1) = |J|  - (k-1-\ellm- \nu_1) \ge |J| - \frac{|J|}{k} = |J|\left(1-\frac1k\right),\]
where the first equality follow from \eqref{eq:t-nu}. The inequality follows by observing that $(k-1-\ellm- \nu_1) 
\leq |J|/k$ for the following reason.
By \clmref{clm:a} the initial potential is $k(k-1-\ellm)$, and the total decrease in the potential from \clmref{clm:b} can be at most $\nu_1k$, and hence   
\[ |J|= \CE(A^{(T)},B^{(T)}, L^{(T)}) \ge \CE(A^{(0)},B^{(0)}, L^{(0)}) - \nu_1 k = k(k-1-\ellm) - \nu_1 k = k(k-1-\ellm - \nu_1).\]
This gives the lower bound on $T$. 

The statement regarding the structure of the list $L$ follows directly from \clmref{clm:c}, so all that remains to prove is \eqref{eqn:finalconf}.

Note the weight of the initial configuration  $(A^{(0)},B^{(0)})$ consists of a product of $k-1$ terms of the form $\frac{H_{i_\kappa,j_{\kappa+1}}}{1+t_\kappa}$ where $\kappa \in [k-1]$. By the definition of weight of a child node, at each step of the branching process, we gain exactly one $H_{i_q,j_{q+1}}t_q$ factor in the weight. It follows that the weight of the leaf configuration $(A^{(T)},B^{(T)}, L^{(T)})$ equals
\begin{equation}\label{eqn:poly1}
    \ \wgt(A^{(T)},B^{(T)}, L^{(T)}) = \epsilon \cdot p(t_1,\dots,t_{k-1}) \cdot \left(\frac{1}{\sqrt{N}}\right)^{T+(k-1)},
\end{equation}
 where $\epsilon$ is the sign of corresponding products of Hadamard entries and $p(t)$ is a non-negative function of $t$ always bounded by $1$. Next we note that 
\begin{equation}\label{eqn:poly}
    \ \Gamma(A^{(T)},B^{(T)}, L^{(T)}) =  \BE[\chi_{S\setminus J}(\B{X}(t) \diamond \B{Y}(t)) \cdot \mu(t, \B{U}, \B{V}),]
\end{equation}
where $\mu(t,u,v)$ is the product of at most $2|J|$ functions in the list $L^{(T)}$. Using \clmref{clm:c}, each function in the list is either some derivative of $\gamma$ or $\Psi_{t_q}$ where the total order $m \le 2k$ (note that derivatives of order zero are $\gamma$ or $\Psi_{t_q}$ and they are always bounded by one). Using \eqref{eqn:hermitebound} and \lref{lem:intbypartsphi}, it then follows that $\max_{t,u,v} |\mu(t,u,v)| \le m^{m/2} \le (4k)^{2k}$. 

Defining $\theta(t,u,v) = \epsilon \cdot p(t)\cdot \mu(t,u,v)$ and plugging in the lower bound of $T+(k-1) \ge |J|\left(1-\frac1k \right)$ in \eqref{eqn:poly} gives us \eqref{eqn:finalconf} in the statement of the lemma.
\end{proof}

\paragraph{Bounding the Total Contribution.} Using \pref{prop:total} and \lref{lem:potential}, we get that the left hand side in \eqref{eqn:start} equals
\begin{equation}\label{eqn:finalbound}
    \  \sum_{\ell=k}^{k(k-1)}  \left(\frac{1}{\sqrt{N}}\right)^{\ell\left(1-\frac1k\right)} \cdot \sum_{\substack{J \subseteq S\\|J|=\ell}} \BE[\chi_{S\setminus J}(\B{X}(t) \diamond \B{Y}(t)) \cdot \theta_{S,J}(t, \B{U}, \B{V}) ], 
\end{equation}
where $\theta_{S,J}(t, u,v)$ is a function which is determined by the collection of paths in all the branching trees that lead to a leaf configuration $(A,B, L) \in \CL(S)$ satisfying $A \cup B = J$. Moreover, the maximum value of $|\theta_{S,J}(t,u,v)|$ is bounded by times the number of such paths times the factor $(4k)^{2k} \le (4k)^{2|J|}$. 

To finish the proof, we argue that $\theta_{S,J}$ only depends on $J$ and we also bound how many leaf configurations correspond to the set $J$ via an encoding argument.

\begin{lemma}\label{lem:enc}
    $\theta_{S,J}(t)$ depends only on $J$ and not on $S$. Moreover, there are at most $(4k)^{12|J|}$ leaf configurations $(A,B, L)$ in the multiset $\CL(S)$ for which $A \cup B = J$ and hence, we have that $\max_{t,u,v} |\theta_J(t,u,v)| \le (4k)^{14|J|}$.
\end{lemma}

\begin{proof} 

    Let us write $J = J_1 \sqcup J_2 \sqcup \ldots \sqcup J_k$ where $J_r \subseteq S_r$ for each $r \in [k]$. First, we note that the branching process only adds elements to the configuration and never removes them. Therefore, all the paths that lead to leaf configurations $(A,B, L)$ satisfying $A \cup B = J$ can only contain intermediate configurations $(A',B', L')$ where the sets $A'_r, B'_r \subseteq J_r$. Moreover, since there is a branching tree for all initial configurations $(A^{(0)},B^{(0)}, L^{(0)})$ satisfying $|A^{(0)}_r|=|B^{(0)}_{r+1}| = 1$ for $r \in [k-1]$ and $|A^{(0)}_k|=|B^{(0)}_1|=0$ where $A_r,B_r \in S_r$, it follows that to determine the collection of paths that lead to a leaf configuration $(A,B) \in \CL(S)$ satisfying $A \cup B = J$, we can assume without any loss of generality that $S_r=J_r$ for every $r \in [k]$. This proves that $\theta_{S,J}(t,u,v)$ only depends on $J$ and not on $S$ as it is determined by this collection of paths.

    Next, we bound the number of paths in this collection. We will describe an encoding that stores at most $\log_2((4k)^{4|J|})$ bits and uniquely determines the entire path of the branching process from the initial configuration $(A^{(0)}, B^{(0)}, L^{(0)})$ to the final leaf configuration $(A^{(T)},B^{(T)}, L^{(T)})$ for which $A^{(T)} \cup B^{(T)}=J$. From this, it follows that the number of leaf configurations $(A,B, L)$ for which $A \cup B=J$ is at most $(4k)^{4|J|}$.

    For the encoding, we initialize a bit-string of length $2|J|$ to the indicator vectors of the subsets $A^{(0)}_1, \cdots, A^{(0)}_k$ and $B^{(0)}_1, \cdots, B^{(0)}_k$ where the initial configuration is $(A^{(0)},B^{(0)}, L^{(0)})$. We only need $2|J|=2\sum_{r=1}^k|J_r|$ bits as we only need to store subsets of each $J_r$ and the list $L^{(0)}$ is also determined as it only consists of the function $\gamma$ indexed by the elements of the subsets. This bit-string will be updated at every step of the branching process, along with some auxiliary information.

At time $\tau \in [T]$, the configuration is updated from $(A^{(\tau-1)}, B^{(\tau-1)})$ to $(A^{(\tau)},B^{(\tau)})$ using either \lref{lem:br1} or \lref{lem:br2}. In each case, for exactly one $r \in \{2, \cdots, k-1\}$, $A_r$ is updated to $A_r \cup \{i_r\}$ and $B_{r+1}$ is updated to $B_{r+1} \cup \{j_{r+1}\}$. To reconstruct this information, we store the following:
\begin{itemize}
    \item We update the two $|J|$-length bit-strings to store the indicator vectors of the configuration $A^{(\tau)},B^{(\tau)}$ at time $\tau$. This requires changing a zero bit to a one bit in each of the two bit-strings as we only ever add elements to the sets. 
    \item We also store the indices of the two locations where the above bit-strings were updated. This requires exactly $2\lceil \log_2 |J| \rceil$ bits. Note that which set $A_r$ or $B_{r'}$ was updated is also determined by the indices. Moreover, we record the $2\lceil \log_2 |J| \rceil$ indices in order, so the exact time $\tau$ when the bits were written is also determined by this information. Finally how the list $L$ was updated (as per \eqref{eqn:typetwo1} or \eqref{eqn:typetwo2}, or, as per \eqref{eqn:typeone1} or \eqref{eqn:typeone2}) is also determined by this information. 
\end{itemize}

Overall, given the above information, one can uniquely determine the exact path from $(A^{(0)},B^{(0)})$ to $(A^{(T)},B^{(T)})$. The total number of bits of information is at most $2|J| + T (2 + 2\lceil \log_2 J \rceil)$.

    Now, from \lref{lem:potential}, it follows that $T \le 3|J|$ and that $|J|\le k(k-1) \le k^2$, so the total number of bits information is at most 
\[ 6|J|\log_2 |J| + 14|J| \le 12|J|\log_2 k + 14|J| = \log_2 (2^{14|J|}k^{12|J|}) \le \log_2((4k)^{12|J|}). \qedhere\]
\end{proof}

Using the above lemma in conjunction with \eqref{eqn:finalbound} completes the proof of \lref{lemma:monomial} for an arbitrary $k$.

\subsection{Fourier Weight under Biased Measures}

We use a random restriction argument to prove \thmref{thm:fwt}. Recall the basic notation about random restrictions introduced in \secref{sec:fourier}.

\fwt*

\begin{proof}
   Define the following product distribution over restrictions $\rho \in \{-1, 1,  \star\}^m$,
    \begin{align*}
        \rho_i = \begin{cases} \star &\text{ with probability }~ {(1-\mu_i^2)}/2 = \sigma_i^2/2,\\
                 1   &\text{ with probability } ~{(1+\mu_i)^2}/4,\\ 
                 -1   &\text{ with probability } ~{(1-\mu_i)^2}/4.  
        \end{cases}
    \end{align*}
    For notational convenience, let us abbreviate $\sigma_W = \prod_{i \in W} \sigma_i$ and $\mu_W = \prod_{i \in W} \mu_i$ for $W \subseteq [m]$. Then, the Fourier coefficient of $f_\rho$ under the uniform measure and of $f$ under the bias $\mu$ are related by the following claim.
\begin{claim}\label{claim:restriction}
    Let $S \subseteq [m]$. Then, we have $ \BE[\fhat_\rho(S)] = 2^{-|S|}\sigma_S \cdot  \fhat^\mu(S)$ where the expectation is taken over $\rho$.
\end{claim}
    Given the above claim, we can finish the proof of \corref{cor:wt} as follows. Let us define $\wgt^\mu_\ell(f,\theta) := \sum_{|S|=\ell} \theta_S \fhat^\mu(S)$ for any sequence of signs $\theta := (\theta_S)_{|S|=\ell}$. Then, using \clmref{claim:restriction} and taking expectation over $\rho$, we obtain
    \begin{align*}
         \ \wgt^\mu_\ell(f,\eps) = \sum_{|S|=\ell} \theta_S \cdot 2^{\ell}\sigma_S^{-1} \cdot \BE[\fhat_\rho(S)] &= \BE\bigg[ \sum_{|S|=\ell} \theta_S \cdot 2^{\ell}\sigma_S^{-1} \cdot \fhat_\rho(S)\bigg] \\
         \ & \le \BE\bigg[ \sum_{|S|=\ell}  2^{\ell}\sigma_S^{-1} \cdot |\fhat_\rho(S)|\bigg] \le  4^{\ell} \cdot \BE[\wgt_\ell(f_\rho)] \le 4^\ell w.
    \end{align*}
 The second last inequality above follows since $\sigma_i = \sqrt{1-\mu_i^2} \ge 1/2$ as $\mu \in [-1/2,1/2]^m$, and the last inequality uses our assumption on the Fourier weight of the restricted function $f_\rho$ under the uniform measure. Since the above is true for an arbitrary sequence of signs $\theta$, it follows that 
    \[ \wgt^\mu_\ell(f) \le 4^\ell w.\]
This finishes the proof assuming \clmref{claim:restriction} which we prove next.
\end{proof}

\begin{proof}[Proof of \clmref{claim:restriction}]
        We note that for any subset $S \subseteq [m]$,
    \[ \fhat_\rho(S) = \sum_{T: T \supseteq S} \fhat(T)  \cdot \ind[{S \subseteq \free(\rho) \text{ and } T\setminus S \subseteq \fix(\rho)}] \cdot \chi_{T \setminus S}(\rho).\]
    Taking expectation over the random restriction $\rho$, we get that
    \begin{equation}\label{eqn:fhatrho}
          \ \BE[\fhat_\rho(S)] = \sum_{T: T \supseteq S} \fhat(T)  \cdot \prod_{i \in S} \frac{\sigma_i^2}{2} \cdot  \prod_{i \in T\setminus S} \Big(\frac{(1+\mu_i)^2}{4} - \frac{(1-\mu_i)^2}{4}\Big) = \sum_{T: T \supseteq S} \fhat(T) \cdot 2^{-|S|}\sigma_S^2 \cdot \mu_{T\setminus S}.
    \end{equation}
    Next, recalling that the Fourier basis with respect to bias $\mu$ is given by $\phi_S(x) = \prod_{i\in S} \frac{(x_i - \mu_i)}{\sigma_i}$, we have
     \[ \fhat^\mu(S) = \BE_{p_\mu}[f(X) \phi_S(X)] =  \sum_{T \subseteq [m]} \fhat(T) \cdot \sigma_S^{-1} \cdot \BE_{p_\mu}\Big[\chi_T(X) \cdot \prod_{i \in S} (X_i - \mu_i)\Big].\]
     Since $\BE_{p_\mu}[X_i]=\mu_i$, it follows that all the terms above where $S \setminus T$ is not the empty set are zero. Moreover, $\BE_{p_\mu}[X_i (X_i-\mu_i)] = 1 - \mu_i^2 = \sigma_i^2$. Therefore, 
     \begin{equation}\label{eqn:fhatmu}
         \ \fhat^\mu(S) =  \sum_{T: T \supseteq S} \fhat(T) \cdot \sigma_S \cdot \mu_{T\setminus S}.
     \end{equation}
      Comparing \eqref{eqn:fhatrho} and \eqref{eqn:fhatmu} gives us the claim.
\end{proof}

\subsection{Proof of Main Lower Bound: \thmref{thm:lower} and \corref{cor:smallerr}}

Given \corref{cor:wt} and \thmref{thm:level}, the proof is straightforward. 

\begin{proof}[Proof of \thmref{thm:lower}]
        Given a randomized decision tree of depth $d$ that has advantage $\eta$, we first amplify the success probability of the decision tree to $1-\delta$, by making $\tau =\Theta(\eta^{-2}\log(1/\delta))$ repetitions and taking the majority vote. Since the error of this randomized decision of $\Theta(d \tau)$ depth is at most $\delta$ on each valid input, we have that for large enough $N$,
        \begin{equation}\label{eqn:advlb}
         \ \left|\BE_{p_1}[f(Z)] - f(0)\right| \ge 6\delta - \frac{\delta^2}{4N} - 2\delta \ge \delta, \\
        \end{equation} 
        because of \thmref{thm:input}.\\
        
       Next, we will show a contradiction to the above statement if the depth $d$ was too small. In particular, applying \thmref{thm:level} and \corref{cor:wt} to the decision tree of depth $d_1=\Theta(d\tau)$, we obtain
       \begin{align*}
           \ \left|\BE_{p_1}[f(Z)] - f(0)\right| &\le  \sup_{\mu \in \left[-\frac12,\frac12\right]^{kN}} \sum_{\ell=k}^{k(k-1)} \left(\frac{1}{\sqrt{N}}\right)^{\ell\left(1-\frac1k\right)}\cdot (8k)^{14\ell} \cdot \wgt^{\mu}_{\ell}(f) \\
            \ &\le \sum_{\ell=k}^{k(k-1)} \left(\frac{1}{\sqrt{N}}\right)^{\ell\left(1-\frac1k\right)} \cdot (8k)^{14\ell} \cdot \sqrt{d_1^{\ell}\log^{\ell-1}(kN)}\\
            \ & \le \sum_{\ell=k}^{k(k-1)} \left(cd_1k^{28} \cdot \left(\frac{\log(kN)}{N}\right)^{1-1/k}\right)^{\ell/2},
       \end{align*}
       for a universal constant $c$. Thus, if 
       $$d_1 < C \cdot \frac{1}{k^{28}}\cdot \left(\frac{N}{\log (kN)}\right)^{1-1/k},$$
       for a suitable constant $C$, then the advantage of the decision tree on the input distribution $p(Z)$ is at most $2^{-5k}/4 = \delta/4$ which contradicts \eqref{eqn:advlb}. This gives us that $d$ must be at least $\Omega\left(\frac{1}{\tau k^{28}}\left(\frac{N}{\log(kN)} \right)^{1-1/k}\right)$ giving us the bound in the statement of the theorem after substituting the value of $\tau$. \qedhere
\end{proof}

\corref{cor:smallerr} can be obtained analogous to the above.

\section*{Acknowledgements}

We thank Ronald de Wolf for discussions throughout the course of this work and for providing helpful feedback on the writing. We also thank Avishay Tal for very useful comments and for encouraging us to extend our results from the setting where $\delta = 1/\polylog^k(N)$ to one where $\delta = 2^{-O(k)}$. We thank Arkadev Chattopadhyay and Suhail Sherif as well, for pointing out the applications to communication complexity.

\bibliographystyle{alpha}
{\bibliography{references.bib}}

\end{document}